%% file: pencat0.tex
\documentclass[a4paper]{article}
\usepackage{color}
\usepackage{graphicx}
\usepackage{psfrag}
\usepackage{amsmath,amssymb,amsthm}
\usepackage{eufrak}
\renewcommand{\thesection}{\Roman{section}}
\newcommand{\N}{{\mathbb N}}
\newcommand{\R}{{\mathbb R}}
\newcommand{\Z}{{\mathbb Z}}
\newcommand{\T}{{\mathbb T}}
\newcommand{\C}{{\mathbb C}}
\newcommand{\Q}{{\mathbb Q}}

\DeclareMathOperator{\Int}{Int}
\DeclareMathOperator{\Ran}{Ran}
\DeclareMathOperator{\id}{id}
\DeclareMathOperator{\diag}{diag}
\DeclareMathOperator{\Rank}{Rank}
\newcommand{\norm}[1]{\lVert#1\rVert}
\newcommand{\blic}[1]{\raisebox{-2pt}{$\bullet \negthickspace\!
\overset{#1}{- \negthickspace -} \negthickspace\! \circ$}}
\newcommand{\clic}[1]{\raisebox{-2pt}{$\circ \negthickspace\!
\overset{#1}{- \negthickspace -} \negthickspace\! \circ$}}
\newcommand{\barb}[1]{\raisebox{-2pt}{$\bullet \negthickspace\! \overset
{#1} {\to \negthickspace \negthickspace -} \negthickspace\! \bullet$}}
\newcommand{\carc}[1]{\raisebox{-2pt}{$\circ \negthickspace\! \overset
{#1} {\to \negthickspace \negthickspace -} \negthickspace\! \circ$}}
\newtheorem{theorem}{Theorem}
\newtheorem{statement}[theorem]{Statement}
\newtheorem{lemma}[theorem]{Lemma}
\newtheorem{definition}[theorem]{Definition}

\begin{document}
\begin{titlepage}

\title{Penrose Tilings, Chaotic Dynamical Systems and Algebraic $K$-Theory}

\author{Tam\'as Tasn\'adi\thanks{E\"otv\"os University, Department
of Solid State Physics, H--1117 Budapest, P\'azm\'any P\'eter
s\'et\'any 1/A, Hungary. (On the leave of the Research Group
for Statistical Physics of the Hungarian Academy of Sciences.)}}

\date{April 09, 2002}

\maketitle
\thispagestyle{empty}

\begin{abstract}
In this article we initiate the use of noncommutative geometry in the
theory of dynamical systems.

After investigating by examples the unusual and striking elementary
properties of the Penrose tilings and the Arnold cat map, we associate a
finite symbolic dynamics with finite grammar rules to each of them. Instead
of studying these Markovian systems with the help of set-topology, which
would give only pathological results, a noncommutative approximately
finite $C^*$-algebra is associated to both systems. By calculating the
$K$-groups of these algebras it is demonstrated that this noncommutative
point of view gives a much more appropriate description of the phase space
structure of these systems than the usual topological approach.

With these specific examples it is conjectured that the methods of
noncommutative geometry could be successfully applied to a wider class of
dynamical systems.

\end{abstract}

\vspace{0.5cm}
\begin{center}
\includegraphics[scale=0.4]{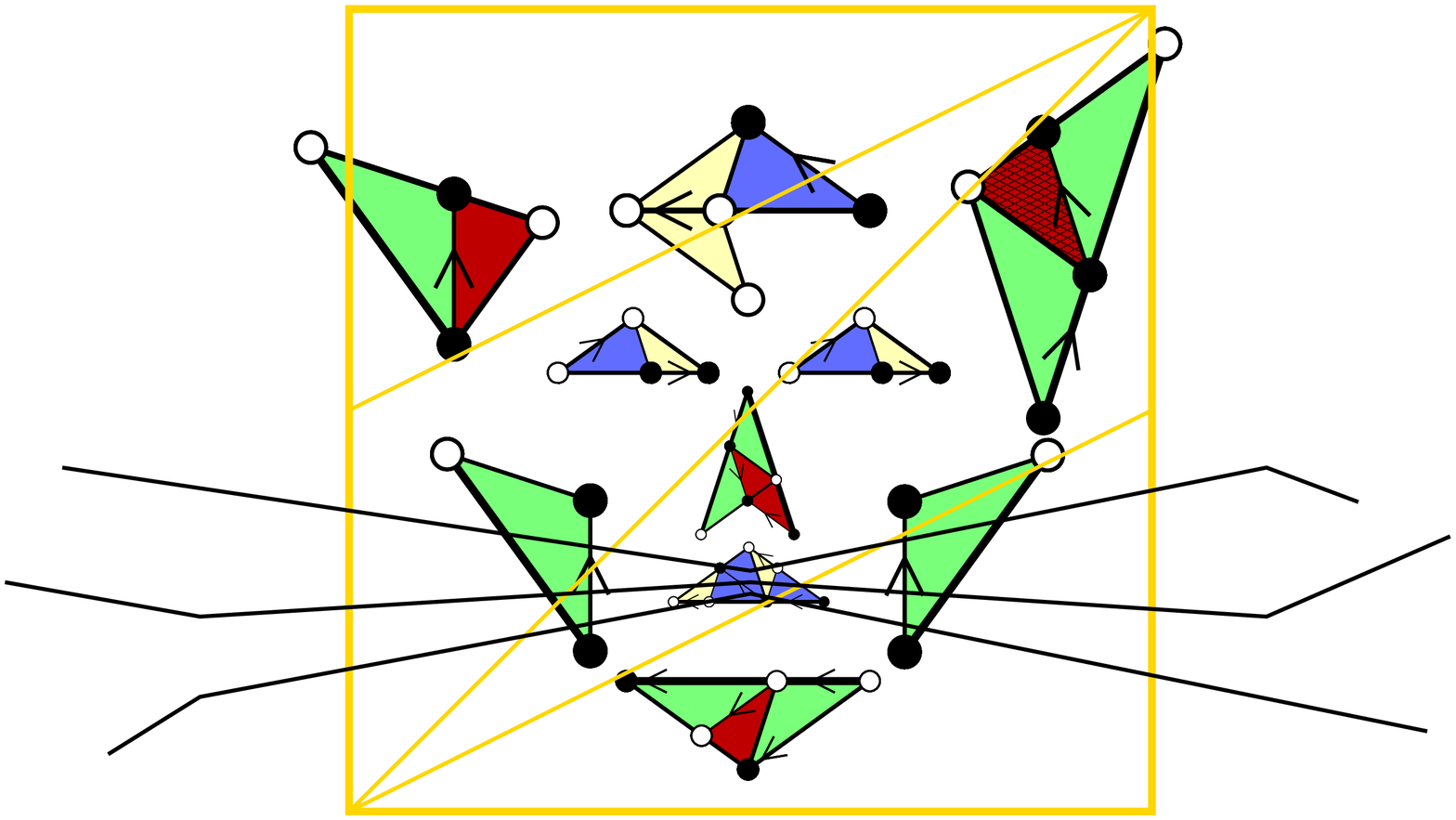}
\end{center}

\end{titlepage}

\pagebreak
\tableofcontents
\pagebreak

\include{pencat1}

\include{pencat2}

\include{pencat3}

\include{pencat4}
\include{pencat5}

\renewcommand{\theequation}{\thesection \arabic{equation}}
\setcounter{equation}{0}
\include{pencata}

\setcounter{equation}{0}
\include{pencatb}

\addcontentsline{toc}{section}{References}
\bibliographystyle{alpha}
\bibliography{pencat}

%\clearpage
%\listoffigures

\end{document}

%% file: pencat1.tex
\section{Introduction}
%=====================
The most important aim of this work is to initiate the use of the methods
of noncommutative geometry \cite{Con0} in the theory of dynamical systems
\cite{A_A,CoFoSi82,Walt82}.

As a preliminary step, in this article we investigate the properties of two
specific, well known and intensively studied systems, which hallmark the
above mentioned two areas of mathematics and mathematical physics, and we
highlight the deep structural similarity between these systems. The first
one is the universe of `Penrose tilings' \cite{Pen74}, which appears as one
of the introductory examples in the book of A.\ Connes \cite{Con0}, and the
second system is Arnold's cat map, which is one of the simplest basic
example for uniformly hyperbolic chaotic systems \cite{A_A}. The connecting
bridge between these at first sight so far from each other lying systems is
the fact that the structure of both ones can be encoded with symbolic
sequences of letters from a finite alphabet, obeying a finite number of
grammar rules.

This Markovian property enables us to associate a noncommutative,
approximately finite dimensional $C^*$-algebra to these systems, and using
one of the powerful weapons from the arsenal of noncommutative geometry,
namely the algebraic $K$-theory \cite{WeggOl93,Dav96}, we demonstrate that
these $C^*$-algebras, indeed, carry important information about the
structure of the original systems.

In the second section we set off by exploring the surprisingly rich universe
of aperiodic Penrose tilings \cite{GruS89}. First the basic properties of
the tilings (as local indistinguishability) are summarized and proved with
elementary methods. Then the construction of the associated symbolic
sequences are explained. It turns out that the topology of the space of all
Penrose tilings is rather pathologic considered from the point of view of
ordinary set-topology.

In the third section similar investigations are performed for the cat map;
first its elementary properties are studied, and then a Markov partition of
the phase space is constructed, and the grammar rules of the symbolic
dynamics are given explicitely. This symbolic coding clearly reveals the
basic similarity between the cat map and the Penrose universe.
Particularly, the phase space structure of the cat map, with the embedded
stable and unstable manifolds determined by the dynamics, is topologically
just as much pathological as the universe of Penrose tilings. This
topological defect manifests itself, on the one hand, in the local
isometric property of the Penrose tilings, and on the other hand, in the
chaotic (uniformly hyperbolic) dynamics of the cat map.

Although the above introduced spaces are ill-behaved as ordinary
topological spaces, from the point of view of noncommutative geometry they
are very interesting spaces with nontrivial properties. This approach is
the subject of the fourth section. First a noncommutative, approximately
finite dimensional $C^*$-algebra is associated to the Penrose universe as
well as to the cat map, and then an important invariant, the $K_0$ group
(with its scale and order structure) is explicitely calculated for both
systems. Comparing these results with the elementary properties of the
systems discussed in the second and third sections, we see that in both
cases these groups do reveal important topological invariants of the
original systems considered.

We conclude by expressing the hope that the methods of noncommutative
geometry, demonstrated here only for the simplest systems, could also be
applied for more difficult cases as not uniformly hyperbolic chaotic
systems or symbolic sequences with pruning (i.e. with infinitely many
grammar rules).

Since the fourth section is technically more demanding than the previous
ones, two appendices help to understand the properties of AF
$C^*$-algebras as well as the  basic concepts and methods of
algebraic $K$-theory.

%% file: pencat2.tex
\section{Aperiodic tilings\label{SPen}}
%======================================
%color codes for the figures in XFIG:
%user color 32: 676FFF (blue)
%user color 33: FFFFB7 (light yellow)
%user color 34: BA0000 (dark red)
%user color 35: 7CFF7C (light green)

A {\it tiling} or {\it tessellation} of the Euclidean plane $\R^2$ {\it of
type} ${\mathcal T} =\{ T_1, T_2 \dots T_n \}$ is an infinite partition of
the plane into pieces congruent to one of the {\it prototiles} $\{ T_i
\}_{i=1}^n$ \cite{GruS89}. We stress that the set ${\mathcal T}$ of
prototiles is always a finite set, the tiles must not overlap and there
should not remain any uncovered area of the plane. We do not require strict
congruence, i.e., the tiles must have the same shape and size as the
prototiles, but they can be reflected. Sometimes it is convenient to color
the vertices or to direct the edges of the prototiles and so impose {\it
matching conditions} for the tessellations investigated. A given tiling is
{\it aperiodic} if it does not possess translational symmetry.

In the first part of this section we summarize the basic properties of the
Penrose tilings \cite{Gar77,Pen78} of type ${\mathcal P} =\{ L, S \}$, which 
were introduced by R.~M.~Robinson in 1975 (see references in \cite{GruS89}),
motivated by the work of Penrose \cite{Pen74},
and then it is shown how the tessellations of type $\mathcal P$ can be
encoded with symbolic sequences. This coding scheme is the starting point
of the noncommutative geometrical investigations of Section \ref{Snoncom}.

\subsection{Elementary properties of the Penrose tilings\label{SPen_a}}
\begin{figure}
\centerline{\includegraphics[scale=0.4]{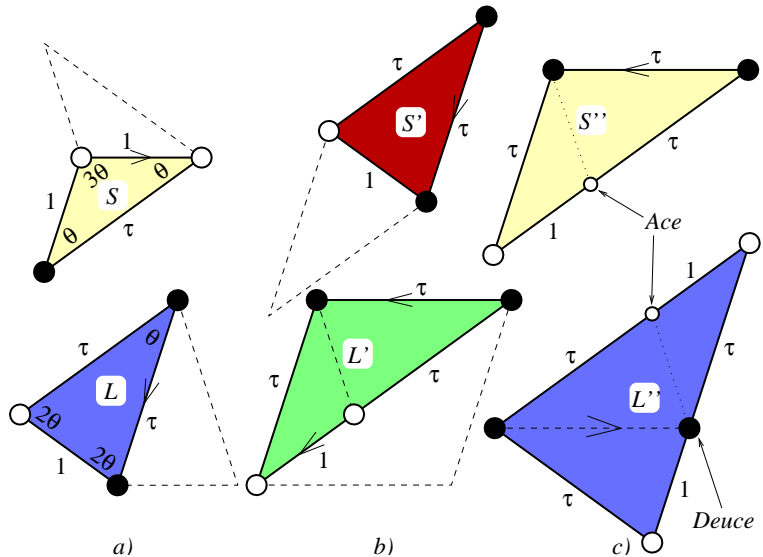}}
\caption{\label{FPen1} {\it a)} The prototiles of the Penrose tiling of
type $\mathcal P$ ($\theta =\frac{\pi}{5}$, $\tau =\frac{1+\sqrt{5}} {2}$);
{\it b)-c)} The prototiles obtained after successive compositions.}
\end{figure}

%----------------------------------------------------------------------
Let us consider the tilings of type ${\mathcal P} =\{ L, S\}$, the
prototiles of which are two isosceles triangles depicted in Fig.
\ref{FPen1}.{\it a)}. (The angles are multiples of $\theta=\frac{\pi}
{5}$, and the ratio of the length of the edges is the golden mean
$\tau= (1+\sqrt{5}) /2$.) As it is shown in the figure, there are matching
conditions for the vertices and for some of the edges. A finite portion of
a possible tiling of type $\mathcal P$ is pictured in Fig.~\ref{FPen6}.{\it
a)}.

\begin{figure}
\centerline{\includegraphics[scale=0.5]{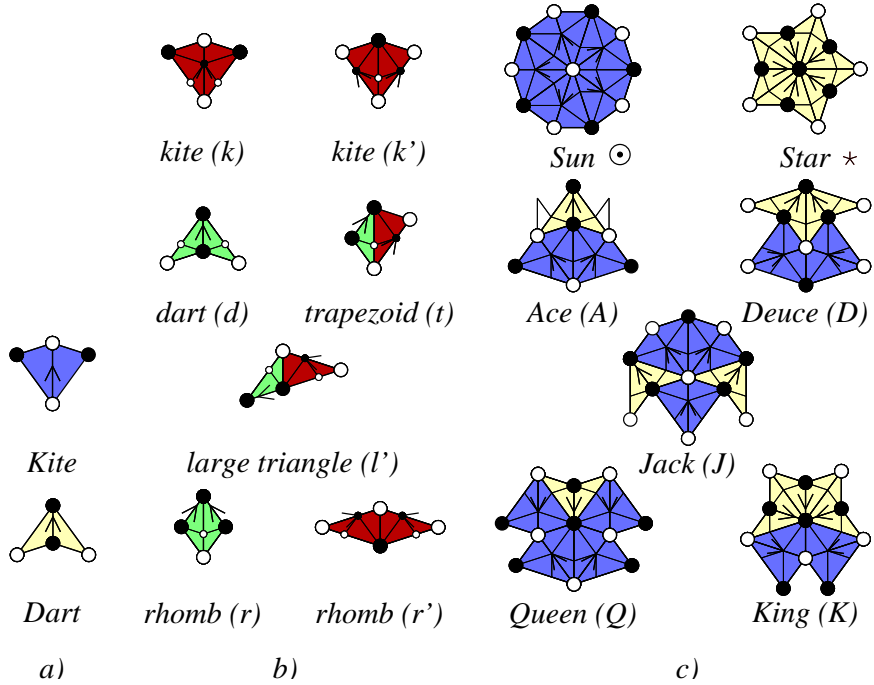}}
\caption{\label{FPen2} {\it a)} The {\it Kite} and {\it Dart} motifs. {\it
b)} The seven {\it edge neighborhoods}. {\it c)} The seven {\it vertex
neighborhoods}. (The thin lines denote the result of a double
decomposition.)}
\end{figure}

It is easy to observe that the directed edges of the tiles match only with
themselves, so the prototiles $L$ and $S$ always occur in pairs, forming a
{\it Kite} and a {\it Dart} like figure (Figs. \ref{FPen1}.{\it a)},
\ref{FPen2}.{\it a)}). The matching conditions for the colors also strongly
restrict the possibilities. It can easily be verified that in the immediate
vicinity of a given edge resp. vertex only seven basically different
arrangements occur. These {\it edge} resp. {\it vertex neighborhoods} are
shown in Fig.~\ref{FPen2}.{\it b)-c)} with their fantasy names and symbols.

As a consequence of these restrictions, only the large triangle $L$ can be
put beside the (nondirected) edge \blic{1}\ of the small triangle $S$, as
it is shown in Fig.~\ref{FPen1}.{\it b)}. It means that by {\it composing}
the tile $S$ together with its neighboring tile $L$ everywhere in a given
tiling of type ${\mathcal P}$ (i.e., by erasing the edges \blic{1}\ between
the tiles $S$ and $L$) a new tessellation of type ${\mathcal P}' =\{ S',L'
\}$ is obtained, where the smaller triangle $S'=L$ has acute angle and the
larger one $L' =S \cup L$ has obtuse angle (Fig. \ref{FPen1}.{\it b)}).

Now the edge \blic{1}\ of $S'$ as well as the edge \blic{1+\tau}\ of $L'$
fit only with themselves forming the so called {\it Penrose rhombs} (Fig.
\ref{FPen1}.{\it b)}). Keeping this fact in mind it is simple to verify
that only the triangle $L'$ can be put beside the (directed) edge
\barb{\tau}\ of $S'$, so the previous composition argument can be repeated.
By composing the tiles $S'$ and $L'$ along their common edge \barb{\tau}\
again a new tessellation of type ${\mathcal P}'' =\{ S'', L'' \}$ is
obtained (Fig. \ref{FPen1}.{\it c)}), where the prototiles are similar to
the original prototiles $S$, $L$ by the ratio $\tau$, and the color coding
of the vertices is just reversed.

\begin{figure}
\centerline{\includegraphics[scale=0.5]{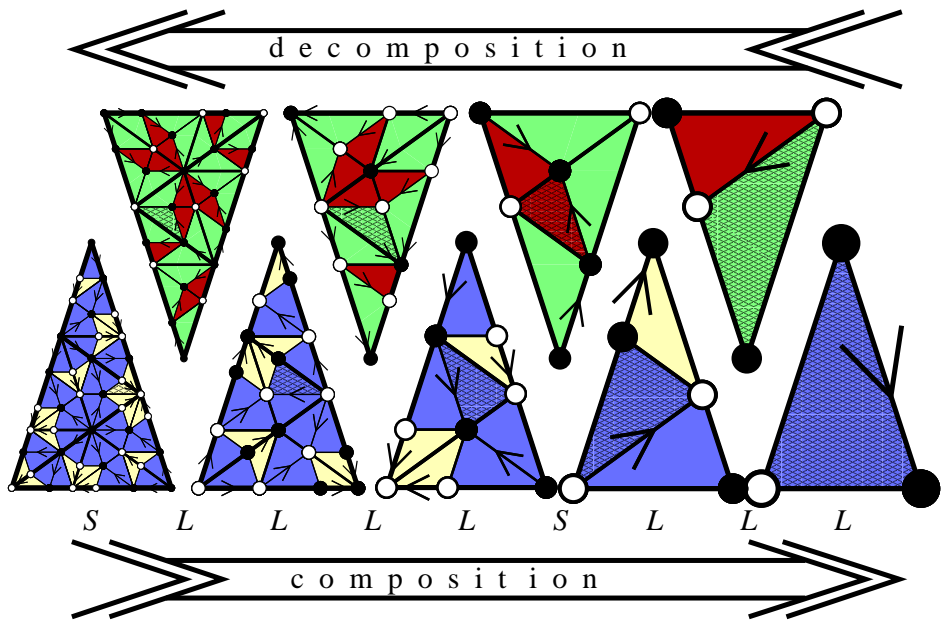}}
\caption{\label{FPen3} Successive {\it compositions} and {\it
decompositions} of a triangle. The letters $L$, $S$ denote the symbolic
sequence associated to the marked triangle.}
\end{figure}

The inverse process, the {\it decomposition} can also be uniquely defined
(Fig.~\ref{FPen3}). Given a tessellation of type ${\mathcal P}''$, by
dividing every tile of type $L''$ into $S'$ and $L'$ a new, finer tiling of
${\mathcal P}'$ is obtained, and another decomposition $L'\to S \cup L$
results in a tessellation of type ${\mathcal P}$, where the prototiles $S$
and $L$ are $\tau$ times smaller than $S''$, $L''$, and the color marks are
reversed.

\begin{figure}
\centerline{\includegraphics[scale=0.5]{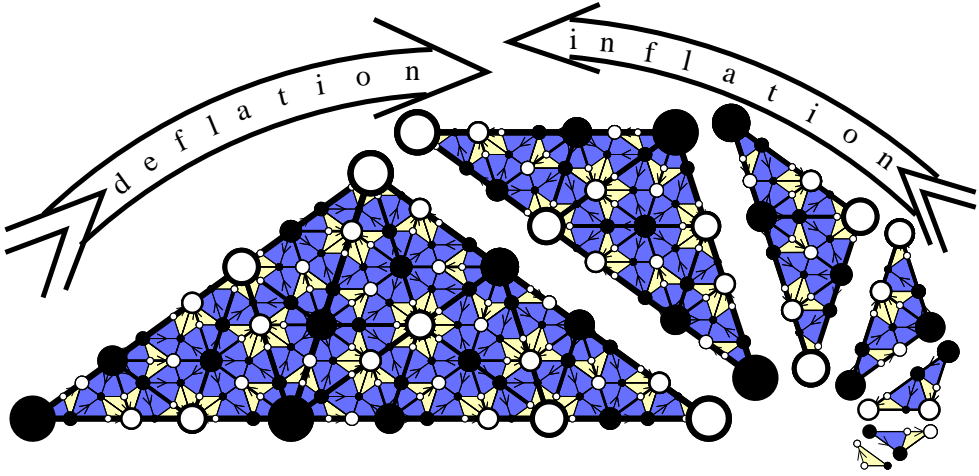}}
\caption{\label{FPen4} Successive {\it inflations} and {\it
deflations} of a triangle.}
\end{figure}

This {\it scaling property} is very characteristic to aperiodic tilings,
most of their unusual features result from this fact. The transformation
consisting of an enlarging by factor $\tau$ succeeded by a double
decomposition (${\mathcal P}'' \to {\mathcal P}$) is called {\it
inflation}, while the inverse process (double composition and reduction by
factor $\tau$) is {\it deflation} (Fig. \ref{FPen4}).

Now it is already clear that an arbitrarily large area of the plane can be
covered with a tiling of type ${\mathcal P}$. Indeed, successively
inflating one of the prototiles of ${\mathcal P}$ larger and larger
tessellated patches are obtained the linear measure of which is
enlarged by $\tau$ in every step (Fig.~\ref{FPen4}). Fig~\ref{FPen6}.{\it
a)} demonstrates the result of five successive inflations applied to the
prototile $L$.

It is also easy to see that no tiling of type ${\mathcal P}$ can be
periodic. Suppose, on the contrary that an infinite tiling possesses a
translational symmetry described by the vector $\mathbf v$. Then the
tessellation obtained by deflation would have a periodicity of ${\mathbf
v}/\tau$, since composition preserves the symmetry (edges of the same type
are erased), and the reduction scales the symmetry vector by $1/\tau$. A
repeated application of the deflation process would yield a tiling of the
same type $\mathcal P$ with an arbitrarily small symmetry vector ${\mathbf
v}/\tau^n$, what is nonsense, since the magnitude of the symmetry vector
must be greater than the linear size of the prototiles. Thus the prototiles
${\mathcal P} =\{ S,L \}$ (Fig. \ref{FPen1}.{\it a)}) admit only aperiodic
tessellations of the plane.

An even more interesting and shocking property of the Penrose tilings is
the fact that although we have considerable freedom in constructing
different infinite tessellations (we shell see it in detail in the next
subsection), every possible finite tiling patch $P$ occurs in every
infinite tessellation infinitely many times, and the {\it ratio of
appearance} of $P$ is fixed. This ratio depends only on the finite patch
$P$ itself, and not on the infinite tiling investigated
(Statement~\ref{StPen}). The possibly overlapping occurrences of $P$ are
distinguished by the position of a preferred prototile of $P$, so the {\it
frequency of appearance} is defined precisely as follows.

\begin{definition}\label{Dfa}
The {\bf number of appearance} $N_P (T)$ of a finite patch $P$ (with a
preferred prototile $p$ in it) in a finite tessellation $T$ of type
$\mathcal P$ is the number of prototiles $t$ in $T$, in the neighborhood of
which the motif $P$ appears in such a way that the preferred tile $p$ of
$P$ coincides with $t$ of $T$. The {\bf frequency} or {\bf ratio of
appearance} $\kappa_P (T)$ of the patch $P$ in the finite tiling $T$ is the
ratio $\kappa_P (T)=\frac{N_P (T)}{N(T)}$, where $N(T)$ is the number of
all prototiles (of both types) in $T$. For infinite tilings $I$ the {\bf
frequency of appearance} of $P$ is defined by the limit $\kappa_P
=\lim_{T\to I} \kappa_P (T)$, where $T$ is a finite portion of $I$ that
extends to the whole tessellation $I$.
\end{definition}

It is easy to see that the number $N_P (T)$ depends neither on the type nor
on the exact choice of the preferred prototile $p$ in the patch $P$. We
shall see later on that the ratio $\kappa_P$ does not depend either on the
infinite tiling $I$ investigated, that is the reason why $I$ is not
involved in the notation. The limit $T\to I$ can be made in any `reasonable
way', i.e., the finite tessellation $T$ should extend `uniformly in all
directions' to infinity.

As a preparation for the main Statement~\ref{StPen} we investigate in three
lemmas the ratios of appearance of the prototiles (Fig.~\ref{FPen1}.{\it
a)}), the edge- as well as the vertex neighborhoods (Fig.~\ref{FPen2}.{\it
b)-c)}).

\begin{lemma}\label{LPen1}{\bf (Frequency of appearance of prototiles)}
In every infinite tiling of type $\mathcal P$ the ratio of appearance of
the prototiles $L$ resp. $S$ is
\begin{align}\label{EfaLS}
\kappa_L &=\tau-1,& &\text{resp.}& \kappa_S &=2-\tau.
\end{align}
\end{lemma}

\begin{proof}
First we calculate the ratio $\lambda =\frac {\kappa_L (T)} {\kappa_S (T)}
=\frac {N_L (T)} {N_S (T)}$ of the number $N_L (T)$, $N_S (T)$ of the
prototiles $L$ and $S$ in finite patches obtained by repeated inflations of
a given finite patch $T$, and then we argue that this ratio has to be the
same in every infinite tiling as well. Indeed, let $N_L =N \kappa_L =
N\frac{\lambda} {1+\lambda}$ resp. $N_S =N\kappa_S =N\frac{1} {1+\lambda}$
be the number of prototiles $L$ resp. $S$ in the finite patch $T$ of $N$
tiles. (The argument $T$ of $N$, $N_L$ and $N_S$ has been omitted for
simplicity.) Since in every double decomposition step of the inflations the
triangles $L''$ resp. $S''$ are subdivided into two samples of $L$ and one
of $S$ resp. one sample of $L$ and $S$ (Fig. \ref{FPen1}.{\it c)}), the
number $N_L '$, $N_S '$ of the prototiles $L$, $S$ after an inflation
process is related to the initial values via the formulas $N_L ' =2N_L
+N_S$, $N_S '= N_L +N_S$, which give the recursion $\lambda_{n+1}
=\frac{2\lambda_n +1} {\lambda_n +1}$ for the ratio $\lambda_n$ of the
number of the prototiles $L$ and $S$ after the $n$-th inflation process. It
is easy to see that for any nonnegative initial value $\lambda_0$ the
series $\lambda_n$ converges to the $\lim_{n \to \infty} \lambda_n =
(1+\sqrt{5}) /2 =\tau$ golden mean.

This observation helps us to prove that the ratio of the two prototiles
have to be the same number $\tau$ in any infinite tiling of type $\mathcal
P$. Indeed, by applying the double composition sufficiently many times, we
obtain an (infinite) tiling of huge triangles $S''{}^{\dots}{}'$ and
$L''{}^{\dots}{}'$, in which the ratio of the original small prototiles $S$
and $L$ can be made to be arbitrarily close to $\tau$. It means that (using
any reasonable definition for the limit process $T \to I$) the ratio of the
prototiles in every infinite Penrose tiling $I$ has the same value $\tau$,
and $\kappa_L =\frac{\lambda}{1+\lambda} =(-1+\sqrt{5}) /2 =\tau-1$,
$\kappa_S =\frac{1} {1+\lambda} =(3-\sqrt{5}) /2 =2-\tau$.
\end{proof}

In the next lemma we investigate the frequencies of appearance of the seven
edge neighborhoods (Fig.~\ref{FPen2}.{\it b)}).

\begin{lemma}\label{LPen1.5}
{\bf (Frequency of appearance of edge neighborhoods)}
In every infinite tiling of type $\mathcal P$ the ratios of appearance of
the seven edge neighborhoods $k$, $k'$, $d$, $t$, $l'$, $r$ resp. $r'$
(Fig.~\ref{FPen2}.{\it b)}) are
\begin{subequations}\label{Efaen}
\begin{align}
\kappa_k &= -1+\tau,&
\kappa_{k'} &= -6+4\tau,&
\kappa_d &= 2-\tau,\\
\kappa_t &= 5-3\tau,&
\kappa_{l'} &= 2-\tau,&
\kappa_r &= -3+2\tau,\\
&\text{resp.}&
\kappa_{r'} &= -3+2\tau.&&
\end{align}
\end{subequations}
\end{lemma}

It is worth noting that all these frequencies have the form $a+b\tau$,
where $a,b \in \Z$ are integers!

\begin{proof}
In this proof it is more convenient to use the notions `number of
appearance' and `frequency of appearance' in a bit modified way, i.e.,
referring to the occurrences of the edge neighborhoods $e$ by the position
of their preferred (middle) edge, instead of using a preferred tile. These
altered quantities are distinguished by a `hat' ($\Hat{\hspace{6pt}}$), so
$\Hat{N}_e (T)$ denotes the number of inner edges in the finite
tessellation $T$ with edge neighborhood $e \in \{ k,k',d,t,l',r,r' \}$, and
$\Hat{\kappa}_e (T) =\frac{\Hat{N}_e (T)} {\Hat{N} (T)}$ is the ratio of
appearance of $e$, in proportion to the number $\Hat{N} (T)$ of all edges
in $T$. (Of course, the edges lying at the boundary of the finite
tessellation $T$ have no well defined edge neighborhood, but their number
is negligible compared to the number of all edges, as $T \to I$ extends to
infinity.)

There is, however, a simple connection between the above defined quantities
$\Hat{N}_e (T)$, $\Hat{N} (T)$ resp. $\Hat{\kappa}_e$ and the general
notions $N_e (T)$, $N(T)$ resp. $\kappa_e$ of Definition~\ref{Dfa}. All
prototiles have three edges, and every edge belongs to two tiles, thus
$\lim_{T \to I} \frac{\Hat{N} (T)} {N(T)} =\frac{3}{2}$, i.e., in an
infinite tessellation the number of edges is half as much again as the
number of tiles. For edge neighborhoods $e$ with reflectional symmetry the
preferred tile of $e$ can be put to both sides of the preferred edge, so in
this case $N_e (T) =2 \Hat{N}_e (T)$ and
\begin{subequations}\label{Ekekte}
\begin{align}\label{Ekekte_a}
\kappa_e &= \lim_{T \to I} \frac{N_e (T)} {N(T)} =3\lim_{T \to I}
\frac{\Hat{N}_e (T)} {\Hat{N}_e (T)} =3 \Hat{\kappa}_e,&
\text{if } e&\in \{k,k',d,r,r' \}. \\
\intertext{For the two other edge neighborhoods without symmetry $N_e (T)
=\Hat{N}_e (T)$, thus}
\label{Ekekte_b}
\kappa_e &= \lim_{T \to I} \frac{N_e (T)} {N(T)} =\frac{3}{2} \lim_{T \to
I} \frac{\Hat{N}_e (T)} {\Hat{N}_e (T)} =\frac{3}{2} \Hat{\kappa}_e,&
\text{if } e&\in \{ t,l' \}.
\end{align}
\end{subequations}

We go on likewise in the previous lemma, and investigate how the numbers
$\Hat{N}_e (T)$ of the seven edge neighborhoods $e \in \{ k,k',d,t,l',r,r'
\}$ change under successive inflations of the finite tessellation $T$. For
example the middle edge \clic{} in the {\it kite} motif is divided into two
edges with $r$ and $r'$ edge neighborhoods, and there are four new edges
created inside the tiles with edge neighborhoods $k$, $k$, $l'$ and $l'$
(Fig~\ref{FPen2}.{\it b)}). To avoid over-counting these four new edges have
to be counted by one third, since they were created inside a triangular
prototile, so they are considered three times with the three edges of the
triangle. Thus the transformation of an edge neighborhood $k$ under an
inflation process is described schematically
\begin{equation}\label{Ekto...}
k\to r+r'+\frac{1}{3} (2k+2l').
\end{equation}

The transformations of the other edge neighborhoods can be similarly
derived, and the final result can be expressed in a matrix equation
\begin{equation}\label{ENevmate}
\begin{bmatrix}
\Hat{N}'_k \\ \Hat{N}'_{k'} \\ \Hat{N}'_d \\ \Hat{N}'_t \\
\Hat{N}'_{l'} \\ \Hat{N}'_r \\ \Hat{N}'_{r'}
\end{bmatrix}
=
\begin{bmatrix}
2/3 & 2/3 & 1 & 1/3 & 1/3 & 0 & 2/3 \\
0 & 1 & 0 & 1 & 0 & 1 & 1 \\
0& 1& 0& 1& 0& 1& 0\\
0& 0& 0& 0& 1& 0& 0\\
2/3& 2/3& 2/3& 2/3& 2/3& 2/3& 2/3 \\
1& 0& 0& 0& 0& 0& 0 \\
1& 0& 0& 0& 0& 0& 0
\end{bmatrix}
\begin{bmatrix}
\Hat{N}_k \\ \Hat{N}_{k'} \\ \Hat{N}_d \\ \Hat{N}_t \\
\Hat{N}_{l'} \\ \Hat{N}_r \\ \Hat{N}_{r'}
\end{bmatrix}
\end{equation}
relating the numbers $\Hat{N}'_e$ of the different edge neighborhoods $e$
after the inflation to their values $\Hat{N}_e =\Hat{N}_e (T)$ before the
inflation. (The arguments $T$ have been omitted for simplicity.)

The characteristic polynomial of the matrix above is $\lambda (\lambda^2
+1) (9\lambda^2 +6\lambda +2) (\lambda^2 -3\lambda +1)$, with roots
$\lambda_1 =0$, $\lambda_{2,3} =\pm i$, $\lambda_{4,5} =(-1\pm i)/3$ and
$\lambda_{6,7} =(3\pm \sqrt{5}) /2$. The eigenvalue with the largest
magnitude is the real number $\lambda_7 =(3+\sqrt{5})/2$, thus its
eigenvector $\boldsymbol{\Hat{\kappa}}$ describes the ratios of appearance
$\Hat{\kappa}_e$ belonging to the different edge neighborhoods $e$ in the
limit of infinitely many inflations applied to the finite tessellation $T$.
The components of $\boldsymbol{\Hat{\kappa}}$ are:
\begin{subequations}\label{Eenrat}
\begin{align}
\Hat{\kappa}_k &=\frac{-1+\tau}{3},&
\Hat{\kappa}_{k'} &=\frac{-6+4\tau}{3},&
\Hat{\kappa}_{d} &=\frac{2-\tau}{3},\\
\Hat{\kappa}_{t} &=\frac{10-6\tau}{3},&
\Hat{\kappa}_{l'} &=\frac{4-2\tau}{3},&
\Hat{\kappa}_{r} &=\frac{-3+2\tau}{3},\\
&&\Hat{\kappa}_{r'} &=\frac{-3+2\tau}{3}.&&
\end{align}
\end{subequations}
(These ratios are normalized, i.e., $\sum_e \Hat{\kappa}_e =1$.)

This result can be extended to infinite tilings using the same arguments as
in the previous proof. Indeed, after sufficiently many (double)
compositions, the infinite tiling consists of huge triangles of type
$L''{}^{\dots}{}'$, $S''{}^{\dots}{}'$, and increasing the number of
compositions applied (thus the size of the huge triangles) the frequency of
appearance of the different edge neighborhoods in each of the finite
triangles $L''{}^{\dots}{}'$, $S''{}^{\dots}{}'$ can be brought arbitrarily
close to the ideal values~\eqref{Eenrat}.

Using the connections~\eqref{Ekekte} the statements~\eqref{Efaen} of
Lemma~\ref{LPen1.5} are obtained.
\end{proof}

For proving similar assertions for arbitrary finite patches of tiling we
need to investigate the ratio of occurrence of the seven vertex
neighborhoods (Fig.~\ref{FPen2}.{\it c)}) in infinite tilings.

\begin{lemma}\label{LPen2}
{\bf (Frequency of appearance of vertex neighborhoods)}
In every infinite tiling of type $\mathcal P$ the ratios of appearance of
the seven vertex neighborhoods $\odot, \star, A, D, J, Q$ resp. $K$ (Fig.
\ref{FPen2}.{\it c)}) are
\begin{subequations}\label{Eksvn}
\begin{align}
\kappa_\odot &=-11+7\tau,& \kappa_\star &=-29+18\tau,& \kappa_A &=2-\tau,\\
\kappa_D &=3+2\tau,& \kappa_J &=5-3\tau,& \kappa_Q &=-8+5\tau\\
&\text{resp.}& \kappa_K &=13-8\tau.&&
\end{align}
\end{subequations}
\end{lemma}

Please notice the `miracle' that all these frequencies are again elements
of the dense subgroup $\Z +\tau \Z \subset \R$ of the additive real group!

\begin{proof}
In this proof we have to use the notions `number of appearance'
$\Tilde{N}_V (T)$ and `frequency of appearance' $\Tilde{\kappa}_V (T)$ of a
vertex neighborhood $V\in \{\odot, \star,$\hspace{0pt}$ A,$\hspace{0pt}$
D,$\hspace{0pt}$ J,$\hspace{0pt}$Q, K \}$ in a finite tiling $T$ again in a
bit altered way. For vertex neighborhoods it is more convenient to label
their appearances with the position of a preferred vertex, namely the
middle vertex, than with the place of a preferred prototile of them. To
every vertex inside $T$ there corresponds a well defined vertex
neighborhood, so let $\Tilde{N}_V (T)$ be the number of inner vertices of
$T$ with vertex neighborhood $V$, and let $\Tilde{\kappa}_V (T)=\frac
{\Tilde{N}_V (T)} {\Tilde{N}(T)}$, where $\Tilde{N} (T)$ is the number of
inner vertices of $T$. (The vertex neighborhoods are not necessarily well
defined for the points lying on the boundary of $T$, but with the expansion
of $T$ the ratio of these boundary vertices becomes negligible, so this
boundary effect can be ignored.)

There is, however, a straightforward connection between the above defined
quantities $\Tilde{N} (T)$, $\Tilde{N}_V (T)$, $\Tilde{\kappa}_V$ and
the general notions $N(T)$, $N_P (T)$, $\kappa_P$ of Definition~\ref{Dfa}.
It is easy to check that $\lim_{T\to I} \frac{N(T)} {\Tilde{N} (T)} =2$,
i.e., in a tessellation $T$ extending to infinity there are two times as
many prototiles as vertices. (Indeed, the angles of all prototiles in $T$
add up to $\pi N(T)$ resp. to $2\pi \Tilde{N} (T)$ if they are counted by
prototiles resp. by vertices.) Since the vertex neighborhoods $A$, $D$,
$J$, $Q$, $K$ has a reflectional symmetry, the preferred prototile of them
can be mapped to two different tiles of $T$, whilst the image of the
middle vertex is unaltered, so in this case $N_V (T) =2\Tilde{N}_V (T)$,
and
\begin{subequations}\label{Ekkt}
\begin{align}\label{Ekkt_a}
\kappa_V &=\lim_{T\to I} \frac{N_V (T)}{N(T)} =
\lim_{T\to I} \frac{2\Tilde{N}_V (T)} {2\Tilde{N}(T)} =\Tilde{\kappa}_V,&
\text{if } V&\in \{A,D,J,Q,K\}.\\
\intertext{The motifs $\odot$ and $\star$ have an additional five-fold
rotational symmetry, so $N_V (T)=10\Tilde{N}_V (T)$, and} \label{Ekkt_b}
\kappa_V &=\lim_{T\to I} \frac{N_V (T)} {N(T)} =\lim_{T\to I}
\frac{10\Tilde{N}_V (T)} {2\Tilde{N}(T)} =5\Tilde{\kappa}_V,&
\text{if } V&\in \{ \odot, \star \}.
\end{align}
\end{subequations}

We proceed similarly as in the previous proofs; first we investigate
the evolution of the numbers $\Tilde{N}_V (T)$ of the different vertex
neighborhoods $V \in\{\odot, \star,$\hspace{0pt}$
A,$\hspace{0pt}$ D,$\hspace{0pt}$ J,$\hspace{0pt}$ Q,K\}$ in a finite
tiling $T$ under successive inflations of $T$. The closest neighborhood of
every vertex determines its future under inflation, as Fig.
\ref{FPen2}.{\it c)} shows, and it is easy to check that the evolution of
the vertex neighborhoods of the existing vertices takes place according to
the following diagram:
\begin{equation}\label{Eev} \unitlength 0.8mm
\raisebox{-6.4mm}{ \begin{picture}(118,16) \thinlines
\put(6,13){\vector(1,0){10}}
\put(22,13){\vector(1,0){10}}
\put(38,13){\vector(2,-1){10}}
\put(54,8){\vector(1,0){10}}
\put(70,8){\vector(1,0){10}}
\put(86,8){\vector(1,0){10}}
\put(102,8){\vector(1,0){10}}
\put(22,3){\vector(1,0){10}}
\put(38,3){\vector(2,1){10}}
\put(3,13){\makebox(0,0){$A$}}
\put(19,13){\makebox(0,0){$J$}}
\put(35,13){\makebox(0,0){$K$}}
\put(19,3){\makebox(0,0){$D$}}
\put(35,3){\makebox(0,0){$Q$}}
\put(51,8){\makebox(0,0){$\odot$}}
\put(67,8){\makebox(0,0){$\star$}}
\put(83,8){\makebox(0,0){$\odot$}}
\put(99,8){\makebox(0,0){$\star$}}
\put(115,8){\makebox(0,0){$\dots$}}
\end{picture}}
\end{equation}

Thus after a few inflation steps every vertex ends up its ephemeral life in
a state flashing between the two celestial forms with the perfectness of
five-fold rotational symmetry. It does not mean, however, that the other
vertex neighborhoods would die out, since in every double decomposition
process new vertices come into the world, in a state of {\it Ace} or {\it
Deuce}. Indeed, the node being born on the \blic{1+\tau}\ edge of $S''$
or $L''$ is determined by its predecessors to start its life as an {\it
Ace}, and the vertex created on the \carc{1+\tau}\ edge of $L''$ can not
be but a {\it Deuce}, considering its vertex neighborhood (Fig.
\ref{FPen1}.{\it c)}).

Comprehending all these transmutations, we are capable now to summarize the
evolution of the numbers of the different vertex neighborhoods during a
single inflation process in a matrix equation:
\begin{equation}\label{ENevmat}
\begin{bmatrix}
\Tilde{N}'_\odot \\ \Tilde{N}'_\star \\ \Tilde{N}'_A \\ \Tilde{N}'_D \\
\Tilde{N}'_J \\ \Tilde{N}'_Q \\ \Tilde{N}'_K
\end{bmatrix}
=
\begin{bmatrix}
0 & 1 & 0 & 0 & 0 & 1 & 1 \\
1 & 0 & 0 & 0 & 0 & 0 & 0 \\
5/2& 5/2& 1/2& 1/2& 3/2& 3/2& 2\\
5/2& 0& 0& 1& 3/2& 0& 0\\
0& 0& 1& 0& 0& 0& 0 \\
0& 0& 0& 1& 0& 0& 0 \\
0& 0& 0& 0& 1& 0& 0
\end{bmatrix}
\begin{bmatrix}
\Tilde{N}_\odot \\ \Tilde{N}_\star \\ \Tilde{N}_A \\ \Tilde{N}_D \\
\Tilde{N}_J \\ \Tilde{N}_Q \\ \Tilde{N}_K
\end{bmatrix}
\end{equation}
(The newly created nodes are counted with the two vertices of the edge they
are lying on. To avoid over-counting, the matrix elements describing the new
vertices have to be divided by two.)

The characteristic polynomial of the matrix above is $\lambda (2\lambda^2
+\lambda +1)(\lambda^2 +\lambda +1/2)(\lambda^2 -3\lambda +1)$ with roots
$\lambda_1 =0$, $\lambda_{2,3} =(-1\pm i\sqrt{7})/4$, $\lambda_{4,5}
=(-1 \pm i)/2$ and $\lambda_{6,7} =(3 \pm \sqrt{5})/2$. Since
only the real eigenvalue $\lambda_7 =(3+\sqrt{5})/3$ has magnitude
greater than unity, its eigenvector $\boldsymbol{\Tilde{\kappa}}$ describes
the limiting ratio of the different vertex neighborhoods after many
inflations applied to the finite patch $P$, the components of which are:
\begin{subequations}\label{Evnrat}
\begin{align}
\Tilde{\kappa}_\odot &= \frac{-11+7\tau}{5},&
\Tilde{\kappa}_\star &= \frac{-29+18\tau}{5},&
\Tilde{\kappa}_A &= 2-\tau, \\
\Tilde{\kappa}_D &= -3+2\tau,& \Tilde{\kappa}_J &= 5-3\tau,&
\Tilde{\kappa}_Q &= -8+5\tau,\\
&& \Tilde{\kappa}_K &= 13-8\tau. &&
\end{align}
\end{subequations}
(These probabilities are normalized in such a way that $\sum_{V}
\Tilde{\kappa}_V =1$.)

This result can be extended to infinite tilings using the same arguments as
in the previous two proofs, and using the connections~\eqref{Ekkt} the
statements~\eqref{Eksvn} of Lemma~\ref{LPen2} are obtained.
\end{proof}

We remark that there are certain linear relations between the results of
the last three lemmas, i.e., between the ratios of appearance of the
prototiles \eqref{EfaLS}, the edge neighborhoods \eqref{Efaen} and the
vertex neighborhoods \eqref{Eksvn}. These connections can be deduced by
considering the vertex neighborhoods as unions of edge neighborhoods, and
those as unions of prototiles. We do not give the details here.

All what is needed from the above results in the followings is the fact
that the ratios \eqref{EfaLS}, \eqref{Efaen} and \eqref{Eksvn} are members
of the subgroup $\Z+\tau\Z$ of the additive group of real numbers $\R$,
which is dense in $\R$, but does not agree with it.

Before the proof of the main Statement~\ref{StPen} one more fact
has to be observed.

\begin{lemma}\label{LPen3}
Let $P$ be a finite patch of type $\mathcal P$, and let $\Hat{P}$ denote
the motif obtained by inflating $P$. Then
\begin{equation}\label{ELPen3}
\kappa_{\Hat{P}} =(2-\tau)\kappa_P.
\end{equation}
\end{lemma}

\begin{proof}
Let $T$ be a finite tiling of type $\mathcal P$, and let $\varPhi :\Hat{P} \to
T$ be an isometry which maps the inflated motif $\Hat{P}$ onto a portion
of $T$ congruent to the motif $\Hat{P}$. Since the composition is a
uniquely defined local process, $\varPhi :\Hat{P}'' \to T''$ remains a
congruence between the motif $\Hat{P}''$ and the infinite tiling $T''$
obtained from $\Hat{P}$ resp. $T$ by double composition. The inverse
statement is also true, namely whenever there is a congruence $\varPhi
:\Hat{P}'' \to T''$ mapping the (composed) motif $\Hat{P}''$ onto one of
its appearances in $T''$, by double decomposition a congruence $\varPhi :
\Hat{P} \to T$ is obtained. Thus the occurrences of a finite motif
in a tiling (namely $\Hat{P}$ in $T$) and the (doubly) composed motif
in the (doubly) composed tiling (namely the appearances of $\Hat{P}''$ in
$T''$) are in one to one correspondence, so the numbers $N_{\Hat{P}}
(T)=N_{\Hat{P}''} (T'')$ are equal. But, by definition, $\Hat{P}''$ is
similar to the original motif $P$, by the ratio $\tau$, so
$\kappa_{\Hat{P}''} =\kappa_P$. That means that

\begin{equation}\label{Ekphkp}
\kappa_{\Hat{P}} =
\lim_{T \to I} \frac{N_{\Hat{P}} (T)} {N(T)} =
\lim_{T\to I} \Big(\frac{N_{\Hat{P}''} (T'')} {N(T'')}
\frac{N(T'')} {N(T)} \Big)=
\kappa_P \lim_{T \to I} \frac{N(T'')} {N(T)}.
\end{equation}

We need the inverse of the factor $\lim_{T \to I} \frac{N(T)} {N(T'')}$ by
which the number of tiles increases in a double decomposition process in
infinitely large tilings $I$. Since in a double decomposition every
prototile of type $L''$ is decomposed into three new pieces, and the tiles
$S''$ are subdivided into two (Fig.~\ref{FPen1}.{\it c)}), using the
results of Lemma~\ref{LPen1} we get

\begin{equation}\label{ElNTNT''}
\lim_{T \to I} \frac{N(T)} {N(T'')} =3\kappa_L +2\kappa_S
=3(\tau-1)+2(2-\tau) =1+\tau.
\end{equation}
So inserting its inverse $\frac{1}{1+\tau} = 2-\tau$ into \eqref{Ekphkp} we
obtain the statement~\eqref{ELPen3}.
\end{proof}

Now we turn to the main assertion of this subsection.

\begin{statement}\label{StPen}
{\bf (Local isometry of Penrose tilings)}

{\it i)}
Every finite motif $P$ of a Penrose tiling of type ${\mathcal
P}$ occurs infinitely many times in every infinite tessellation $I$.

{\it ii)}
Moreover, the frequency $\kappa_P$ of occurrence of $P$ in $I$
is fixed, i.e., it does not depend on the infinite tiling $I$, it is
determined solely by the motif $P$.

{\it iii)}
Finally, $\kappa_P$ is an element of the subgroup $\Z+\tau \Z \subset \R$
of the additive group of reals.
\end{statement}

We remark that different appearances of the finite patch $P$ are allowed
to overlap in $I$, the occurrences of $P$ are distinguished by the position
of a preferred tile in $P$, according to Definition~\ref{Dfa}.

\begin{proof}
Since the motif $P$ is finite, after sufficiently many (say $k\in
\N$) double compositions applied to $I$ the size of the composed triangles
$S''{}^{\dots}{}'$ and $L''{}^{\dots}{}'$ surpasses the size of $P$, what
assures that any occurrence of $P$ is fully covered by a vertex
neighborhood in $I''{}^{\dots}{}'$. (Here $I''{}^{\dots}{}'$ denotes the
infinite tiling of type ${\mathcal P}''{}^{\dots}{}'$ obtained after $k$
successive compositions.) Since every vertex neighborhood appears in any
infinite tessellation infinitely many times, the motif $P$ appears also
infinitely many times in $I$, which is assertion~{\it i)} of the
Statement.

\begin{figure}
\centerline{\includegraphics[scale=0.5]{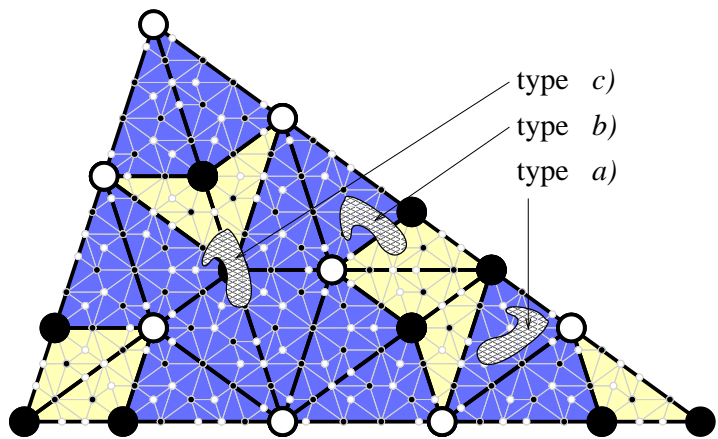}}
\caption{\label{FPen5} The possible positions of the finite motif $P$ in
the composed tiling $I''{}^{\dots}{}'$.}
\end{figure}

Unfortunately it is not true that a single prototile $S''{}^{\dots}{}'$ or
$L''{}^{\dots}{}'$ would cover every occurrence of $P$, since the motif $P$
can appear at the vertices or edges of the composed triangles, no matter
how big the triangles are (Fig.~\ref{FPen5}). But it is for sure that any
occurrence of $P$ is fully covered either {\it a)} by a single prototile
$S''{}^{\dots}{}'$, $L''{}^{\dots}{}'$, or {\it b)} by an edge neighborhood
$e$, in such a way that $P$ intersects the middle edge of $e$, or {\it
c)} by a vertex neighborhood $V$ of $I''{}^{\dots}{}'$, in such a way
that $P$ contains the middle vertex of $V$ (Fig.~\ref{FPen5}). Moreover,
the above classification of the occurrences of the motif $P$ in the
composed tessellation $I''{}^{\dots}{}'$ is unambiguous, thus the ratio of
appearance $\kappa_P$ can be expressed with the basic frequencies of the
prototiles [equation~\eqref{EfaLS}], edge- and vertex neighborhoods
[equations~\eqref{Efaen}, \eqref{Eksvn}] already determined.

More formally, let the nonnegative integers $n_L$, $n_S$, $n_e$ resp.
$n_V$ denote the number of appearance of the motif $P$ in the finite
tilings obtained by $k$ successive inflations of the prototiles $L$, $S$,
the edge neighborhoods $e\in \{k, k',$\hspace{0pt}$ d,$\hspace{0pt}$
t,$\hspace{0pt}$ l',$\hspace{0pt}$ r,r' \}$ resp. the vertex neighborhoods
$V\in \{\odot, \star,$\hspace{0pt}$ A,$\hspace{0pt}$ D,$\hspace{0pt}$
J,$\hspace{0pt}$ Q,K \}$. In $n_L$, $n_S$ we count only the appearances of
$P$ in the interior of $L''{}^{\dots}{}'$, $S''{}^{\dots}{}'$, in $n_e$
only the occurrences of type~{\it b)} are considered, and in $n_V$ the
occurrences of type~{\it c)} are counted (Fig.~\ref{FPen5}). If the
edge- or vertex neighborhood $e$ or $V$ possesses a symmetry than the
numbers $n_e$, $n_V$ should be reduced, since these symmetries had already
been accounted for in the calculation of $\kappa_e$ and $\kappa_V$.

With these notations the ratio of appearance of the motif $P$ in any
infinite tessellation is
\begin{equation}\label{EkapP}
\kappa_P =(2-\tau)^k \Big(\kappa_L n_L +\kappa_S n_S +
\sum_e \kappa_e n_e + \sum_V \kappa_V n_V \Big),
\end{equation}
where the result of Lemma~\ref{LPen3} was taken into consideration with the
prefactor $(2-\tau)^k$.

This formula proves the last two assertions~{\it ii)} and {\it iii)} of the
Statement. Indeed, it is easy to see that $\kappa_P \in \Z +\tau \Z$, since
the $n$'s are integers and the $\kappa$'s are members of the additive group
$\Z +\tau\Z$ and $\tau^2 =1+\tau$.
\end{proof}

We remark that the proof of the above Statement~\ref{StPen} without its
last assertion~{\it iii)} (stating that $\kappa_P \in \Z +\tau \Z$) would
have been much simpler, since then the ratios $\kappa_e$ and $\kappa_V$
[equations~\eqref{Efaen}, \eqref{Eksvn}] need not have been determined
exactly.

\subsection{Symbolic dynamics associated to the
%----------------------------------------------
Penrose tilings\label{SPen_b}}
%-----------------------------

In this subsection a symbolic coding scheme is presented, with the help of
which every infinite tiling of type $\mathcal P$ can be encoded. The coding
is not unique, in the sense that different symbolic sequences describe the
same (i.e., isometric) infinite tessellations, and this non-uniqueness can
also be characterized in terms of the coding.

\begin{figure}
\centerline{\includegraphics[scale=0.5]{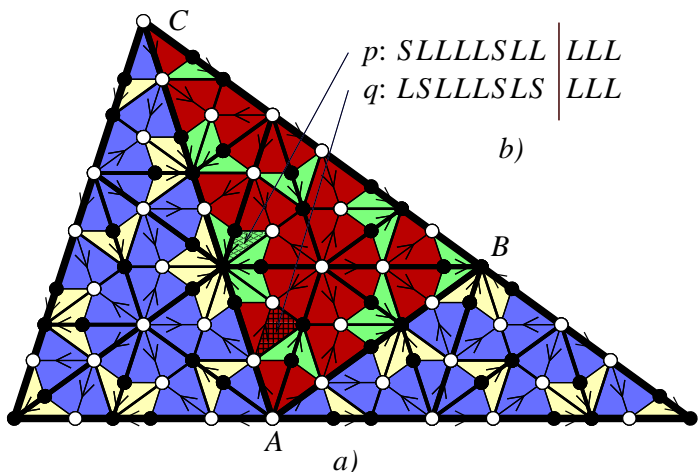}}
\caption{\label{FPen6} {\it a)} The result of five successive inflations
applied to the prototile $L$. The thicker lines denote the tilings obtained
after successive compositions. {\it b)} The symbolic sequences associated
to the two marked prototiles $p$ and $q$.}
\end{figure}

Given an infinite tessellation $I$ of type $\mathcal P$  with a preferred
prototile $p\in I$ in it, the associated symbolic sequence $\Bar{x}(I) =\{
x_i \}_{i\in \N} \in \{ L,S\}^{\N}$ of letters $x_i \in \{ L,S\}$ is
constructed in the following way. If the preferred tile $p\in I$ is of type
$L$ than $x_0 =L$, otherwise (i.e., if $p$ is of type $S$) $x_0 =S$. Then
make a (single) composition $I\to I'$ in the tessellation, and let $p'\in
I'$ be the composed tile (of type $L'$ or $S'$) in which $p$ is contained.
The next letter $x_1$ of the symbolic sequence is $L$ resp. $S$ according
to the type of $p'$. This process should be repeated infinitely, i.e., the
consecutive letters of the sequence $\Bar{x}$ give the type of the
prototile in which the preferred tile $p$ is contained after consecutive
compositions. The Figures~\ref{FPen3} and \ref{FPen6} demonstrate this
process graphically.

In the next statement the properties of this coding are investigated.

\begin{statement}\label{StPencod}
{\bf (Symbolic coding of Penrose tilings)}

{\it i)}
The encoding process described above renders every infinite
tessellation $I$ of type $\mathcal P$, with a preferred prototile $p \in I$
into a unique symbolic sequence $\Bar{x} =\{ x_i \} \in \{ L,S\}^{\N}$
which satisfies the following `grammar rules' corresponding to the
transition matrix $T_{\mathrm{P}}$:
\begin{align}\label{EPensym}
\unitlength 0.8mm
\raisebox{-8mm}{
\begin{picture}(40,22)
\thinlines
\put(15,16){\vector(0,-1){10}}
\put(15,16){\vector(2,-1){20}}
\put(35,16){\vector(-2,-1){20}}
\put(10,16){\makebox(10,6){$L$}}
\put(30,16){\makebox(10,6){$S$}}
\put(10,0){\makebox(10,6){$L$}}
\put(30,0){\makebox(10,6){$S$}}
\put(0,16){\makebox(10,6){$x_{i} :$}}
\put(0,0){\makebox(10,6){$x_{i+1} :$}}
\end{picture}}&&
T_{\mathrm{P}} &=
\begin{bmatrix} 1&1\\1&0 \end{bmatrix}
\end{align}
i.e., $x_{i+1} =L$ whenever $x_i =S$, for all $i\in \N$. (If the $(i,j) \in
\{L,S\} ^2$ component of $T_{\mathrm{P}}$ is one then the $j\to i$
transition is allowed.)

{\it ii)}
If the position of the preferred tile $q\in I$ is changed then the new
symbolic sequence $\Bar{y} \in \{ L,S\}^{\N}$ has the same infinite tail as
$\Bar{x}$, i.e., there exists a finite $n<\infty$ such that $x_i =y_i$ for
all $i>n$.

{\it iii)}
Reversely, every infinite sequence $\Bar{x} \in\{ L,S\}^{\N}$ satisfying
the grammar rules~\eqref{EPensym} determines a Penrose tiling.

{\it iv)}
If two infinite sequences $\Bar{x}, \Bar{y} \in \{ L,S\}^{\N}$ satisfying
the grammar rules~\eqref{EPensym} have the same infinite tail (i.e.,
$\exists n<\infty$ such that $(\forall i>n)$ $x_i =y_i$) than the Penrose
tilings determined by the two sequences are isometric.
\end{statement}

Thus there is a bijective correspondence between the (congruence classes
of) infinite Penrose tilings and the (equivalence) classes of series in
$\{L,S\}^{\N}$ satisfying~\eqref{EPensym} and having the same infinite
tail.

\begin{proof}
The uniqueness in assertion {\it i)} of the Statement is an evident
consequence of the definition of the symbolic sequences. To prove the fact
that a symbol $S$ can not be followed again by an $S$ we have to observe
Fig~\ref{FPen1}. Indeed, in every composition step the smaller prototiles
$S$ or $S'$ are always composed with $L$ or $L'$, forming a triangle of
type $L'$ or $L''$, respectively.

It is also easy to see that this coding process can be reversed, i.e.,
given a sequence $\Bar{x} \in \{L,S\}^{\N}$ satisfying the
rules~\eqref{EPensym}, an (up to isometry) unique infinite tessellation $I
=\lim_{n\to \infty} T_n$ with preferred tile $p\in I$ can be constructed,
the symbolic sequence of which agrees with $\Bar{x}$. Indeed, let $T_0 =L$,
if $x_0 =L$, and $T_0 =S$, if $x_0 =S$. Then given $T_n \in \{
\Hat{L}^n, \Hat{S}^n \}$, let $T_{n+1} :=T_n =\Hat{S}^{n+1}$, if $x_{n+1} =
S$, and otherwise (if $x_{n+1}=L$) $T_{n+1} :=\Hat{L}^{n+1} =\Hat{L}^n \cup
\Hat{S}^n$, where $\Hat{L}^n$ resp. $\Hat{S}^n$ denote the finite
triangular tilings of type $\mathcal P$, obtained by decomposing the tiles
$L''{}^{\dots}{}'$ resp. $S''{}^{\dots}{}'$ with $n$ `primes' of
Fig.~\ref{FPen1}. Thus consecutively reading the letters of $\Bar{x}$,
encountering an $S$ we leave $T_n$ unaltered, and encountering an $L$ we
enlarge $T_n$ by joining to it another triangle of different type. The
infinite tessellation $I =\lim_{n\to \infty} T_n$ with preferred prototile
$p =T_0$ has the prescribed symbolic sequence $\Bar{x}$ by construction,
what proves assertion {\it iii)} of the Statement (Fig~\ref{FPen3}).

For proving assertion {\it ii)} we have to notice that after sufficiently
many compositions of $I$ the two preferred prototiles $p$, $q$ of type
$\mathcal P$ will fall into the same composed tile of type $\mathcal
P''{}^{\dots}{}'$, and the symbolic sequences coincide from this point.
Figure~\ref{FPen6} demonstrates this phenomenon, where after seven
compositions the prototiles $p$ and $q$ lie in the same (dark) triangle
$ABC$, and from this point on the positions of the preferred tiles $p$ and
$q$ are indistinguishable.

Assertion {\it iv)} is a consequence of the first three statements.
The first $n$ letters of $\Bar{x}$ and $\Bar{y}$ determine isometric finite
tessellations, only the position of the preferred tile is different, and
from that point on the sequences coincide.
\end{proof}

We remark that not every infinite tessellation determined by a sequence
$\Bar{x}$ extends to the whole plane. There are `exceptional' sequences,
the tiling corresponding to which covers only a half-plane or an infinite
domain bounded by an angle. We do not address here the question of
classifying these `exceptional' sequences.

For convenience, let the set of all possible symbolic series satisfying the
grammar rules~\eqref{EPensym} be denoted with
\begin{equation}\label{EMPen}
M_{\mathrm{P}} =\bigl\{ \Bar{x} \in \{L,S\} ^{\N} \: \vert\: (\forall i\in
\N) \: x_i =S \Rightarrow x_{i+1} =L \bigr\},
\end{equation}
and let the notation $\Bar{x} \sim \Bar{y}$ resp. $[\Bar{x} ]$ be
introduced for the equivalence resp. equivalence class of series having the
same infinite tail. The `universe of all Penrose tilings', i.e., the set
$X_{\mathrm{P}}$ of (equivalence classes of) non-isometric infinite, type
$\mathcal P$ tessellations is the factor space $X_{\mathrm{P}}
=M_{\mathrm{P}}/\sim$.

The set $M_{\mathrm{P}} \subset \{L,S\}^{\N}$ with its natural subspace
topology inherited from $\{L,S\}^{\N}$ as well as the ambient space $\{
L,S\}^{\N}$ with its product topology are both totally disconnected
compact spaces homeomorphic to the dyadic Cantor set. Indeed, according to
the construction of the Cantor set, its points can be uniquely labeled
with an infinite sequence of two letters, say $l$ and $s$. The space of
such sequences $\{l,s\}^{\N}$ with the product topology is compact by the
theorem of Tychonov. The sets of sequences having prescribed values at
finitely many points are open-closed sets which separate the points of $\{
l,s\}^{\N}$, thus the space of two-letter-sequences with the product
topology is totally disconnected. The homeomorphism between the
spaces $M_{\mathrm{P}}$ and $\{ l,s\}^{\N}$ can be established by
the recoding $l \leftrightarrow L$, $s \leftrightarrow SL$ of the letters
in the sequences. (According to this homeomorphism the
set $M_{\mathrm{P}}$ could be substituted by the space $\{l,s\}^{\N}$,
which has much simpler coding because of the loss of the grammar rules. The
reason, however, for not doing this is the fact that the description of the
equivalence $\sim$ would be much more difficult in terms of the
symbols $l$ and $s$.)

It is easy to see that for any symbolic sequence $\Bar{x} \in
M_{\mathrm{P}}$ its equivalence class $[\Bar{x}]$ is dense in
$M_{\mathrm{P}}$, so the `Penrose universe' $X_{\mathrm{P}}
=M_{\mathrm{P}} /\sim$ with its natural factor topology is pathologic
from the point of view of ordinary topology. (The closure of every nonempty
subset of it is the whole space $X_{\mathrm{P}}$ itself, so
$X_{\mathrm{P}}$ does not differ much from the one-point topological
space.) The main aim of Section~\ref{Snoncom} is to present more
appropriate methods capable to grab the structure of the factor space
$X_{\mathrm{P}} =M_{\mathrm{P}} /\sim$.

Finally it is worth highlighting the close relationship between Penrose
tilings and Markov shifts. Indeed, given an infinite tessellation $I^0$ of
type $\mathcal P$ with a preferred prototile $p^0 \in I^0$, a `discrete
time dynamics' can be associated to it by applying successive compositions.
After the $n^{\text{th}}$ step an infinite tiling
$I^n= I''{}^{\dots}{}'$ (with $n$ primes) of type $\mathcal
P''{}^{\dots}{}'$ is obtained with preferred tile $p^n\in I^n$ of type
$\mathcal P''{}^{\dots}{}'$, containing $p^0$. The Markov shift
$\varPhi_{\mathrm{P}}$ is constructed by translating the infinite `pointed'
tessellations $(I^n, p^n)$ into binary sequences $\Bar{x}^n \in
M_{\mathrm{P}}$. Thus its phase space is $M_{\mathrm{P}}$, and the
dynamics $\varPhi_{\mathrm{P}}: M_{\mathrm{P}} \to M_{\mathrm{P}}$,
$\Bar{x}^n \mapsto \Bar{x}^{n+1}$ is the unilateral left shift $x^{n+1}_i
=x^n _{i+1}$. Two tessellations $I \sim I^*$ are isometric if and only if
the associated symbolic sequences $\Bar{x} \sim \Bar{x}^*$ have the same
tail, which means that they lie on the same stable manifold of the Markov
shift.

In the next section a well known dynamical system is investigated in order
to point out that in chaotic systems one encounters ill-behaved topological
spaces similar to $X_{\mathrm{P}}$ at every turn.

%% file: pencat3.tex
\section{Chaotic dynamical systems\label{SCat}}
%==============================================
In this section one of the simplest uniformly hyperbolic dynamical system,
Arnold's cat map \cite{A_A} is investigated. First the definition of the
map is given and its stable and unstable manifolds are characterized. Then a
generating Markov partition \cite{CoFoSi82} of the phase space is
presented, and the dynamics as well as the stable and unstable manifolds
are described in terms of the symbolic sequences. This coding scheme will
be the starting point of the investigations in Section~\ref{Snoncom}.

\subsection{Elementary properties of Arnold's cat map\label{SCat_a}}
%-------------------------------------------------------------------
The phase space of the cat map is the two dimensional torus $\T^2
=\R^2/\Z^2$ with its natural Lebesgue measure $\mu$ inherited from the
covering space $\R^2$. The invertible dynamics $\varPhi_A :\T^2 \to \T^2$ is
deduced from a linear map $A:\R^2 \to \R^2$ of the covering space, which is
area preserving, i.e., $\det A =1$, and respects the covering projection
$\pi:\R^2 \to \T^2$, i.e., whenever two points $x,y\in \R^2$ happen to be
on the same fiber [$\pi(x)=\pi(y)$] then their images are also on the same
fiber [$\pi(Ax)=\pi(Ay)$]. This latter condition is satisfied if and only
if the elements of the matrix $A$ are integers and in this case $\varPhi_A
=\pi \circ A \circ \pi^{-1}$ is a well defined continuous automorphism of
the torus $\T^2$ \cite{Walt82}.

\begin{figure}
\centerline{\includegraphics[scale=0.5]{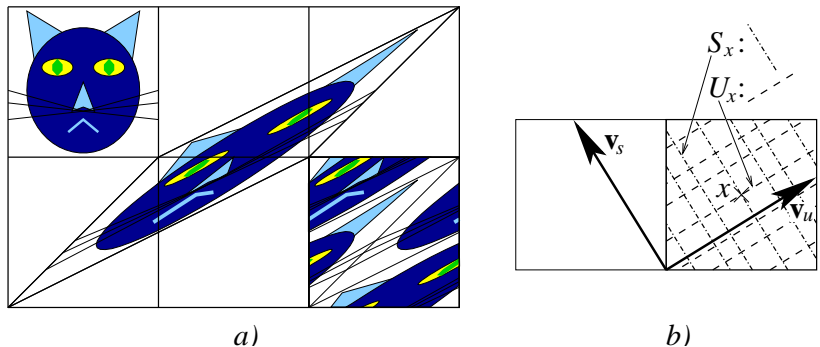}}
\caption{\label{Fcat} {\it a)} The effect of the cat map, illustrated with
the traditional cat portrait; {\it b)} The stable (${\mathbf v}_s$),
unstable (${\mathbf v}_u$) eigenvectors and the stable, unstable manifolds
($S_x$, $U_x$) corresponding to the point $x$.}
\end{figure}

In the followings we fix the matrix $A=\bigl[ \begin{smallmatrix} 2 & 1 \\
1 & 1 \end{smallmatrix} \bigr]$ and omit the subscript of $\varPhi$. The
dynamical system $(\T^2,\mu,\varPhi)$ is called {\it Arnold's cat map} and
its effect on the phase space $\T^2$ as well as on its covering space
$\R^2$ is demonstrated in Fig~\ref{Fcat}.{\it a)}.

The effect of the cat map can be understood via the linear map $A:\R^2 \to
\R^2$ of the covering space. The matrix $A$ has two orthogonal
eigenvectors, ${\mathbf v}_u= \frac{1}{2}\bigl[ \begin{smallmatrix} 2
\\ \sqrt{5}-1 \end{smallmatrix} \bigr]$ and ${\mathbf v}_s=
\frac{1}{2}\bigl[ \begin{smallmatrix} 1-\sqrt{5} \\ 2 \end{smallmatrix}
\bigr]$ corresponding to the eigenvalues $\lambda_u =\frac{3+\sqrt{5}}{2}$
and $\lambda_s =\frac{3-\sqrt{5}}{2}$. The map $A$ stretches by the factor
$\lambda_u >1$ in the {\it unstable direction} ${\mathbf v}_u$ and
contracts by the factor $\lambda_s <1$ in the {\it stable direction}. The
{\it stable} resp. {\it unstable manifold} $S_x$ resp. $U_x$ corresponding
to any phase space point $x\in \T^2$ is the projection by $\pi$ of the line
emanating from $x$ with tangent vector ${\mathbf v}_s$ resp. ${\mathbf
v}_u$, as shown in Fig. \ref{Fcat}.{\it b)}. As a consequence of the above
facts, the cat map is a {\it uniformly hyperbolic} dynamical system.

Since the eigenvalues $\lambda_s$ and $\lambda_u$ are irrational, the lines
$S_x$ and $U_x$ densely fill the whole phase space for any $x\in \T^2$,
i.e., $\overline{S_x} =\overline{U_x} =\T^2$. It is also easy to see that
the disjoint decomposition $\T^2 =\biguplus_{x\in I} S_x$ (resp. $\T^2
=\biguplus_{x\in I} U_x$) of the entire phase space to stable (resp.
unstable) manifolds contains continuously infinite elements, i.e., the
index set $I$ is uncountably infinite.

These properties are very characteristic to a class of strongly chaotic
systems, in the present case they mean that for any two given points $p,f
\in \T^2$ and for any arbitrarily tiny open subset ${\mathcal E}\subset
\T^2$ of the phase space there are infinitely many points $x\in {\mathcal
E}\cap U_p \cap S_f$ which have `the same past' as $p$ and the `same
future' as $f$. This means that the cat map is ergodic and mixing
\cite{A_A,CoFoSi82}.

\subsection{Symbolic dynamics associated to the cat map\label{SCat_b}}
%---------------------------------------------------------------------
In this subsection a generating Markov partition of the phase space of the
cat map is constructed \cite{AdWe67}.

We recall that a finite partition ${\mathcal P} =\{ P_i \}_{i=1}^n$ of the
phase pace $M$ of a uniformly hyperbolic dynamical system $(M,\varPhi,\mu)$
into closed parallelograms $\{P_i\}_{i=1}^n$, the edges of which are
parallel to the stable resp. unstable directions, is {\it Markovian}
\cite{CoFoSi82}, if
\begin{subequations}\label{EDMar}
\begin{gather}
M =\bigcup_{i=1}^n P_i,\qquad \Int(P_i)\bigcap \Int(P_j) = \emptyset,
\qquad\text{if $i\ne j$, and}\\
\varPhi\bigl(\Gamma_s ({\mathcal P}) \bigr) \subset \Gamma_s ({\mathcal P}),
\qquad \varPhi^{-1}\bigl(\Gamma_u ({\mathcal P}) \bigr) \subset \Gamma_u
({\mathcal P}),
\end{gather}
\end{subequations}
\cite{Sin68a,Sin68b,Bow70,CoFoSi82})
where $\Gamma_s ({\mathcal P}) =\bigcup_{i=1}^n \Gamma_s(P_i)$ resp.
$\Gamma_u ({\mathcal P}) =\bigcup_{i=1}^n \Gamma_u(P_i)$ denote the union
of the stable resp. unstable edges of the parallelograms $P_i$, and
$\Int(P)$ is the interior of the set $P$. Moreover, a Markov partition is
{\it generating} \cite{CoFoSi82}, if the intersections
$\varPhi(P_i) \bigcap P_j$ as well as $\varPhi^{-1}(P_i) \bigcap P_j$ (where $i,j
\in \{1,2,\dots n\}$) have at most one connected component (which is again
a parallelogram).

\begin{figure}
\centerline{\includegraphics[scale=0.4]{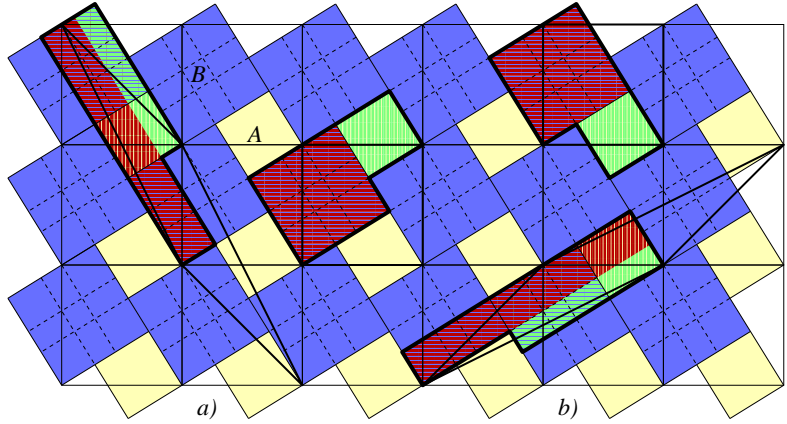}}
\caption{\label{FMar1} The non-generating Markov partition $\T^2 =A\bigcup
B$ of the torus consisting of two squares $A$ and $B$. The backward {\it
a)} and forward iteration {\it b)} of the partition under the cat map
$\varPhi$.}
\end{figure}

\begin{figure}
\centerline{\includegraphics[scale=0.4]{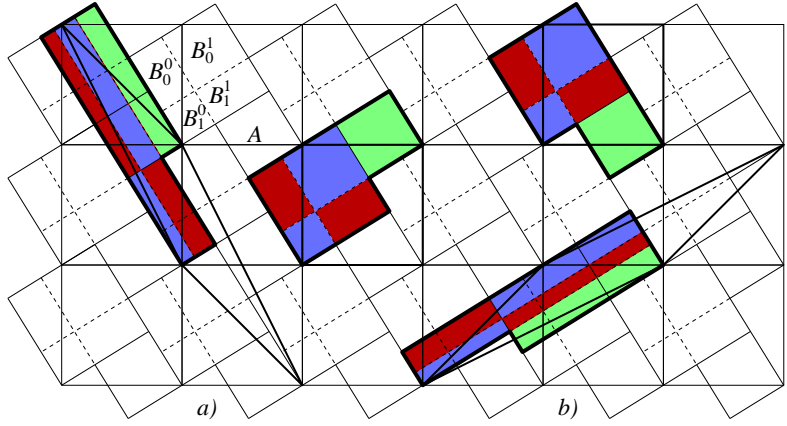}}
\caption{\label{FMar2} The generating Markov partition $P_{\mathrm{C}}
=\{A, B_0^0, B_0^1, B_1^0, B_1^1 \}$ of the torus consisting of five
rectangles. The backward {\it a)} and forward iteration {\it b)} of the
partition $P_{\mathrm{C}}$ under the cat map $\varPhi$.}
\end{figure}

First let us consider the partition of the two dimensional torus $\T^2
=A \bigcup B$ into two disjoint squares $A$ and $B$ whose edges are
parallel to the stable and unstable directions, as it is shown in
Fig.~\ref{FMar1}. It is easy to check that this partition is Markovian,
i.e., fulfills the conditions~\eqref{EDMar}, but it is not generating,
since the intersections $\varPhi(B)\bigcap B$ as well as $\varPhi^{-1}(B)
\bigcap B$ have two disjoint components (Fig.~\ref{FMar1}). This
shortcoming, however, can easily be overcome by subdividing the bigger
square into four smaller rectangles $B_0^0, B_0^1, B_1^0, B_1^1$
shown in Fig.~\ref{FMar2}.

\begin{lemma}\label{LMarpar}
{\bf (Generating Markov partition of the cat map)}
The partition $P_{\mathrm{C}} =\{ A, B_0^0, B_0^1, B_1^0, B_1^1 \}$ is
a generating Markov partition of five elements. The grammar rules
induced by the dynamics are depicted schematically and encoded in a
transition matrix $T_{\mathrm{C}}$ as follows:
\begin{align}\label{Egram}
\unitlength 0.8mm
\raisebox{-9mm}{
\begin{picture}(60,24)
\thinlines
\put(15,17){\vector(0,-1){10}}
\put(15,17){\vector(1,-1){10}}
\put(15,17){\vector(2,-1){20}}
\put(25,17){\vector(2,-1){20}}
\put(25,17){\vector(3,-1){30}}
\put(35,17){\vector(-2,-1){20}}
\put(35,17){\vector(-1,-1){10}}
\put(35,17){\vector(0,-1){10}}
\put(45,17){\vector(0,-1){10}}
\put(45,17){\vector(1,-1){10}}
\put(55,17){\vector(-4,-1){40}}
\put(55,17){\vector(-3,-1){30}}
\put(55,17){\vector(-2,-1){20}}
\put(0,17){\makebox(10,7){$x_i :$}}
\put(10,17){\makebox(10,7){$A$}}
\put(20,17){\makebox(10,7){$B_0^0$}}
\put(30,17){\makebox(10,7){$B_0^1$}}
\put(40,17){\makebox(10,7){$B_1^0$}}
\put(50,17){\makebox(10,7){$B_1^1$}}
\put(0,0){\makebox(10,7){$x_{i+1} :$}}
\put(10,0){\makebox(10,7){$A$}}
\put(20,0){\makebox(10,7){$B_0^0$}}
\put(30,0){\makebox(10,7){$B_0^1$}}
\put(40,0){\makebox(10,7){$B_1^0$}}
\put(50,0){\makebox(10,7){$B_1^1$}}
\end{picture}}&&
T_{\mathrm{C}} &=
\begin{bmatrix}
1& 0& 1& 0& 1\\
1& 0& 1& 0& 1\\
1& 0& 1& 0& 1\\
0& 1& 0& 1& 0\\
0& 1& 0& 1& 0
\end{bmatrix}.
\end{align}
\end{lemma}
(The $(i,j) \in P_{\mathrm{C}}^2$ element of the transition matrix is
$T_{ij} =1$, if the interior of the parallelogram $j$ is mapped by the
dynamics into $i$, i.e., $\Int\bigl( \varPhi(j)\bigr) \bigcap \Int(i) \ne
\emptyset$.)

\begin{proof}
The proof of the first statement as well as the structure of the allowed
transitions can be easily read off from Figure~\ref{FMar2}.
\end{proof}

Almost every phase space point $x$ is uniquely represented by an in both
directions infinite series $\Bar{x}= \{ x_i\}_{i\in \Z} \in
P_{\mathrm{C}}^{\Z}$ of letters from the finite alphabet $P_{\mathrm{C}}$
defined by the partition elements in which the successive forward and
backward iterations of $x$ fall, i.e., $\varPhi^i (x) \in x_i$. The inverse
statement is also true; every symbolic sequence $\Bar{x} =\{ x_i\}_{i\in
\Z} \in P^\Z$, which satisfies the finite grammar rules \eqref{Egram},
defines a unique point of the phase space. Thus, introducing the notation
\begin{equation}\label{EMCat}
M_{\mathrm{C}}=\left\{ \{x_i\}_{i\in \Z} \in P_{\mathrm{C}}^\Z \mid
\{x_i\}_{i \in \Z} \text{ satisfies the grammar rules \eqref{Egram} }
\right\},
\end{equation}
for the (topological) space of all possible series defined by the dynamics,
there is a $\mu$-almost everywhere defined map
\begin{align}\label{Epicat}
&&\pi_{\mathrm{C}}: \T^2 &\rightarrowtail M_{\mathrm{C}},&
x &\mapsto \Bar{x}=\{ x_i\}_{i\in \Z},&
(\text{here }& \varPhi^i (x) \in x_i \in P_{\mathrm{C}}),
\end{align}
which is bijective on its domain. This is another way to express that
$P_{\mathrm{C}}$ (Fig.~\ref{FMar2}) is a generating Markov partition of
$\T^2$. (The phase space points $y\in \T^2$ on which $\pi_{\mathrm{C}}$
is not defined are the ones lying on the forward or backward iterations of
the dividing lines of the Markov partition.)

The stable (resp. unstable) manifolds can also be easily described in
terms of the symbolic dynamics; two points $x,y\in \T^2$ lie on the same
stable (resp. unstable) manifold if and only if their symbolic sequences
$\Bar{x}$, $\Bar{y}$ coincide after (resp. before) a threshold index $N\in
\Z$, i.e., $x_i =y_i$ for all $i>N$ (resp. for all $i<N$).

As we did for the Penrose tilings, let us denote the equivalence relation
indicating that the points $x,y\in \T^2$ corresponding to the sequences
$\Bar{x}, \Bar{y} \in M_{\mathrm{C}}$ are on the same stable resp.
unstable manifold by $\Bar{x}\sim_s \Bar{y}$ resp. $\Bar{x}\sim_u \Bar{y}$.
This means that $\Bar{x}\sim_s \Bar{y}$ (resp. $\Bar{x}\sim_u \Bar{y}$)
holds if and only if $S_x =S_y$ (resp. $U_x =U_y$), and the equivalence
classes $[x]_s$ (resp. $[x]_u$) are the $S_x$ stable (resp. $U_x$ unstable)
manifolds.

The topology of the factor spaces $X_{\mathrm{C}}^s :=M_{\mathrm{C}}
/\sim_s$ (resp. $X_{\mathrm{C}}^u :=M_{\mathrm{C}}
/\sim_u$) are again ill behaved with their natural factor topology, since
the equivalence classes of $\sim_s$ (resp. $\sim_u$) are dense sets in
$M_{\mathrm{C}}$.

The structural similarity between the universe of Penrose tilings and
Arnold's cat map, which is a uniformly hyperbolic dynamical system, is
quite clear by now. It is also understood that this similarity is due to a
Markov shift hiding in the background. To make this parallelism more
tight, we give the counterpart of Statement~\ref{StPencod} for the cat map
$(\T^2, \mu, \varPhi)$.

\begin{statement}\label{StCat}
{\bf (Local indistinguishability of symbolic trajectories in the cat map)}

{\it i)}
Every finite symbolic section $\Bar{f} =\{f_i\}_{i=0}^n$ (where $n<\infty$)
satisfying the grammar rules~\ref{Egram} (with $f$ instead of $x$) appears
infinitely many times in the infinite sequence $\Bar{x} \in
M_{\mathrm{C}}$ of $\mu$-almost every phase space point $x\in \T^2$.

{\it ii)}
Moreover, the frequency $\kappa_{\Bar{f}}$ of occurrence of $\Bar{f}$ in
$\Bar{x}$ is a fixed number depending solely on $\Bar{f}$, but not on the
selected point $x$.

{\it iii)}
Finally, the frequency of appearance $\kappa_{\Bar{f}}$ is an element of
the dense subgroup $\zeta \Z+\eta \Z \subset \R$ of the additive group of
reals, where $\zeta =\frac{1}{2}-\frac{\sqrt{5}}{10}$ and $\eta
=\frac{\sqrt{5}}{5}$.
\end{statement}

The exclusion of a zero measure set in the statement is necessary for two
reasons. First, the mapping $\pi_{\mathrm{C}}$ (given
in~\ref{Epicat}) is defined only on a subset of maximal measure of
$\T^2$, and second, the ergodic theorem, which we going to utilize in the
proof, is also valid only with the exclusion of a zero measure set.

\begin{proof}
The first two statements~{\it i)}, {\it ii)} are simple consequences of the
ergodicity of the cat map. Indeed, denoting with $X(\Bar{f})$ the set
$X(\Bar{f}) =\big\{ x\in \T^2 | (\pi_{\mathrm{C}} x)_i =x_i =f_i \text{
for all } i\in\{0,1\ldots n\} \big\} \subset \T^2$ of phase space points
having a symbolic sequence starting with $\Bar{f}$, and applying the
ergodic theorem \cite{A_A,CoFoSi82,Walt82} for the characteristic function
$\chi_{X(\Bar{f})}$ of $X(\Bar{f}) \subset \T^2$, we get
\begin{equation}\label{Ekapcat}
\kappa_{\Bar{f}} =\mu\big(X(\bar{f}) \big).
\end{equation}

To prove assertion~{\it iii)} of the statement the area $\mu\big(X(\bar{f})
\big)$ has to be determined. Elementary calculations yield that
\begin{subequations}\label{EmuMar}
\begin{gather}
\mu(A)= \mu(B_0^1 )=\frac{5-\sqrt{5}}{10}= \zeta,
\qquad
\mu(B_0^0 )= \mu(B_1^1 )= \frac{3\sqrt{5}-5}{10} =2\eta- \zeta,
\\
\mu(B_1^0 )= \frac{5-2\sqrt{5}}{5} =2\zeta-\eta.
\end{gather}
\end{subequations}
It is also straightforward to determine the ratio $a_{ij} =\frac{\mu\big(
j \bigcap \varPhi^{-1} (i) \big)} {\mu(j)}$ of the phase space points
in the parallelogram $j\in P_{\mathrm{C}}$, whose first iterate
falls into $i\in P_{\mathrm{C}}$. In matrix notation
\begin{equation}\label{Eaijcat}
[a_{ij}] =
\begin{bmatrix}
\lambda_s   & 	0& 	\lambda_s&    	0& 	\lambda_s\\
1-2\lambda_s& 	0& 	1-2\lambda_s& 	0& 	1-2\lambda_s\\
\lambda_s&    	0& 	\lambda_s&    	0& 	\lambda_s\\
0&        	\lambda_s& 	0&   	\lambda_s& 	0\\
0& 		1-\lambda_s& 	0&	1-\lambda_s& 	0
\end{bmatrix},
\end{equation}
where $\lambda_s =\frac{3-\sqrt{5}}{2}$ is the stable eigenvalue of the
map.

The phase space area $\mu\big(X(\Bar{f})\big)$ corresponding to the finite
symbolic section $\Bar{f} =\{f_i\}_{i=0}^n$ is clearly
\begin{equation}\label{EmuXf}
\mu\big(X(\Bar{f})\big) =
a_{f_n f_{n-1}} \cdots a_{f_3 f_2} a_{f_2 f_1} a_{f_1 f_0} \mu(f_0).
\end{equation}
For finishing the proof of assertion~{\it iii)} of the statement we have to
notice only that the multiplication with $\lambda_s$ does not lead out of
the set $\zeta \Z +\eta \Z$. Indeed, elementary calculations yield that
\begin{align}\label{Eelmcalc}
\lambda_s (\zeta n+\eta m) &=\zeta (2n-m) +\eta (2m -n)&&
\text{for all $n,m\in \Z$.}
\end{align}
Since the matrix elements $a_{ij}$ have the form $k+\lambda_s l$
(by equation~\eqref{Eaijcat}) and $\mu(f_0) = \zeta n +\eta m$
(by equation~\eqref{EmuMar}), with $k,l,m,n \in \Z$ integers, the
value~\eqref{EmuXf} of $\mu\big(X(\Bar{f})\big)$ is also in the group
$\zeta\Z +\eta \Z$ for arbitrary finite symbolic section $\Bar{f}$.
\end{proof}

In the next section we demonstrate how the methods of noncommutative
geometry can be utilized in the study of the two systems (Penrose tilings,
cat map) described above.

%% file: pencat4.tex
\section{Noncommutative geometrical approach\label{Snoncom}}
%===========================================================
The previous two sections revealed the intimate relation between certain
aperiodic tilings and chaotic dynamical systems. In both cases there
naturally appears a topological space ($M_{\mathrm{P}}$ resp.
$M_{\mathrm{C}}$) with an equivalence relation ($\sim$ resp. $\sim_s$,
$\sim_u$), and one urges to study the factor spaces ($X_{\mathrm{P}}=
M_{\mathrm{P}} /\sim$ resp. $X_{\mathrm{C}}^s=  M_{\mathrm{C}}/ \sim_s$,
$X_{\mathrm{C}}^u=  M_{\mathrm{C}}/ \sim_u$), which are pathologic as
topological spaces with their inherited factor topologies. In practice, it
means that it is equally impossible to establish the equality of two
Penrose tilings knowing only finite (but arbitrarily large) patches of
them, as it is to make statements about the ultimate fate of phase space
points in chaotic systems, if the position of the points is not known with
an infinite precision.

In this section we demonstrate how the methods of noncommutative geometry
\cite{Con0} can be exploited to give a more appropriate description of the
factor space $X=M/\sim$. (In the general considerations the subscripts
`P' as well as `C' are omitted.) Since this section is mathematically
more demanding than the two previous ones, two appendices have been
included for the sake of better intelligibility, which summarize the basic
properties of approximately finite dimensional (AF) $C^*$-algebras and the
most important constructions of their $K$-theory.

In the first part of this section a noncommutative $C^*$-algebra $C^*
(M,\sim)$ is associated to the factor space $X=M/\sim$, which is, due to
the Markov property of the investigated systems and the finiteness of the
grammar rules, an {\it approximately finite dimensional} (AF)
$C^*$-algebra. (See Appendix~\ref{Aa4}.) It is worth noting that the
commutative $C^*$-algebra $C(X)$ of continuous complex valued functions,
which unambiguously describes well-behaved (i.e., locally compact
Hausdorff) topological spaces (see Appendix~\ref{Aa2}), constitute the
center of $C^* (M,\sim)$, thus it is obvious that the noncommutative
algebra $C^* (M,\sim)$ carries more information about the structure of
$X=M/\sim$ than the commutative one $C(X)$. Not to mention the fact that in
the case of our two systems, due to the non separable topology of $X$, the
commutative algebras $C(X_{\mathrm{P}}) \cong C(X_{\mathrm{C}}) \cong
\C$ are trivial. (Indeed, the only continuous functions on
$X_{\mathrm{P}}$ or $X_{\mathrm{C}}$ are the constant ones.)

In the second part of the section, just to demonstrate the force of the
noncommutative theory as opposed to the usual topology, we calculate the
$K_0$ group of the noncommutative algebras $C^* (M_{\mathrm{P}}, \sim)$
and $C^* (\tilde{M}_{\mathrm{C}}, \sim)$, which is a complete invariant in
the case of AF algebras. (See Appendix~\ref{Ab3}.)

A part of this section (the material concerning the Penrose tilings)
is strongly motivated by (and partially overlaps with) the sections II.2
and II.3 of~\cite{Con0}.

\subsection{The $C^*$-algebra associated to factor spaces\label{SAF}}
%--------------------------------------------------------------------

Our first task is to define the noncommutative algebra $C^* (M, \sim)$
associated to the space $M$, which possesses two structures, a topology and
a partition into equivalence classes. These two structures are many times,
like in our concrete cases, inconsistent to each other. (The equivalence
classes are everywhere dense sets; every open set contains elements from
all equivalence classes.)

The most satisfactory way would be to define the operation $C^*$ as a
functor from the category of certain topological spaces with equivalence
relations to the category of $C^*$ algebras, but this approach is far
beyond our present aim. We content ourselves with the simplest
definition valid for finite spaces, since the topological spaces
$M_{\mathrm{P}}$ resp. $M_{\mathrm{C}}$ can be presented as projective
limits of finite spaces.

So let us forget about the topology, more precisely, let us suppose
that $M$ is a finite space with discrete topology, e.g. $M=\{ m_1, m_2,
m_3, m_4, m_5\}$ with equivalence classes $m_1 \sim m_2 \sim m_3$ and $m_4
\sim m_5$. One algebraic way for the description of the space $X=M/\sim$ is
to consider the (continuous) complex valued functions on $M$ which are
constant on the equivalence classes. This construction yields the
commutative algebra $C(X)$ which is in our present example $\C^2$, since
$M$ has two equivalence classes.

Another algebraic way to describe the factorization $M/\sim$ is to consider
the matrix algebra
\begin{equation}\label{EC*}
C^*(M,\sim) =\left\{ \left.
\begin{bmatrix} a& b& c& 0& 0\\
                d& e& f& 0& 0\\
                g& h& i& 0& 0\\
                0& 0& 0& j& k\\
                0& 0& 0& l& m
\end{bmatrix}
\right| a,b\ldots m \in \C \right\} \cong {\mathcal M}_3 \oplus {\mathcal
M}_2 \subset {\mathcal M}_5,
\end{equation}
with the usual (noncommutative) matrix product and adjunction. Roughly, we
associate a full matrix algebra to each equivalence class, the rank of
which agrees with the number of elements in the equivalence class, and then
we take the direct sum of these algebras. There is a natural inclusion
\begin{align}\label{ECX}
C(X)\cong \C^2 &\hookrightarrow C^*(M,\sim) \cong {\mathcal M}_3 \oplus
{\mathcal M}_2, &
(a,b) &\mapsto \diag [a,a,a,b,b],
\end{align}
which is a bijection between $C(X)$ and the center of $C^*(M,\sim)$.

Another clever way of looking at the algebra $C^*(M,\sim)$ is to consider
its elements as complex valued functions on the set $R_{\sim}$ of
equivalent pairs in $M$, i.e.,
\begin{subequations}\label{EC*R}
\begin{align}\label{EC*R:a}
C^*(M,\sim) &= \big\{ \text{(certain) $R_{\sim} \to \C$ functions}\big\}
\intertext{where}
\label{EC*R:b}
R_\sim &=\big\{ (m,n) \in M^2 \;|\; m\sim n \big\} \subset M^2.
\end{align}
\end{subequations}
(For finite $M$ all $R_{\sim} \to \C$ functions are allowed, but for
infinite sets there are restrictions for the continuity and norm-finiteness
of the functions, for details see \cite{Con0}.) The product of two such
functions, $f,g$ at $(m,n) \in R_\sim$ is given by
\begin{equation}\label{Ef*g}
(f \cdot g) (m,n) =\sum_{\substack{k \\ k\sim m\sim n}} f(m,k) g(k,n).
\end{equation}

For finite, discrete $M$ the two definitions~\eqref{EC*} and \eqref{EC*R}
of the algebra \mbox{$C^*(M, \sim)$} are clearly equivalent, a function
$f:R_{\sim} \to \C$ in definition~\eqref{EC*R}, evaluated at the point
$(i,j)\in R_{\sim}$ just gives the $(i,j)$ matrix element of the
corresponding matrix in definition~\eqref{EC*}. For infinite sets with
nontrivial topology the second definition can be more easily generalized
\cite{Con0}, but it is beyond our present needs.

\subsubsection{The AF $C^*$-algebra of the Penrose tilings\label{SAFPen}}
%'''''''''''''''''''''''''''''''''''''''''''''''''''''''''''''''''''''
In this part the noncommutative AF algebra $C^*(M_{\mathrm{P}},\sim)$ is
constructed, by giving its Brattely diagram.

It is a simple topological fact that the totally disconnected
topological space $M_{\mathrm{P}}$ is the projective limit
$M_{\mathrm{P}}= \varprojlim_{i\in \N_+} M_i$ of the finite (discrete)
spaces $M_i \subset \{ L,S\}^i$, which consist of finite series (of length
$i$) satisfying the grammar rules~\eqref{EPensym}. The surjections $\pi_i
:M_{i+1} \twoheadrightarrow M_i$ are given by omitting the last symbol of
the sequences in $M_{i+1}$. Recall that, by definition, the projective
limit means that there exist (continuous) surjections $\sigma_i
:M_{\mathrm{P}} \twoheadrightarrow M_i$ such that the upper
(unprimed) part of the diagram bellow commutes, and $M_{\mathrm{P}}$ is
universal, i.e., whenever given another candidate $M'$ and $\sigma'_i :M'
\twoheadrightarrow M_i$ with the same properties, then there is a {\it
unique} (continuous) map $\varphi: M' \to M$ making the whole diagram
commuting.

\begin{equation}\label{DprojPen}
\unitlength 0.5mm
\raisebox{-11mm}{
\begin{picture}(135,48)
\thinlines
\put(4,24){\makebox(0,0){$M_1$}}
\put(32,24){\makebox(0,0){$M_2$}}
\put(60,24){\makebox(0,0){$M_3$}}
\put(88,24){\makebox(0,0){$M_4$}}
\put(124,43){\makebox(0,0){$M_{\mathrm{P}}$}}
\put(124,5){\makebox(0,0){$\color{blue}M'$}}
\put(99,24){\makebox(0,0){$\cdots$}}
\put(18,27){\makebox(0,0){$\pi_1$}}
\put(46,27){\makebox(0,0){$\pi_2$}}
\put(74,27){\makebox(0,0){$\pi_3$}}
\put(27,24){\vector(-1,0){18}}\put(27,24){\vector(-1,0){17}}
\put(55,24){\vector(-1,0){18}}\put(55,24){\vector(-1,0){17}}
\put(83,24){\vector(-1,0){18}}\put(83,24){\vector(-1,0){17}}
\put(8,33){\vector(-1,-1){4}}\put(8,33){\vector(-1,-1){3}}
\qbezier(5,30)(18,43)(116,43)
\put(36,33){\vector(-1,-1){4}}\put(36,33){\vector(-1,-1){3}}
\qbezier(33,30)(46,43)(116,43)
\put(64,33){\vector(-1,-1){4}}\put(64,33){\vector(-1,-1){3}}
\qbezier(61,30)(74,43)(116,43)
\put(92,33){\vector(-1,-1){4}}\put(92,33){\vector(-1,-1){3}}
\qbezier(89,30)(102,43)(116,43)
{\color{blue}
\put(8,15){\vector(-1,1){4}}\put(8,15){\vector(-1,1){3}}
\qbezier(5,18)(18,5)(116,5)
\put(36,15){\vector(-1,1){4}}\put(36,15){\vector(-1,1){3}}
\qbezier(33,18)(46,5)(116,5)
\put(64,15){\vector(-1,1){4}}\put(64,15){\vector(-1,1){3}}
\qbezier(61,18)(74,5)(116,5)
\put(92,15){\vector(-1,1){4}}\put(92,15){\vector(-1,1){3}}
\qbezier(89,18)(102,5)(116,5)}
{\color{red}
\put(124,10){\vector(0,1){28}}
\put(130,24){\makebox(0,0){$\exists ! \varphi$}}}
\put(8,38){\makebox(0,0){$\sigma_1$}}
\put(104,34){\makebox(0,0){$\sigma_4$}}
\put(8,10){\makebox(0,0){$\color{blue}\sigma'_1$}}
\put(104,14){\makebox(0,0){$\color{blue}\sigma'_4$}}
\end{picture}}
\end{equation}
This abstract and elegant definition of $M_{\mathrm{P}}=
\varprojlim_{i\in \N_+} M_i$ really yields the same topological space as
the concrete one~\eqref{EMPen} given in Section~\ref{SPen_b}.

The equivalence relation $\sim$ on $M_{\mathrm{P}}$ can also be described
in a similar manner. Two series $\Bar{x}, \Bar{y} \in M_{\mathrm{P}}$ are
equivalent, $\Bar{x} \sim \Bar{y}$, if their tails coincide either from the
first position or from the second or from the third or \dots from the
$n^{\rm th}$ position ($n \in \N$). This trivial observation, disguised in
a mathematical formulation, is the statement that the equivalence relation
$\sim$ on $M_{\mathrm{P}}$ is the injective limit (increasing union)
$(M_{\mathrm{P}}, \sim) =\varinjlim_{i\in \N_+}
(M_{\mathrm{P}},\sim_i)$, where the equivalence $\Bar{x} \sim_i \bar{y}$
means that the series $\Bar{x}$, $\Bar{y}$ coincide from the $i^{\rm th}$
position on, i.e., $x_j =y_j$ for all $j\ge i$. [The injections in the
limit are the identity maps $\id_{M_{\mathrm{P}}}
:\mbox{$(M_{\mathrm{P}},\sim_i)$} \to
\mbox{$(M_{\mathrm{P}},\sim_{i+1})$}$, which preserve the relations, i.e.,
$\Bar{x} \sim_i \Bar{y} \Rightarrow \Bar{x} \sim_{i+1} \Bar{y}$.] The
diagram corresponding to the above injective limit construction is the
following:
\begin{equation}\label{DinjPen}
\unitlength 0.5mm
\raisebox{-11mm}{
\begin{picture}(182,48)
\thinlines
\put(16,24){\makebox(0,0){$(M_{\mathrm{P}},\sim_1)$}}
\put(66,24){\makebox(0,0){$(M_{\mathrm{P}},\sim_2)$}}
\put(116,24){\makebox(0,0){$(M_{\mathrm{P}},\sim_3)$}}
\put(138,24){\makebox(0,0){$\cdots$}}
\put(166,5){\makebox(0,0){$(M_{\mathrm{P}},\sim)$}}
\put(172,24){\makebox(0,0){$\color{red}\exists ! \varphi$}}
\put(166,43){\makebox(0,0){$\color{blue}(M',\sim')$}}
\put(32,24){\vector(1,0){18}}
\put(82,24){\vector(1,0){18}}
{\color{red}\put(166,10){\vector(0,1){28}}}
{\color{blue}
\put(146,43){\vector(1,0){4}}\qbezier[1000](16,29)(30,43)(150,43)
\qbezier(66,29)(80,43)(150,43)
\qbezier(116,29)(130,43)(150,43)}
\put(146,5){\vector(1,0){4}}\qbezier[1000](16,19)(30,5)(150,5)
\qbezier(66,19)(80,5)(150,5)
\qbezier(116,19)(130,5)(150,5)
\put(41,27){\makebox(0,0){$\id$}}
\put(91,27){\makebox(0,0){$\id$}}
\put(32,8){\makebox(0,0){$\rho_1$}}
\put(32,40){\makebox(0,0){\color{blue}$\rho'_1$}}
\put(132,14){\makebox(0,0){$\rho_3$}}
\put(132,34){\makebox(0,0){\color{blue}$\rho'_3$}}
\end{picture}}
\end{equation}

(Well, it is not really an `increasing' union, since the sets $M_{\mathrm
P}$ are the same in all terms of the sequence, so nothing really increases.
The relations $\sim_i$, however, are not the same, so the construction does
make sens.) The arrows $\rho_i$ are injective, relation preserving
mappings. (They are simply the identity maps $\rho_i =\id_{M_{\mathrm{P}}}:
M_{\mathrm P} \to M_{\mathrm P}$, because of the above mentioned
speciality.) The injective limit \mbox{$(M_{\mathrm{P}},\sim)$} has again
the universal property, i.e., given any other candidate \mbox{$(M',\sim')$}
for the injective limit and morphisms $\rho'_i
:\mbox{$(M_{\mathrm{P}},\sim_i)$} \to \mbox{$(M',\sim')$}$ (with the same
properties as $\rho_i$ have), there exists a {\it unique} relation
preserving map $\varphi: \mbox{$(M_{\mathrm{P}},\sim)$} \to
\mbox{$(M',\sim')$}$ for which the whole diagram commutes.

This diagram is quite similar to the previous one~\eqref{DprojPen}, just
the arrows are reversed. Perhaps this fact is responsible for the
incompatibility between the topology of $M_{\mathrm{P}}$ and the
relation $\sim$ on it.

With the help of the projections $p_i :M_{\mathrm P} \twoheadrightarrow
M_i$, which omit the symbols after the $i^{\textrm{th}}$ position of the
finite sequence [i.e., $p_i :(x_1,x_2 \ldots x_i, x_{i+1} \ldots) \mapsto
(x_1,x_2 \ldots x_i )$], the (unprimed parts of the) two diagrams can be
merged together.

\begin{equation}\label{DPen}
\unitlength 0.5mm
\raisebox{-16mm}{
\begin{picture}(182,67)
\thinlines
\put(16,24){\makebox(0,0){$(M_{\mathrm{P}},\sim_1)$}}
\put(66,24){\makebox(0,0){$(M_{\mathrm{P}},\sim_2)$}}
\put(116,24){\makebox(0,0){$(M_{\mathrm{P}},\sim_3)$}}
\put(138,24){\makebox(0,0){$\cdots$}}
\put(16,43){\makebox(0,0){$(M_1,\sim^1)$}}
\put(66,43){\makebox(0,0){$(M_2,\sim^2)$}}
\put(116,43){\makebox(0,0){$(M_3,\sim^3)$}}
\put(138,43){\makebox(0,0){$\cdots$}}
\put(16,29){\vector(0,1){9}} \put(16,29){\vector(0,1){8}}
\put(21,33){\makebox(0,0){$p_1$}}
\put(66,29){\vector(0,1){9}} \put(66,29){\vector(0,1){8}}
\put(71,33){\makebox(0,0){$p_2$}}
\put(116,29){\vector(0,1){9}} \put(116,29){\vector(0,1){8}}
\put(121,33){\makebox(0,0){$p_3$}}
\put(166,5){\makebox(0,0){$(M_{\mathrm{P}},\sim)$}}
\put(166,62){\makebox(0,0){$M_{\mathrm{P}}$}}
%\put(170,34){\makebox(0,0){$\id$}}
%\put(166,10){\vector(0,1){47}}\put(166,57){\vector(0,-1){47}}
%
\put(32,24){\vector(1,0){18}}
\put(82,24){\vector(1,0){18}}
\put(41,27){\makebox(0,0){$\id$}}
\put(91,27){\makebox(0,0){$\id$}}
\put(50,43){\vector(-1,0){18}}\put(50,43){\vector(-1,0){17}}
\put(100,43){\vector(-1,0){18}}\put(100,43){\vector(-1,0){17}}
\put(41,46){\makebox(0,0){$\pi_1$}}
\put(91,46){\makebox(0,0){$\pi_2$}}
\put(20,52){\vector(-1,-1){4}}\put(21,53){\vector(-1,-1){4}}
\qbezier[1000](17,49)(30,62)(150,62)
\put(70,52){\vector(-1,-1){4}}\put(71,53){\vector(-1,-1){4}}
\qbezier(67,49)(80,62)(150,62)
\put(120,52){\vector(-1,-1){4}}\put(121,53){\vector(-1,-1){4}}
\qbezier(117,49)(130,62)(150,62)
\put(146,5){\vector(1,0){4}}\qbezier[1000](16,19)(30,5)(150,5)
\qbezier(66,19)(80,5)(150,5)
\qbezier(116,19)(130,5)(150,5)
\put(32,8){\makebox(0,0){$\id$}}
\put(32,59){\makebox(0,0){$\sigma_1$}}
\put(132,14){\makebox(0,0){$\id$}}
\put(132,53){\makebox(0,0){$\sigma_3$}}
\end{picture}}
\end{equation}

The top row of the diagram describes the topology of $M_{\mathrm P}$ while
the bottom row is related to the equivalence relation $\sim$. With the help
of the projections $p_i$ the relations $\sim_i$ can be `pushed
forward' to the finite sets $M_i$, the resulting relations are
denoted by $\sim^i$. The relation $\Bar{x} \sim^i \Bar{y}$ holds, by
definition, between two finite symbolic series $\Bar{x}
\sim^i \Bar{y}$ (here $\Bar{x}, \Bar{y} \in M_i$)
if and only if there exist two $\sim_i$-equivalent infinite series
$\Bar{x}' \sim_i \Bar{y}' \in M_{\mathrm P}$ such that they project to
the finite sequences, i.e., $p_i (\Bar{x}') =\Bar{x}$ and $p_i (\Bar{y}')
=\Bar{y}$. In our case it means simply that the last symbols of $\Bar{x}$
and $\Bar{y}$ agree, i.e., $x_i =y_i$. From this it is also clear that
$\sim^i$ is an equivalence relation on $M_i$, too. We note that the
surjections $\pi_i$ do not preserve this equivalence.

Up to this point our investigations were purely `commutative', we used only
classical topological concepts, and the machinery of noncommutative
geometry was not exploited. Now we associate noncommutative $C^*$-algebras
to the spaces $(M_i, \sim^i)$ appearing in the first row of the
diagram~\eqref{DPen}, and substitute injective $C^*$-algebra homomorphisms
for the continuous surjections $\pi_i$ between them.

The noncommutative $C^*$-algebra \mbox{$C^*(M_i, \sim^i)$} associated to
the finite, partitioned space \mbox{$(M_i,\sim^i)$} is constructed
according to the method demonstrated at the beginning of the section [see
\eqref{EC*} and \eqref{EC*R}]. Since there are two equivalence classes in
every set $M_i$ (labeled by the last symbols of the series), this algebra
is isomorphic to the direct sum of two matrix algebras, $\mbox{$C^*(M_i,
\sim^i)$} \cong {\mathcal M}_{f_i} \oplus {\mathcal M}_{f_{i+1}}$, where
the dimensions $f_i$ are just the Fibonacci numbers $1,1,2,3,5,8\ldots$.
Indeed, let $f_i$ be the number of series $\Bar{x} \in M_i$ with last
symbol $x_i =S$. It agrees with the number of series in $M_{i-1}$ ending
with $L$, since every $S$ is preceeded by an $L$, according to the grammar
rules~\eqref{EPensym}. Thus the number of sequences in $M_i$ with a last
symbol $L$ is $f_{i+1}= f_{i-1}+ f_i$, since it follows either an $S$ or an
$L$ in $M_{i-1}$.

The construction of the $C^*$-algebra homomorphisms
$C^*(\pi_i): \mbox{$C^*(M_i, \sim^i)$} \to \mbox{$C^*(M_{i+1},
\sim^{i+1})$}$ is also natural, but it is a bit more complicated.
First we have to consider the mappings incorporated in the following
diagram
\begin{equation}\label{DPen2}
\unitlength 0.5mm
\raisebox{-8mm}{
\begin{picture}(212,35)
\thinlines
\put(20,5){\makebox(0,0){$R^1$}}
\put(100,5){\makebox(0,0){$R^2$}}
\put(180,5){\makebox(0,0){$R^3$}}
\put(206,5){\makebox(0,0){$\cdots$}}
\put(20,30){\makebox(0,0){$M_1^2$}}
\put(100,30){\makebox(0,0){$M_2^2$}}
\put(180,30){\makebox(0,0){$M_3^2$}}
\put(206,30){\makebox(0,0){$\cdots$}}
\put(20,12){\oval(4,4)[b]}\put(22,12){\vector(0,1){13}}
\put(27,18){\makebox(0,0){$\iota_1$}}
\put(100,12){\oval(4,4)[b]}\put(102,12){\vector(0,1){13}}
\put(107,18){\makebox(0,0){$\iota_2$}}
\put(180,12){\oval(4,4)[b]}\put(182,12){\vector(0,1){13}}
\put(187,18){\makebox(0,0){$\iota_3$}}
\put(85,30){\vector(-1,0){50}}\put(85,30){\vector(-1,0){48}}
\put(60,33){\makebox(0,0){$\pi_1 \times \pi_1$}}
\put(165,30){\vector(-1,0){50}}\put(165,30){\vector(-1,0){48}}
\put(140,33){\makebox(0,0){$\pi_2 \times \pi_2$}}
\end{picture}}
\end{equation}
where $\pi_i \times \pi_i :M_{i+1}^2 \twoheadrightarrow M_i^2$ is the
surjection $(\Bar{x}, \Bar{y}) \mapsto \big( \pi_i (\Bar{x}), \pi_i
(\Bar{y}) \big)$, the symbol $R^i =\big\{ (\Bar{x}, \Bar{y}) \in M_i^2
\;|\; \Bar{x} \sim^i \Bar{y} \big\} \subset M_i^2$ denotes the subset
representing the $\sim^i$-equivalent pairs, and $\iota_i :R^i
\hookrightarrow M_i^2$ is the natural inclusion.

Let us apply the pull-back functor to this diagram, i.e., instead of the
spaces $M_i^2$ resp. $R^i$ let us consider the set of (continuous) complex
valued functions $C(M_i^2)$ resp. $C(R^i)$ with the appropriate pull-back
mappings between them. (Since the spaces $M_i^2$ and $R^i$ are finite
spaces with discrete topology, the continuity of the functions means no
restriction.)
\begin{equation}\label{DCPen2}
\unitlength 0.5mm
\raisebox{-8mm}{
\begin{picture}(212,35)
\thinlines
\put(20,5){\makebox(0,0){$C^*(M_1,\sim^1)$}}
\put(100,5){\makebox(0,0){$C^*(M_2,\sim^2)$}}
\put(180,5){\makebox(0,0){$C^*(M_3,\sim^3)$}}
\put(206,5){\makebox(0,0){$\cdots$}}
\put(20,30){\makebox(0,0){$C(M_1^2)$}}
\put(100,30){\makebox(0,0){$C(M_2^2)$}}
\put(180,30){\makebox(0,0){$C(M_3^2)$}}
\put(206,30){\makebox(0,0){$\cdots$}}
\put(17,25){\vector(0,-1){15}}\put(17,25){\vector(0,-1){13}}
\put(13,18){\makebox(0,0){$\iota_1^*$}}
\put(97,25){\vector(0,-1){15}}\put(97,25){\vector(0,-1){13}}
\put(93,18){\makebox(0,0){$\iota_2^*$}}
\put(177,25){\vector(0,-1){15}}\put(177,25){\vector(0,-1){13}}
\put(173,18){\makebox(0,0){$\iota_3^*$}}
%BLUE
\put(23,12){\color{blue}\oval(4,4)[b]}
\put(25,12){\color{blue}\vector(0,1){13}}
\put(30,18){\makebox(0,0){\color{blue}$\lambda_1$}}
\put(103,12){\color{blue}\oval(4,4)[b]}
\put(105,12){\color{blue}\vector(0,1){13}}
\put(110,18){\makebox(0,0){\color{blue}$\lambda_2$}}
\put(183,12){\color{blue}\oval(4,4)[b]}
\put(185,12){\color{blue}\vector(0,1){13}}
\put(190,18){\makebox(0,0){\color{blue}$\lambda_3$}}
%RED
\put(42,5){\color{red}\oval(4,4)[l]}
\put(42,3){\color{red}\vector(1,0){38}}
\put(60,8){\makebox(0,0){\color{red}$C^*(\pi_1)$}}
\put(122,5){\color{red}\oval(4,4)[l]}
\put(122,3){\color{red}\vector(1,0){38}}
\put(140,8){\makebox(0,0){\color{red}$C^*(\pi_2)$}}
\put(37,30){\oval(4,4)[l]}\put(37,28){\vector(1,0){48}}
\put(60,33){\makebox(0,0){$(\pi_1 \times \pi_1)^*$}}
\put(117,30){\oval(4,4)[l]}\put(117,28){\vector(1,0){48}}
\put(140,33){\makebox(0,0){$(\pi_2 \times \pi_2)^*$}}
\end{picture}}
\end{equation}
According to the definition~\eqref{EC*R}, the space $C(R^i)$ is just the
algebra \mbox{$C^*(M_i, \sim^i)$}, as it is denoted in the diagram. The
pull-back mappings act in reverse direction, and the pull-back of a
surjective (resp. injective) mapping is injective (resp. surjective). The
pull-back $\iota_i^*$ of the inclusion $\iota_i :R^i \hookrightarrow M_i^2$
is just the restriction of the functions $M_i^2 \to \C$ to $R^i \subset
M_i^2$. The pull-back of the surjection $\pi_i \times \pi_i$ is the
injective mapping $(\pi_i \times \pi_i)^* : C(M_i^2) \hookrightarrow
C(M_{i+1}^2)$, $f \mapsto f \circ (\pi_i \times \pi_i )$, thus $(\pi_i
\times \pi_i )^* f(\Bar{x}, \Bar{y}) =f\big( \pi_i(\Bar{x}), \pi_i(\Bar{y})
\big)$. In addition to these pull-back maps there are the injections
$\lambda_i$ defined by
\begin{gather}\label{Elam}
\lambda_i : C^*(M_i, \sim^i) \hookrightarrow C(M_i^2),\qquad\qquad
f \mapsto \tilde{f}, \qquad\qquad \text{where}
\\ \nonumber
\tilde{f}(\Bar{x}, \Bar{y}) =
\begin{cases}
f(\Bar{x}, \Bar{y}),& \text{if } (\Bar{x}, \Bar{y})\in R^i;\\
0,& \text{if } (\Bar{x}, \Bar{y})\not\in R^i.
\end{cases}
\end{gather}
We note that $\iota_i^* \circ \lambda_i =\id_{C(R^i)}$, but $\lambda_i
\circ \iota_i^*$ is a proper projection of the algebra $C(M_i^2)$.

\begin{figure}
\centerline{\includegraphics[scale=0.5]{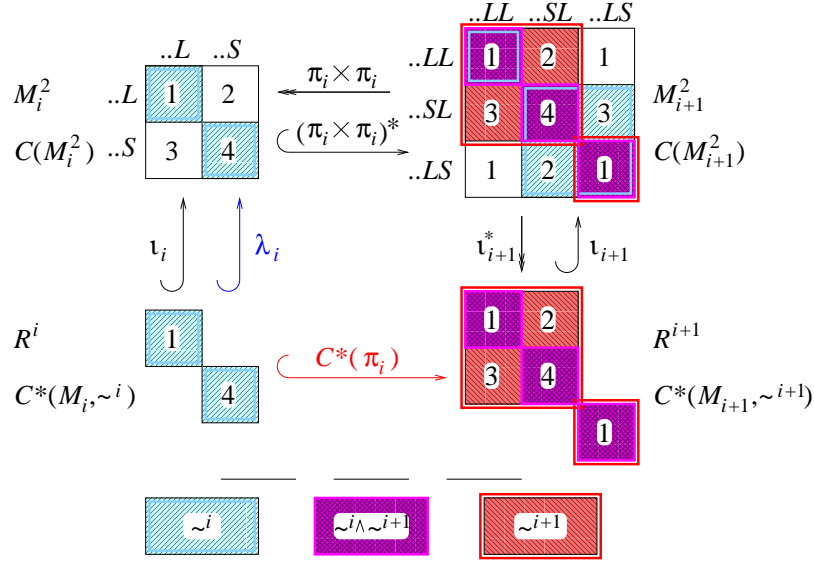}}
\caption{\label{FPDiag} The graphical representation of the building
blocks of the diagrams~\eqref{DPen2} as well as \eqref{DCPen2}.}
\end{figure}

A building block of the diagrams~\eqref{DPen2} as well as \eqref{DCPen2}
is graphically represented in figure~\ref{FPDiag}.

The direct product spaces $M_i^2$ as well as $M_{i+1}^2$ are represented by
squares, which also resemble to the form of the matrices in the
definition~\eqref{EC*}. The squares are partitioned into smaller squares
according to the relations $\sim^i$, $\sim^{i+1}$ and
\mbox{$\pi_i^*(\sim^i)$}. (This latter relation holds between two elements
$\Bar{x}, \Bar{y} \in M_{i+1}$ if and only if $\pi_i (\Bar{x}) \sim^i \pi_i
(\Bar{y})$, i.e., if the last but one symbol of $\Bar{x}$ and $\Bar{y}$ is
the same, $x_i =y_i$.) The numbers in the small squares help us to keep
track of the point-to-point as well as the pull-back mappings of the
diagrams~\eqref{DPen2} and \eqref{DCPen2}, while the shading (coloring) and
the patterns designate the different equivalence relations.

The mappings $C^*(\pi_i) : \mbox{$C^*(M_i, \sim^i)$} \hookrightarrow
\mbox{$C^*(M_{i+1}, \sim^{i+1})$}$ in the second row of the
diagram~\eqref{DCPen2} are defined by the composition
\begin{equation}\label{EC*pi}
C^*(\pi_i)= \iota_{i+1}^* \circ (\pi_i \times \pi_i)^* \circ \lambda_i,
\qquad\qquad
\text{for all $i \in \N_+$}
\end{equation}
and from the figure~\eqref{FPDiag} it is clear that $C^*(\pi_i)$ is an {\it
injective $C^*$-algebra homomorphism} from \mbox{$C^*(M_i, \sim^i)$} to
\mbox{$C^* (M_{i+1}, \sim^{i+1})$}. [Indeed, $C^*(\pi_i)$ is injective,
since both blocks `1' and `4' constituting \mbox{$C^*(M_i, \sim^i)$} appear
in \mbox{$C^*(M_{i+1}, \sim^{i+1})$}. It preserves multiplication, since it
is a mapping between block-diagonal matrices, according to
definition~\eqref{EC*} of \mbox{$C^*(M_i, \sim^i)$}.]

With these constructions we are almost ready with the definition of the
$C^*$-algebra \mbox{$C^*(M_{\mathrm P}, \sim)$} associated to the universe
of the Penrose tilings. Remember that in the first row of
diagram~\eqref{DPen} the topological space $M_{\mathrm P}$ was defined as
the projective limit of the spaces \mbox{$(M_i, \sim^i)$}:
\begin{equation}\label{DPen'}
\unitlength 0.5mm
\raisebox{-6mm}{
\begin{picture}(182,29)
\thinlines
\put(16,5){\makebox(0,0){$(M_1,\sim^1)$}}
\put(66,5){\makebox(0,0){$(M_2,\sim^2)$}}
\put(116,5){\makebox(0,0){$(M_3,\sim^3)$}}
\put(138,5){\makebox(0,0){$\cdots$}}
\put(166,24){\makebox(0,0){$M_{\mathrm{P}}$}}
\put(50,5){\vector(-1,0){18}}\put(50,5){\vector(-1,0){17}}
\put(100,5){\vector(-1,0){18}}\put(100,5){\vector(-1,0){17}}
\put(41,8){\makebox(0,0){$\pi_1$}}
\put(91,8){\makebox(0,0){$\pi_2$}}
\put(20,14){\vector(-1,-1){4}}\put(21,15){\vector(-1,-1){4}}
\qbezier[1000](17,11)(30,24)(150,24)
\put(70,14){\vector(-1,-1){4}}\put(71,15){\vector(-1,-1){4}}
\qbezier(67,11)(80,24)(150,24)
\put(120,14){\vector(-1,-1){4}}\put(121,15){\vector(-1,-1){4}}
\qbezier(117,11)(130,24)(150,24)
\put(32,21){\makebox(0,0){$\sigma_1$}}
\put(132,15){\makebox(0,0){$\sigma_3$}}
\end{picture}}
\end{equation}
and using the technique of noncommutative geometry the `commutative' spaces
\mbox{$(M_i,\sim^i)$} as well as the topological mappings $\pi_i$ have
been changed for the noncommutative algebras \mbox{$C^*(M_i, \sim^i)$}, as
well as for the injective $C^*$-homomorphisms $C^*(\pi_i)$:
\begin{equation}\label{DCPen2'}
\unitlength 0.5mm
\raisebox{-2mm}{
\begin{picture}(212,13)
\thinlines
\put(20,5){\makebox(0,0){$C^*(M_1,\sim^1)$}}
\put(100,5){\makebox(0,0){$C^*(M_2,\sim^2)$}}
\put(180,5){\makebox(0,0){$C^*(M_3,\sim^3)$}}
\put(206,5){\makebox(0,0){$\cdots$}}
\put(42,5){\oval(4,4)[l]}\put(42,3){\vector(1,0){38}}
\put(60,8){\makebox(0,0){$C^*(\pi_1)$}}
\put(122,5){\oval(4,4)[l]}\put(122,3){\vector(1,0){38}}
\put(140,8){\makebox(0,0){$C^*(\pi_2)$}}
\end{picture}}
\end{equation}
(It is the bottom row of the diagram~\eqref{DCPen2}.)

It is quite natural to desire that the above passage from commutative to
noncommutative structures should respect the projective/injective limit
constructions. In the light of this we have obtained the following result:

\begin{statement}\label{StC*Pen}
{\bf (The $C^*$-algebra associated to the Penrose tilings)}
The approximately finite dimensional (noncommutative) algebra associated to
the universe of Penrose tilings is defined as the injective limit
\begin{subequations}\label{EStPen}
\begin{equation}\label{EinjlimPen}
C^*(M_{\mathrm P}, \sim) =\varinjlim_{i\in \N} C^*(M_i, \sim^i)
\end{equation}
with the injections $C^*(\pi_i): \mbox{$C^*(M_i, \sim^i)$} \hookrightarrow
\mbox{$C^*( M_{i+1}, \sim^{i+1})$}$ given in~\eqref{EC*pi}. (For $i=0$ the
map $C^*(\pi_0): \C \hookrightarrow \mbox{$C^* (M_1, \sim^1)$} \subset
{\mathcal M}_2$ is $z \mapsto \bigl[ \begin{smallmatrix} z & 0 \\ 0 & z
\end{smallmatrix} \bigr]$.) The multiplicity matrices $A_0^{\mathrm P}$,
$A_i^{\mathrm P}$ (for $i \in \N_+$) and the Bratteli diagram of the AF
algebra $C^*(M_{\mathrm P}, \sim)$ is
\begin{align}\label{EBrPen}
A_0^{\mathrm P} &= \begin{bmatrix} 1 \\ 1 \end{bmatrix},&
A_i^{\mathrm P} &= \begin{bmatrix} 1& 1\\ 1& 0 \end{bmatrix};&&
\unitlength 0.7mm \raisebox{-7mm}{
\begin{picture}(90,22)
\thinlines
\put(8,16){\vector(1,0){10}}
\put(8,16){\vector(1,-1){10}}
\put(5,16){\makebox(0,0){$1$}}
\put(24,6){\vector(1,1){10}}
\put(24,16){\vector(1,0){10}}
\put(24,16){\vector(1,-1){10}}
\put(21,16){\makebox(0,0){$1$}}
\put(21,6){\makebox(0,0){$1$}}
\put(40,6){\vector(1,1){10}}
\put(40,16){\vector(1,0){10}}
\put(40,16){\vector(1,-1){10}}
\put(37,16){\makebox(0,0){$2$}}
\put(37,6){\makebox(0,0){$1$}}
\put(56,6){\vector(1,1){10}}
\put(56,16){\vector(1,0){10}}
\put(56,16){\vector(1,-1){10}}
\put(53,16){\makebox(0,0){$3$}}
\put(53,6){\makebox(0,0){$2$}}
\put(72,6){\vector(1,1){10}}
\put(72,16){\vector(1,0){10}}
\put(72,16){\vector(1,-1){10}}
\put(69,16){\makebox(0,0){$5$}}
\put(69,6){\makebox(0,0){$3$}}
\put(85,16){\makebox(0,0){$\dots$}}
\put(85,6){\makebox(0,0){$\dots$}}
\end{picture}}
\end{align}
\end{subequations}
\end{statement}

\noindent
(For the definition of approximately finite dimensional algebras
and related notions see Appendix~\ref{Aa}.)

\begin{proof}
The algebra~\eqref{EinjlimPen} is nothing but the injective limit of the
sequence~\eqref{DCPen2'}. It is clear from figure~\ref{FPDiag} that the
Bratteli diagram of this algebra is really~\eqref{EBrPen}.
\end{proof}

This algebra~\eqref{EStPen} is sometimes called the {\it Fibonacci algebra}
since the ranks of the full matrix algebras in the injective limit
construction are the Fibonacci numbers \cite{Dav96}.

It is worth noticing that the Bratteli diagram~\eqref{EBrPen} is built up
of the blocks~\eqref{EPensym} (rotated into horizontal position),
which describe the grammar rules belonging to the symbolic coding of the
Penrose tilings, and the transition matrix $T_{\mathrm P}$ in
equation~\eqref{EPensym} is the same as the multiplicity matrix
$A_i^{\mathrm P}$ in~\eqref{EBrPen}. Thus the simplest practical way of
obtaining the final result~\eqref{EStPen}, the (Bratteli
diagram~\eqref{EBrPen} of the) $C^*$-algebra \mbox{$C^*(M_{\mathrm P},
\sim)$} would have been to draw the diagrams~\eqref{EPensym} representing
the allowed transitions for the symbolic coding into a single chain.

Going through the steps of the injective limit
construction~\eqref{DCPen2'} again, it is easy to see that the above
observation is generally true for any finite Markov chain. In the next
subsection we use this short cut to determine the AF algebra associated to
the cat map.

\subsubsection{The AF $C^*$-algebra of the cat map\label{SAFCat}}
%'''''''''''''''''''''''''''''''''''''''''''''''''''''''''''''
In this subsection the constructions presented just before for the Penrose
system are adapted to the cat map. The only relevant differences between
the two systems are that in the latter case the symbolic
sequences~\eqref{EMCat} are infinite in both directions, and there are two
equivalence relations~$\sim_s$ and $\sim_u$ corresponding to forward and
backward iterations, respectively. These difference, however, can be easily
eliminated by simple tricks.

Let $\sim$ be the equivalence relation on the set $M_{\mathrm C}$ obtained
by merging $\sim_u$ and $\sim_s$, i.e.,
\begin{align}\label{Ecatsim}
\Bar{x} &\sim \Bar{y}&
\text{if and only if}&&
(\Bar{x} \sim_u \Bar{y}) &\wedge (\Bar{x} \sim_s \Bar{y}),&
\Bar{x}, \Bar{y} &\in M_{\mathrm C}.
\end{align}
Thus $\Bar{x} \sim \Bar{y}$ means that the initial and final tail of
$\Bar{x}$ and $\Bar{y}$ coincide, so the two series differ from each other
only in a finite number of letters.

The other difficulty, namely the fact that the symbolic sequences
associated to the cat map are infinite in both directions, can be overcome
by `stretching the finite sequences in both directions at the same time'.
More precisely, instead of the symbol set $P_{\mathrm C} =\{ A_0^0, B_0^0,
B_0^1, B_1^0, B_1^1 \}$ of five elements (see Lemma~\ref{LMarpar}) let us
introduce the set $\tilde{P}_{\mathrm C} =P_{\mathrm C}^2$ of ordered
pairs, which contains 25 elements. As phase space let us use the set
\begin{equation}\label{EMtC}
\tilde{M}_{\mathrm C} =\left\{
\Bar{X} =\big\{ (x_i^+, x_i^- ) \big\}_{i \in \N} \in
\tilde{P}_{\mathrm C}^{\N} \left|
\parbox{5cm}
{$x_0^- =x_0^+$, and for all $i \in \N$ the transitions $x_{i+1}^- \to
x_i^-$ and $x_i^+ \to x_{i+1}^+$ are allowed by the grammar
rules~\eqref{Egram}}
\right. \right\},
\end{equation}
which is already a set of series infinite only in one direction. There is a
straightforward bijection between $M_{\mathrm C}$ and $\tilde{M}_{\mathrm
C}$ given by
\begin{subequations}\label{EMtM}
\begin{align}\label{EMtM_a}
M_{\mathrm C} &\to \tilde{M}_{\mathrm C},&
\Bar{x} =\{ x_i \}_{i \in \Z} &\mapsto \big\{ (x_{-i}, x_i )\big\}_{i\in
\N};
\\ \label{EMtM_b}
\tilde{M}_{\mathrm C} &\to M_{\mathrm C},&
\Bar{X} =\big\{ (x_i^-, x_i^+) \big\}_{i\in \N} &\mapsto
\ldots x_2^-, x_1^-, x_0^- =x_0^+, x_1^+, x_2^+ \ldots
\end{align}
\end{subequations}
and the relation $\Bar{X} \sim \Bar{Y}$ (considered on $\tilde{M}_{\mathrm
C}$ according to the above bijection $M_{\mathrm C}
\overset{\cong}{\longleftrightarrow} \tilde{M}_{\mathrm C}$) holds between
two elements of $\tilde{M}_{\mathrm C}$ if and only if their tails
coincide, i.e., there exists an $n\in \N$ such that $(x_i^-, x_i^+)
=(y_i^-, y_i^+)$ for all $i \ge n$.

Instead of the set with two equivalence relations $(M_{\mathrm C}, \sim_u,
\sim_s )$ we investigate the object $(\tilde{M}_{\mathrm C}, \sim)$, which
is the phase space of an in one direction infinite Markov chain, with
transition matrix
\begin{align}\label{ETtC}
\tilde{T}_{\mathrm C} &=T_{\mathrm C}^T \otimes T_{\mathrm C},&&
\text{or in components}&
\tilde{T}_{(i,k),(j,l)} &= T_{i,j}^T \cdot T_{k,l},
\end{align}
where $T_{\mathrm C}$ is defined in~\eqref{Egram}. Indeed, the backward
allowed transitions are described by the transpose $T_{\mathrm C}^T$ of the
transition matrix, and because of the tensor product $T_{\mathrm C}^T$ acts
on the first element while $T_{\mathrm C}$ acts on the second element of
the pairs in $\tilde{M}_{\mathrm C}$. The transition $(x^-, x^+) \to
(y^-, y^+)\in \tilde{P}_{\mathrm C}$ is allowed, if and only if
$\tilde{T}_{(y^-, y^+), (x^-, x^+)} =T_{y^-, x^-}^T \cdot T_{y^+, x^+}
=T_{x^-, y^-} \cdot T_{y^+, x^+} =1$ holds for the appropriate matrix
element, i.e., the transitions $y^- \to x^-$ and $x^+ \to y^+$ are both
allowed in the original symbol space $P_{\mathrm C}$.

The space of equivalence classes $\tilde{X}_{\mathrm C} =\tilde{M}_{\mathrm
C} /\sim$ is again pathologic as topological space, what is essentially due
to the fact that every open set in the phase space of the cat map contains
(an infinite number of) points with arbitrarily prescribed initial and final
symbolic tails.

The transition matrix $\tilde{T}_{\mathrm C}$ has the size of $25 \times
25$, which is too big to permit a graphical representation, but still, the
algebra $C^*(\tilde{M}_{\mathrm C}, \sim)$ can be precisely described via
the matrix $\tilde{T}_{\mathrm C}$.

\begin{statement}\label{StC*cat}
{\bf (The $C^*$-algebra associated to the cat map)}
The noncommutative $C^*$-algebra $C^*(\tilde{M}_{\mathrm C}, \sim )$
associated to the cat map is the AF algebra defined by the injective limit
(see Appendix~\ref{Aa4})
\begin{subequations}\label{EC*cat}
\begin{align}\label{EC*cat_a}
C^*(\tilde{M}_{\mathrm C}, \sim) &= \varinjlim_{n \in \N} {\mathcal A}_n,
&&\text{where}&
{\mathcal A_0} &=\C^5,
\end{align}
and the multiplicity matrices $A_n$ (see Appendix~\ref{Aa3}) of the
successive (unital) injections $\varPhi_n :{\mathcal A}_n \hookrightarrow
{\mathcal A}_{n+1}$ are given by
\begin{align}\label{EC*cat_b}
A^0_{(i,j),k} &= T_{i,k}^T \cdot T_{j,k},&&&&\text{and}&&
\\ \label{EC*cat_c}
A_n =A &=\tilde{T}_{\mathrm C} =T_{\mathrm C}^T \otimes T_{\mathrm C},
&& \text{i.e.,}&
A_{(i,k),(j,l)} &= T_{i,j}^T \cdot T_{k,l}&&
\text{for $n \ge 1$},
\end{align}
\end{subequations}
where $T_{\mathrm C}$ is given in \eqref{Egram}.
\end{statement}

\begin{proof}
According to the remark at the end of the previous subsection, the
multiplicity matrices of the inclusions in the injective limit are the same
as the transition matrices describing the grammar rules of the symbolic
sequences. For $n \ge 1$ this matrix is given in \eqref{ETtC},
independently of the value of $n$.

For $n=0$ the algebra ${\mathcal A}_0$ is simply $\C^5$, since the zeroth
letter of a sequence $\Bar{X} \in \tilde{M}_{\mathrm C}$ can have only five
different values, because $x_0^- =x_0^+$. The multiplicity matrix $A^0$ of
the inclusion $\varPhi_0 :{\mathcal A}_0 \hookrightarrow {\mathcal A}_1$ is
clearly the matrix (of size $25 \times 5$) given in \eqref{EC*cat_b}, since
for a given zeroth symbol $x_0^- =x_0^+$ the matrix $T_{\mathrm C}^T$ resp.
$T_{\mathrm C}$ describes the allowed backward resp. forward steps.
\end{proof}

Although the presentation of the AF algebra $C^*(\tilde{M}_{\mathrm C},
\sim)$ is not so direct as it was in the case of Penrose tilings
(Statement~\ref{StC*Pen}), the formulas~\eqref{EC*cat} still unambiguously
characterize the algebra in question, and allow us to calculate its $K_0$
group, which is the subject of the next subsection.

\subsection{The $K_0$ groups of the algebras}
%--------------------------------------------
In this subsection the $K_0$ groups corresponding to the AF
algebras $C^*(M_{\mathrm P}, \sim)$ and $C^*(\tilde{M}_{\mathrm C}, \sim)$
are determined (see formulas~\eqref{EStPen} and \eqref{EC*cat}).

The $K_0$ group is a very important invariant of $C^*$-algebras, especially
of AF $C^*$-algebras, for in this case it is a complete invariant
\cite{Ell76}, and the subclass of commutative groups representing
the possible $K_0$ groups of all AF $C^*$-algebras is also well described
\cite{Dav96}.

In our case, comparing the obtained results for $K_0 \big( C^*(M_{\mathrm
P}, \sim) \big)$ resp. for $K_0 \big( C^*(\tilde{M}_{\mathrm C}, \sim)
\big)$ with the last assertions of the Statements~\ref{StPen}
(on page~\pageref{StPen}) resp. \ref{StCat} (on page~\pageref{StCat}) it
turns out that these groups are indeed important invariants of the systems
investigated.

For the sake of better intelligibility Appendix~\ref{Ab} gives a brief
summary of the necessary part of algebraic $K$ theory. Before going into
the details of the calculations let us recall that the scaled dimension
group of a finite dimensional $C^*$-algebra $\bigoplus_{i=1}^k {\mathcal
M}_{n_i}$ has the form $\big( \Z^k ,\N^k, \prod_{i=1}^k [0,n_i ]\big)$ (see
also formula~\eqref{Esdg} in Appendix~\ref{Ab}), and the functor $K_0$
commutes with direct limit, in the sense that the group $K_0 ({\mathcal
A})$ (as an ordered, scaled group) of the approximately finite dimensional
algebra ${\mathcal A} =\varinjlim_{i \to \infty} {\mathcal A}_i$ (with
unital inclusions $\varPhi_i: {\mathcal A}_i \hookrightarrow {\mathcal
A}_{i+1}$) is $K_0 ({\mathcal A})= \varinjlim_{i\to \infty} K_0({\mathcal
A}_i )$, where the (positive unital) group homomorphisms $K_0 (\varPhi_i)
:K_0({\mathcal A}_i) \to K_0 ({\mathcal A}_{i+1})$ are given by
the multiplicity matrices $A_i$ of the unital injections $\varPhi_i$.

A concrete representation of the direct limit group $K_0 ({\mathcal A})$ is
given by the formulas~\eqref{EilKgen} in the general case, and by the
expressions~\eqref{EinjlimK} for the case when the matrices $A_i$ are
injective (from a threshold index).

\subsubsection{The $K_0$ group associated to the Penrose tilings}
%''''''''''''''''''''''''''''''''''''''''''''''''''''''''''''''''

Let $f_i$ denote the series of Fibonacci numbers, i.e., $f_0 =0$, $f_1 =1$,
$f_2 =1$, $f_3 =2$, $f_4 =3$, $f_5 =5$, $\dots$, $f_{i+1} =f_i +f_{i-1}$.
According to Statement~\ref{StC*Pen}, the $C^*$-algebra associated to the
universe of Penrose tilings is given by the injective limit $C^*(M_{\mathrm
P}, \sim) =\varinjlim_{i\in \N} C^*(M_i, \sim^i)$, where the finite
dimensional algebras $C^*(M_i, \sim^i)$, as well as their $K_0$ groups
(with their order and scale structure) have the form

\begin{subequations}\label{EC*Pi}
\begin{align}\label{EC*Pi_a}
C^*(M_0,\sim^0) &\cong \C,\qquad \qquad
K_0 \big( C^*(M_0,\sim^0) \big) \cong \big(\Z, \N, \{0,1\} \big)
\\ \label{EC*Pi_b}
C^*(M_i, \sim^i) &\cong {\mathcal M}_{f_i}\oplus {\mathcal M}_{f_{i+1}},
\\ \label{EC*Pi_c}
K_0 \big( C^*(M_i, \sim^i) \big) &\cong
\big( Z^2, \N^2, \{0 \dots f_i\} \times \{0 \dots f_{i+1}\} \big)
\qquad \text{for $i\in \N_+$}.
\end{align}
\end{subequations}

In the followings we determine the form of $K_0\big( C^*(M_{\mathrm P},
\sim)\big)$, as the direct limit of the groups~\eqref{EC*Pi_c}.

\begin{statement}\label{StK0Pen}
{\bf (The $K_0$ group of the Penrose universe)}
The scaled dimension group of the AF algebra $C^*(M_{\mathrm P}, \sim)$ has
the form
\begin{subequations}\label{EStKP}
\begin{align}\label{EStKP_a}
K_0\big( C^*(M_{\mathrm P}, \sim) \big) &\cong
( \Z^2, K_0^+, \Gamma ) \cong
\\ \label{EStKP_b}
& \cong \big( \Z + \tau \Z, [0, \infty) \cap (\Z + \tau \Z), [0,1]
\cap (\Z +\tau \Z) \big),
\end{align}
where $\tau =\frac{1+\sqrt{5}}{2}$ is the `golden mean', and
\begin{align}\label{EStKP_c}
K_0^+ &= \big\{ (x,y) \in \Z^2 \;|\; 0 \le \tau x + y \big\},
\\ \label{EStKP_d}
\Gamma &= \big\{ (x,y) \in \Z^2 \;|\; 0 \le \tau x + y \le \tau+1 \big\}.
\end{align}
\end{subequations}
\end{statement}

\begin{proof}
First we prove the row~\eqref{EStKP_a} of the statement. The homomorphisms
in the direct limit construction of the group $K_0 \big( C^*(M_{\mathrm
P},\sim) \big)$ are determined by the multiplicity matrices $A_i^{\mathrm
P}$ given in ~\eqref{EBrPen} (see also equation~\eqref{EK0F} in
Appendix~\ref{Ab3}), so they have the following form ($i \in \N_+$):
\begin{subequations}\label{EAP}
\begin{align}\label{EAP_0}
A_0^P &: K_0(\C) \cong \Z \to K_0 (\C^2) \cong \Z^2,
&
z &\mapsto \begin{bmatrix} z & 0 \\ 0 & z \end{bmatrix},
\end{align}
\begin{multline}\label{EAP_i}
A_i^P : K_0({\mathcal M}_{f_i} \oplus {\mathcal
M}_{f_{i+1}}) \cong \Z^2 \to K_0 ({\mathcal M}_{f_{i+1}} \oplus {\mathcal
M}_{f_{i+2}}) \cong \Z^2,
\\
\begin{bmatrix} x \\ y \end{bmatrix} \mapsto
\begin{bmatrix} 1 & 1 \\ 1 & 0 \end{bmatrix}
\begin{bmatrix} x \\ y \end{bmatrix} =
\begin{bmatrix} x+y \\ x \end{bmatrix}.
\end{multline}
\end{subequations}
(We have used the connections~\eqref{EC*Pi}.)

The map $A_0^{\mathrm P}:\Z \to \Z^2$ is not surjective, but for all other
indices $i\in N_+$ the linear mappings $A_i^{\mathrm P}$ are $\Z^2 \to
\Z^2$ bijections, which means that the $K_0$ group of the injective limit
is also $\Z^2$. (This case is described by the formulas~\eqref{EinjlimK} of
Appendix~\ref{Ab3}.)

To determine the order and scale of the $K_0$ group we need to know the
(``stable'' and ``unstable'') eigenvalues $\lambda_s$, $\lambda_u$ as well
as the corresponding eigenvectors ${\mathbf v}_s$, ${\mathbf v}_u$ of the
(hyperbolic) matrix $A^{\mathrm P} =\big[ \begin{smallmatrix} 1&1 \\ 1&0
\end{smallmatrix}\big]$: \begin{subequations}\label{EAeig}
\begin{align}\label{EAeig1}
\lambda_u &= \tau =\frac{1 +\sqrt{5}}{2},&
{\mathbf v}_u &=\begin{bmatrix} \tau \\ 1 \end{bmatrix};
\\
\lambda_s &= -\frac{1}{\tau} =\frac{1-\sqrt{5}}{2},&
{\mathbf v}_s &=\begin{bmatrix} 1 \\ -\tau \end{bmatrix}.
\end{align}
\end{subequations}
These eigenvectors are illustrated in Fig.~\ref{FAPeig}.

\begin{figure}
\centerline{\includegraphics[scale=0.5]{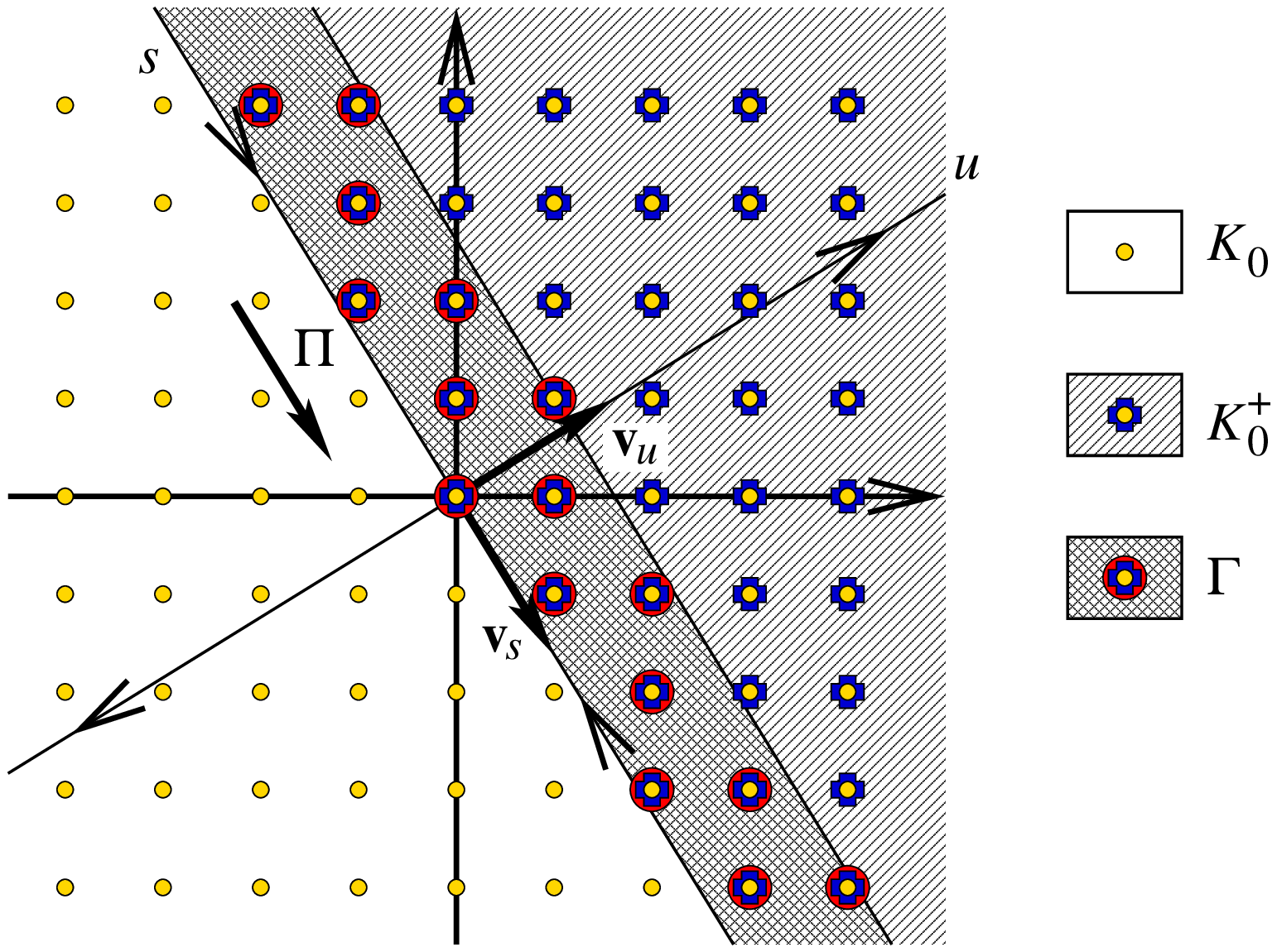}}
\caption{\label{FAPeig} The eigenvectors ${\mathbf v}_s$, ${\mathbf
v}_u$ of the matrix
$A^{\mathrm P} =\big[ \genfrac{}{}{0pt}{1}{1}{1}
\genfrac{}{}{0pt}{1}{1}{0} \big]$, and the positive cone $K_0^+$ as well as
the scale $\Gamma$ belonging to the algebra $C^*(M_{\mathrm P}, \sim)$.}
\end{figure}

The cone $K_0^+$ of positive elements consists of the points of $\mbox{$K_0
\big(C^*(M_1,\sim^1) \big)$} \cong \Z^2$ which after a finite (but
arbitrarily large) number of bijections $A^{\mathrm P}: \Z^2 \to \Z^2$ fall
(and thus remain) in the positive cone $\mbox{$K_0^+ \big( C^*(M_i, \sim^i)
\big)$} \cong \N^2 \subset \Z^2$. (See the formulas~\eqref{EinjlimK} of
Appendix~\ref{Ab3}.) These are exactly the points ${\mathbf u} =\left[
\begin{smallmatrix} x \\ y \end{smallmatrix} \right] \in \Z^2$ for which
the scalar product ${\mathbf u} \cdot {\mathbf v}_u =\tau x +y \ge 0$ is
not negative, as it is stated in \eqref{EStKP_c}.

Similarly, the scale $\Gamma$ consists of the points of $\mbox{$K_0
\big(C^*(M_1,\sim^1) \big)$} \cong \Z^2$ which after the $n^{\rm th}$
application of the bijection $A^{\mathrm P}: \Z^2 \to \Z^2$ fall into
the scale $\Gamma_{n+1} =\{ 0 \ldots f_{n+1} \} \times \{ 0 \ldots
f_{n+2} \}$ for sufficiently large $n \in \N$. Since $A^{\mathrm P} \cdot
\left[ \begin{smallmatrix} f_i \\ f_{i+1} \end{smallmatrix} \right] =
\left[ \begin{smallmatrix} f_{i+1} \\ f_{i+2} \end{smallmatrix} \right]$,
the scale $\Gamma$ is the set of points lying in the strip parallel
to the stable direction $s$, and determined by the elements $(0,0)$, $(1,1)
\in \Z^2$, as it is depicted in Fig.~\ref{FAPeig} and given in
\eqref{EStKP_d}.

To prove assertion~\eqref{EStKP_b} of the statement we have to notice that
the mapping
\begin{align}\label{Etinj}
\Pi : \Z^2 &\to \R,& (x,y) &\mapsto \frac{\tau x+y}{\tau+1} =
(\tau-1)x +(2-\tau)y = \tau(x-y) +2y-x,
\end{align}
which is essentially (up to a scaling factor) the projection along the
stable direction $s$ to the unstable direction $u$, is an {\it injective}
group homomorphism. (The injectivity follows from the fact that $\tau
=\frac{1 +\sqrt{5}}{2}$ is irrational.) The image of the scaled dimension
group $(\Z^2, K_0^+, \Gamma)$ under the projection $\Pi$ has the form given
in \eqref{EStKP_b}.
\end{proof}

Comparing the previous result with assertion {\it iii)} of
Statement~\ref{StPen} (page~\pageref{StPen}) we see that the noncommutative
algebra $C^* (M_{\mathrm P}, \sim)$ associated to the universe of Penrose
tilings do carry nontrivial information about the structure of this space.

\subsubsection{The $K_0$ group associated to the cat map}
%''''''''''''''''''''''''''''''''''''''''''''''''''''''''

In order to determine the scaled dimension group $K_0 \big( C^*
(\tilde{M}_{\mathrm C}, \sim) \big)$ of the (noncommutative) AF algebra
$C^*(\tilde{M}_{\mathrm C},\sim)$ associated to the cat map, we basically
follow the same steps as we did in the previous subsection, investigating
the algebra of the Penrose tilings. The only difference between the two
cases is rather technical; now the allowed transitions are described by the
matrix $\tilde{T}_{\mathrm C}$ of size $25 \times 25$, which is much bigger
than the two by two matrix $A^{\mathrm P}$, and not invertible.
The analytical calculation of the $K_0$ group, however, is still possible,
since the tensor product structure of $\tilde{T}_{\mathrm C} = T^T_{\mathrm
C} \otimes T_{\mathrm C}$ permits its eigenstate decomposition.

\begin{statement}\label{StK0cat}
{\bf (The $K_0$ group of the cat map)}
The scaled dimension group of the AF $C^*$-algebra $C^* (\tilde{M}_{\mathrm
C}, \sim )$ (defined in \eqref{EC*cat}) has the form
\begin{subequations}\label{EStKcat}
\begin{align}\label{EStKcat_a}
K_0 \big( C^*(\tilde{M}_{\mathrm C}, \sim) \big) &\cong
(\Z^{4}, K_0^+, \Gamma) \cong
\\ \label{EStKcat_b}
&\cong \big( \zeta \Z + \eta \Z,
[0,\infty) \cap (\zeta \Z + \eta \Z), [0,1] \cap (\zeta \Z + \eta \Z)\big)
\times \Z^2
\end{align}
\begin{align}\label{EStKcat_c}
&\text{where}&
\zeta&= \frac{1}{2} -\frac{\sqrt{5}}{10},&
\eta &= \frac{\sqrt{5}}{5},&
&\text{and}
\end{align}
\begin{align}\label{EStKcat_d}
K_0^+ &=\left\{ \left. \begin{bmatrix} u&v\\w&z \end{bmatrix} \in
\Z^4 \right|
0 \le u+ \frac{\sqrt{5}-1}{2} (v+w) +\frac{3-\sqrt{5}}{2} z
\right\},
\\ \label{EStKcat_e}
\Gamma &=\left\{ \left. \begin{bmatrix} u&v\\w&z \end{bmatrix} \in
\Z^4 \right|
0\le u+ \frac{\sqrt{5}-1}{2} (v+w) +\frac{3-\sqrt{5}}{2} z \le
\frac{25+11\sqrt{5}}{2} \right\}.
\end{align}
\end{subequations}
\end{statement}

\begin{proof}
The algebra associated to the cat map is defined as the injective limit
$C^* (\tilde{M}_{\mathrm C}, \sim) =\varinjlim_{n \in \N} {\mathcal A}_n$
(see Statement~\ref{StC*cat}, equation~\eqref{EC*cat}), and the scaled
dimension groups of the finite matrix algebras ${\mathcal A}_n$ are clearly
(see Appendix~\ref{Ab3}, equation~\eqref{Esdg}):
\begin{subequations}\label{EKAn}
\begin{align}\label{EKAn_0}
K_0({\mathcal A}_0) &\cong
\big( \Z^5, \N^5, \{0,1\}^5 \big),&&
\\ \label{EKAn_n}
K_0({\mathcal A}_n) &\cong
\Big( \Z^{25}, \N^{25}, {\textstyle \prod_{i,j \in P_{\mathrm C}}}
\big\{0,1 \ldots a^{(n)}_{i,j}\big\} \Big),&\text{for }& n\ge 1,
\end{align}
\end{subequations}
where $a^{(n)}_{i,j}$ is the number of symbolic sequences $\Bar{x} =(
x_{-n} =i, x_{-n+1} \ldots x_{n-1}, x_{n}=j )$ of length $2n-1$, satisfying
the grammar rules~\eqref{Egram}, starting with the symbol $x_{-n}=i \in
P_{\mathrm C}$, at the $-n^{\textrm{th}}$ position and ending with the
symbol $x_n =j \in P_{\mathrm C}$ at the $n^{\textrm{th}}$ position. It is
easy to check that $a^{(n)}_{i,j} \ge 1$ holds for all these numbers ($n
\ge 1$, $i,j \in P_{\mathrm C}$), i.e., in at least two steps every $i \to
\to j$ transition is allowed. (Indeed, the matrix $T_{\mathrm C}^2$ has no
zero entries.) Moreover, since the inclusions $\varPhi_n : {\mathcal A}_n
\hookrightarrow {\mathcal A}_{n+1}$ are unital, and their multiplicity
matrices are given by~(\ref{EC*cat}.b--c), the following recursions are
valid:
\begin{subequations}\label{Erecanij}
\begin{alignat}{2}\label{Erecanij_a}
a_i^{(0)} &= 1, && \text{for all $i\in P_{\mathrm C}$},
\\ \label{Erecanij_b}
a^{(1)}_{i,j} &= \sum_{k \in P_{\mathrm C}}
T^T_{i,k} \cdot T_{j,k} \cdot a_k^{(0)}=
\big[ (T_{\mathrm C} \cdot T_{\mathrm C})^T \big]_{i,j} =&&
\left[ \begin{smallmatrix}
2& 2& 2& 1& 1\\
1& 1& 1& 1& 1\\
2& 2& 2& 1& 1\\
1& 1& 1& 1& 1\\
2& 2& 2& 1& 1
\end{smallmatrix} \right],
\\ \label{Erecanij_c}
a^{(n+1)}_{i,j} &= \sum_{k,l \in P_{\mathrm C}} \tilde{T}_{(i,j),(k,l)}
\cdot a^{(n)}_{k,l}&&
\text{for $n\ge 1$}.
\end{alignat}
\end{subequations}

The $K_0$ group of the algebra $C^*(\tilde{M}_{\mathrm C}, \sim)$ is
the direct limit of the (scaled dimension) groups $K_0 ({\mathcal A}_n)$,
given in \eqref{EKAn}, where the (positive unital) homomorphisms
between the $K_0$ groups are the multiplicity matrices of the corresponding
$\varPhi_n :{\mathcal A}_n \hookrightarrow {\mathcal A}_{n+1}$ algebra
homomorphisms [see (\ref{EC*cat_b}-c)], so
\begin{align}\label{EKFcat}
K_0 (\varPhi_0)_{(i,j),k} &= T_{i,k}^T \cdot T_{j,k},&
K_0 (\varPhi_n) &= \tilde{T}_{\mathrm C} =T_{\mathrm C}^T \otimes
T_{\mathrm C}&
\text{for } n&\ge 1,
\end{align}
where the matrix $T_{\mathrm C}$ is given in \eqref{Egram}.

In order to calculate the direct limit $K_0$ group, we have to know the
eigenspace decomposition of the matrix $T_{\mathrm C}$ and its transpose
$T_{\mathrm C}^T$. The eigenvalues and the corresponding right resp. left
eigenvectors (eigenspaces) of $T_{\mathrm C}$ are:
\begin{subequations}\label{Eeigcat}
\begin{align}\label{Eeigcat_0}
\lambda_0 &=0,&
E_0^R &=\left\{ \left. \left[
\begin{smallmatrix} a \\ b \\ c \\ -b \\ -a-c \end{smallmatrix}
\right] \right| a,b,c \in \R \right\},&
E_0^L &=\left\{ \left. \left[
\begin{smallmatrix} a \\ b \\ -a-b \\ c \\ -c \end{smallmatrix}
\right] \right| a,b,c \in \R \right\};
\\ \label{Eeigcat_s}
\lambda_s &=\frac{3-\sqrt{5}}{2},&
{\mathbf v}_s^R &=
\left[ \begin{smallmatrix} 1 \\ 1 \\ 1 \\ -(1+\sqrt{5})/2 \\
-(1+\sqrt{5})/2 \end{smallmatrix} \right],&
{\mathbf v}_s^L &=
\left[ \begin{smallmatrix} 1 \\ -(1+\sqrt{5})/2 \\ 1 \\
-(1+\sqrt{5})/2 \\ 1 \end{smallmatrix} \right];
\\ \label{Eeigcat_u}
\lambda_u &=\frac{3+\sqrt{5}}{2},&
{\mathbf v}_u^R &=
\left[ \begin{smallmatrix} 1 \\ 1 \\ 1 \\ (\sqrt{5}-1)/2 \\
(\sqrt{5}-1)/2 \end{smallmatrix} \right],&
{\mathbf v}_u^L &=
\left[ \begin{smallmatrix} 1 \\ (\sqrt{5}-1)/2 \\ 1 \\
(\sqrt{5}-1)/2 \\1 \end{smallmatrix} \right].
\end{align}
\end{subequations}
It is clear that the rank of the matrices $T_{\mathrm C}$ and $T_{\mathrm
C}^T$ is two, and their ranges are
\begin{align}\label{ERanTTC}
\Ran T_{\mathrm C} &= \left\{ \left. \left[
\begin{smallmatrix}a \\ a \\ a \\ b \\ b \end{smallmatrix}
\right] \right| a,b \in \R \right\} &
\text{and}&&
\Ran T_{\mathrm C}^T &= \left\{ \left. \left[
\begin{smallmatrix}a \\ b \\ a \\ b \\ a \end{smallmatrix}
\right] \right| a,b \in \R \right\}.
\end{align}

Consequently the rank of $\tilde{T}_{\mathrm C} =T_{\mathrm C}^T \otimes
T_{\mathrm C}$ is four, and its four nonzero eigenvalues with the
corresponding eigenvectors are
\begin{subequations}\label{EeigTtT}
\begin{align}\label{EeigTtT_1}
{\mathbf v}_s^L &\otimes {\mathbf v}_u^R \quad\text{and}\quad
{\mathbf v}_u^L \otimes {\mathbf v}_s^R &&
\text{for} &
\tilde{\lambda}_1 &=\tilde{\lambda}_2 =1,
\\ \label{EeigTtT_s}
{\mathbf v}_s^L &\otimes {\mathbf v}_s^R &&
\text{for} &
\tilde{\lambda}_s &=\lambda_s^2 <1,
\\ \label{EeigTtT_u}
{\mathbf v}_u^L &\otimes {\mathbf v}_u^R &&
\text{for} &
\tilde{\lambda}_u &=\lambda_u^2 >1.
\end{align}
\end{subequations}
The effect of the matrix $\tilde{T}_{\mathrm C}$ on its four dimensional
range is already a bijection, which means, according to the
formulas~(\ref{EilKgen_a}--c) in Appendix~\ref{Ab3}, that the direct limit
group $K_0\big(C^*(\tilde{M}_{\mathrm C}, \sim) \big)$, consisting of the
series $(x_i )_{i \in \N} \in \prod_{i \in \N} K_0 ({\mathcal A}_i)$ with
`predictable tails' is homomorphic to $\Z^4$, as stated in
\eqref{EStKcat_a}.

A convenient way of realizing the bijection between the range of
$\tilde{T}_{\mathrm C}$ and $\Z^4 \cong Z^2 \otimes \Z^2$ is to use the
mapping
\begin{equation}\label{EPdef}
\begin{split}
\tilde{P} &= P^L \otimes P^R : \Z^{25} \cong \Z^5 \otimes \Z^5
\longrightarrow \Z^4 \cong \Z^2 \otimes \Z^2,
\qquad \quad \text{where} \\
P^L &= \begin{bmatrix} 1& 1& 1& 0& 0\\
0& 0& 0& 1& 1 \end{bmatrix} \qquad \quad \text{and} \qquad \quad
P^R =\begin{bmatrix} 1& 0& 1& 0& 1 \\ 0& 1& 0& 1& 0 \end{bmatrix}.
\end{split}
\end{equation}
Indeed, the kernel of $P^L$ resp. $P^R$ is the zero eigenspace $E_0^L$
resp. $E_0^R$ (see \eqref{Eeigcat_0}). It is also easy to see that the
mapping $\tilde{P}$ preserves the natural order on $\Ran T_{\mathrm C}^T
\otimes \Ran T_{\mathrm C}$, for the image of an element ${\mathbf a}
=(a_{i,j})_{i,j \in P_{\mathrm C}} \in \Ran T_{\mathrm C}^T \otimes \Ran
T_{\mathrm C}$ is positive (i.e., $\forall k,l \in \{1,2\}$, $\sum_{i,j \in
P_{\mathrm C}} P_{k,i}^L \cdot P_{l,j}^R \cdot a_{i,j} \ge 0$) if and only
if ${\mathbf a}$ is positive (i.e., $\forall i,j \in P_{\mathrm C}$,
$a_{i,j} \ge 0$).

The effect of $T_{\mathrm C}$ resp $T_{\mathrm C}^T$ on $\Ran T_{\mathrm
C}$ resp. on $\Ran T_{\mathrm C}^T$ can be encoded by the two by two
matrix $Z$, in the sense that for any integer $n \ge 1$
\begin{subequations}\label{EZ}
\begin{gather}\label{EZ_a}
P^R \cdot T_{\mathrm C}^n = Z^n \cdot P^R
\qquad \qquad\text{and} \qquad \qquad
P^L \cdot \big( T_{\mathrm C}^T \big)^n = Z^n \cdot P^L,
\\ \label{EZ_b}
\text{thus} \quad
\tilde{P}\cdot \tilde{T}_{\mathrm C}^n =
(Z \otimes Z)^n \cdot \tilde{P} =(Z^n \otimes Z^n ) \cdot \tilde{P},
\quad \text{where} \quad
Z =\begin{bmatrix} 2& 1\\ 1& 1 \end{bmatrix}.
\end{gather}
\end{subequations}
(The connections~\eqref{EZ_a} can be easily checked by direct
calculations.)

The eigenvalues and the corresponding eigenvectors of the matrix $Z$ are
\begin{subequations}\label{EZeig}
\begin{align}\label{EZ_s}
\lambda_s &= \frac{3 -\sqrt{5}}{2}, &
{\mathbf w}_s &= \begin{bmatrix} 1\\ -\frac{1+\sqrt{5}}{2} \end{bmatrix};
\\ \label{EZ_u}
\lambda_u &= \frac{3 +\sqrt{5}}{2}, &
{\mathbf w}_u &= \begin{bmatrix} 1\\ \frac{\sqrt{5}-1}{2} \end{bmatrix};
\end{align}
\end{subequations}
(cf. (\ref{Eeigcat_s}--c)) and the eigenspace decomposition of $Z \otimes
Z$ is quite similar to \eqref{EeigTtT}:
\begin{subequations}\label{EeigZtZ}
\begin{align} \nonumber
{\mathbf w}_s \otimes {\mathbf w}_u &=
\begin{bmatrix} 1& \frac{\sqrt{5}-1}{2} \\ -\frac{1+\sqrt{5}}{2}& -1
\end{bmatrix} &
&\text{and}&&
\\ \label{EeigZtZ_1}
{\mathbf w}_u \otimes {\mathbf w}_s &=
\begin{bmatrix} 1& -\frac{1+\sqrt{5}}{2} \\ \frac{\sqrt{5}-1}{2}& -1
\end{bmatrix} &
&\text{for} &
\tilde{\lambda}_1 &=\tilde{\lambda}_2 =1,
\\ \label{EeigZtZ_s}
{\mathbf w}_s \otimes {\mathbf w}_s &=
\begin{bmatrix} 1& -\frac{1+\sqrt{5}}{2} \\ -\frac{1+\sqrt{5}}{2}&
\frac{3+\sqrt{5}}{2} \end{bmatrix}&
&\text{for} &
\tilde{\lambda}_s &=\lambda_s^2 <1,
\\ \label{EeigZtZ_u}
{\mathbf w}_u \otimes {\mathbf w}_u &=
\begin{bmatrix} 1& \frac{\sqrt{5}-1}{2} \\ \frac{\sqrt{5}-1}{2} &
\frac{3-\sqrt{5}}{2} \end{bmatrix}&
&\text{for} &
\tilde{\lambda}_u &=\lambda_u^2 >1.
\end{align}
\end{subequations}
Apparently, only the unstable eigenvector ${\mathbf w}_u \otimes {\mathbf
w}_u$ is positive.

According to \eqref{EilKgen_d}, the positive cone $K_0^+ \big( C^*
(\tilde{M}_{\mathrm C}, \sim) \big) \subset \Z^2 \otimes \Z^2$ consists of
the vectors ${\mathbf x} =\left[ \begin{smallmatrix} u& v&\\ w& z
\end{smallmatrix} \right] \in \Z^2 \otimes \Z^2$ whose entries become
positive after sufficiently many iterations of the mapping $Z \otimes Z$,
i.e.,
\begin{equation}\label{EZtZitx}
(Z \otimes Z)^n
\begin{bmatrix}u& v\\ w& z \end{bmatrix} \ge
\begin{bmatrix} 0& 0\\ 0& 0 \end{bmatrix}
\qquad
\text{for $n$ sufficiently large},
\end{equation}
which means that the projection of ${\mathbf x}$ to the (positive) unstable
eigenvector ${\mathbf w}_u \otimes {\mathbf w}_u$ has to be nonnegative,
so
\begin{equation}\label{Exwug0}
{\mathbf x} \cdot ({\mathbf w}_u \otimes {\mathbf w}_u) =
\begin{bmatrix} u& v\\w& z\end{bmatrix} \cdot
\begin{bmatrix} 1& \frac{\sqrt{5}-1}{2} \\ \frac{\sqrt{5}-1}{2} &
\frac{3-\sqrt{5}}{2} \end{bmatrix}=
u+ \frac{\sqrt{5}-1}{2} (v+w) + \frac{3-\sqrt{5}}{2} z \ge 0,
\end{equation}
as stated in \eqref{EStKcat_d}. (Here ${\mathbf x}$ and ${\mathbf w}_u
\otimes {\mathbf w}_u$ were considered as four dimensional vectors, and
the operation `$\;\cdot\;$' is the scalar product between them.)

Using \eqref{Erecanij_b}, the image of the $K_0$-group element
$[\id_{{\mathcal A}_1}]$ under $\tilde{P} =P^L \otimes P^R$ is
\begin{equation}\label{E13885}
\sum_{k,l \in P_{\mathrm C}} P_{i,k}^L \cdot P_{j,l}^R \cdot a_{k,l}^{(1)}
= \begin{bmatrix} 13& 8\\ 8& 5 \end{bmatrix}.
\end{equation}
It means, according to \eqref{Escale} and \eqref{EilKgen_e}, that the scale
$\Gamma\big( C^*({\tilde M}_{\mathrm C}, \sim ) \big)$ consists of the
elements ${\mathbf x} =\left[ \begin{smallmatrix} u& v&\\ w& z
\end{smallmatrix} \right] \in K_0^+ \big( C^*({\tilde M}_{\mathrm C}, \sim
) \big)$, for which after a sufficiently large number $n$ of iterations
\begin{align}\label{EGamcrit}
(Z \otimes Z)^n \cdot
\begin{bmatrix} u& v\\ w& z \end{bmatrix} &\le
(Z \otimes Z)^n \cdot
\begin{bmatrix} 13& 8\\ 8& 5 \end{bmatrix} &
&\text{or}&
(Z \otimes Z)^n \cdot
\begin{bmatrix} 13-u & 8-v\\ 8-w& 5-z \end{bmatrix}
&\ge 0
\end{align}
holds, where the order `$\le$' is the natural one, thus it is valid to each
matrix entry separately. This criterion has the same form as
\eqref{EZtZitx}, and it gives the following inequality for the entries
$u,v,w,z \in \Z$:
\begin{equation}\label{Euvwz}
u+\frac{\sqrt{5}-1}{2}(v+w) +\frac{3-\sqrt{5}}{2} z \le
\frac{25 +11 \sqrt{5}}{2}.
\end{equation}
This proves the statement~\eqref{EStKcat_e}.

In order to obtain the form~\eqref{EStKcat_b} of the dimension group first
let us apply the injective (scaled dimension) group homomorphism
\begin{subequations}\label{Esdghomo}
\begin{gather}\label{Esdghomo_a}
\Pi:(\Z^4, K_0^+, \Gamma) \longrightarrow
\big( \R, [0, \infty), [0,1] \big) \times \Z^2,
\\ \label{Esdghomo_b}
\begin{bmatrix} u& v\\ w& z \end{bmatrix} \longmapsto
\left( \frac{ 2u+ (\sqrt{5}-1) (v+w)+ (3-\sqrt{5})z}{25+11\sqrt{5}}, u, v
\right)=
\\ \label{Esdghomo_c}
= \left( \frac{5u-8(v+w) +13z}{2} +\frac{\sqrt{5}}{10} \big( -11u+18(v+w)
-29z \big), u,v \right),
\end{gather}
\end{subequations}
to the form~\eqref{EStKcat_a} of the dimension group. The first component
of this homomorphism is essentially (up to a scaling factor) the projection
of $\Z^4 \cong \Z^2 \otimes \Z^2$ onto the unstable eigenvector ${\mathbf
w}_u \otimes {\mathbf w}_u$, and the second and third trivial components
of the mapping just ensure injectivity. (The order and scale preserving
property of the mapping follows directly from its construction.)

The range (of the first component) of the homomorphism $\Pi$
(equation~\eqref{Esdghomo}) is clearly in the additive subgroup
$\frac{1}{2} \Z +\frac{\sqrt{5}}{10} \Z \subset \R$, but it does not equal
to it! Indeed, the general element $\alpha =\frac{1}{2}n
+\frac{\sqrt{5}}{10} k$ for $n,k \in \Z$ is in the range of $\Pi$ if and
only if the equations
\begin{equation}\label{Enk}
\begin{split}
n&= 5u-8(v+w)+13z\\
k&=-11u+18(v+w)-29z
\end{split}
\end{equation}
can be satisfied for certain {\it integer} values of $u,v,w,z \in \Z$.
Treating $u$ and $v$ as parameters, and expressing the other two variables
$w$ and $z$ we obtain
\begin{equation}\label{Ewz}
\begin{split}
w&=14n+6k-u-v +\frac{n+k}{2} \\
z&= 9n+4k-u
\end{split}
\end{equation}
which means that $n+k$ has to be even, say $n+k =2m$ where $m\in \Z$. Then
$k=2m-n$, and $\alpha =\frac{5-\sqrt{5}}{10} n+\frac{\sqrt{5}}{5} m$, thus
the homomorphism $\Pi$ is a {\it bijection}
\begin{equation}\label{EPibij}
\Pi:\Z^4 \longrightarrow
\left( \frac{5-\sqrt{5}}{10} \Z +\frac{\sqrt{5}}{5} \Z \right) \times \Z^2
\end{equation}
given explicitely by \eqref{Esdghomo_b}, and its inverse is
\begin{equation}\label{EPibijinv}
\left( \frac{5-\sqrt{5}}{10} n +\frac{\sqrt{5}}{5} m ,u,v \right)
\longmapsto
\begin{bmatrix} u& v\\
8n+13m-u-v& 5n+8m-u \end{bmatrix}.
\end{equation}

This proves the assertion~\eqref{EStKcat_b} of the statement.
\end{proof}

Again, turning back to page~\pageref{StCat}, and comparing the present
result with assertion~{\it iii)} of Statement~\ref{StCat}, we see that the
noncommutative algebra $C^*(\tilde{M}_{\mathrm C}, \sim)$ associated to the
cat map as well as its $K_0$ group are really important invariants of this
dynamical system.

%% file: pencat5.tex
\section{Conclusions, outlook}
%=============================
In the preceeding sections we have successfully applied certain
methods of noncommutative geometry in the study of a specific
uniformly hyperbolic chaotic dynamical system, Arnold's cat map. The whole
investigation was motivated by, and performed along the guiding lines of a
similar but technically less sophisticated example: the isomorphism
classes of Penrose tilings \cite{Con0}.

First a Markov partition of the phase space of the cat map was constructed
(Lemma~\ref{LMarpar}, on page~\pageref{LMarpar}), then a noncommutative,
approximately finite dimensional $C^*$-algebra has been associated to the
system (Statement~\ref{StC*cat}, on page~\pageref{StC*cat}), and finally
its scaled dimension group has been explicitely calculated
(Statement~\ref{StK0cat}, on page~\pageref{StK0cat}). Comparing this result
with the last assertion of Statement~\ref{StCat} (on page~\pageref{StCat}),
we see that the $K_0$ group coincides with the dense additive subgroup of
reals describing the frequency of appearance of finite symbolic sequences
in typical symbolic trajectories.

The result of this investigation clearly demonstrates that the methods of
noncommutative geometry are, indeed, adaptable for the study of chaotic
dynamical systems, and they give better results than the usual topological
methods. It, however, also poses a great amount of further questions to
investigate. First, {\it why} is this apparent coincidence between the
frequency ratio of finite symbolic sections and the $K_0$ group of the
noncommutative algebra? Under what circumstances does it hold?

But there are also deeper, and from pure theoretical point of view more
interesting questions! To what extent does the noncommutative AF algebra
$C^*(\tilde{M}_{\mathrm{C}}, \sim)$ associated to the cat map depend on the
particular choice of the Markov partition of the phase space, used in its
construction? Is it possible to generalize this procedure to a wider class
of dynamical systems, which are not necessarily Markovian? Can one do it in
a functorial way? If `yes', then how? Which other invariants of the
associated noncommutative $C^*$-algebra are important from the point of
view of the original dynamical system?

There is hope to give positive answer to these questions and to build up
a `dictionary' between the notions and constructions in the theory of
dynamical systems and their counterparts in the language of noncommutative
geometry. In this case the noncommutative geometrical approach could shed
light on new aspects of the old theory of dynamical systems. According
to the results presented in this paper one row in this dictionary could be
something like this:

\begin{center}
\rule{12cm}{0.5pt}\\
\parbox[t]{5cm}{ratios of appearances of finite symbolic sections in
typical symbolic series}
\hspace{1cm}
\parbox[t]{5cm}{scaled dimension group of the associated
$C^*$-algebra}
\\ \rule{12cm}{0.5pt}
\end{center}

With this article we would like to encourage the research in this
direction.

\subsection*{Addendum}
%---------------------
\addcontentsline{toc}{subsection}{Addendum}
After finishing the manuscript the author became aware of the fact that an
even simpler chaotic system, the well known baker's map \cite{A_A,
CoFoSi82} also nicely fits into the frame of the present investigations.
The baker's map is a hyperbolic dynamical system, too, possessing a Markov
partition of two elements, and there are no grammar rules at all, thus the
transition matrix is simply $T_{\mathrm B} =\big[ \begin{smallmatrix} 1&1
\\ 1& 1 \end{smallmatrix}\big]$. It means that the frequency of appearance
of any finite symbolic section of length $n \in \N_+$ is $2^{-n}$,
regardless of the actual form of the sequence.

The $C^*$-algebra related to this system is the AF algebra, the Bratteli
diagram of which is built up from the block
\newlength{\mylength}
\settowidth{\mylength}{\mbox{$\nearrow$}}
$\begin{smallmatrix}
2^n &\longrightarrow & 2^{n+1} \\
& \mbox{$\nearrow$} \hspace{-\mylength} \mbox{$\searrow$} & \\
2^n & \longrightarrow & 2^{n+1}
\end{smallmatrix}$.
It is a well known algebra called Canonical Anticommutation Relations (CAR
algebra) \cite{Dav96}, the dimension group of which is the set of diadic
rational numbers $K_0 =\{ \left. \frac{p}{2^n} \right| p\in \Z, n \in \N \}
\subset \R$ with the natural ordering inherited from the set of real
numbers. The scale of the group is simply $[0,1] \cap K_0$.

As a matter of fact, the possible frequencies of appearances of the finite
symbolic sequences are again restricted to the elements of the $K_0$ group!

\subsection*{Acknowledgement}
%----------------------------
\addcontentsline{toc}{subsection}{Acknowledgement}

The author is thankful to Professor P\'eter Sz\'epfalusy, who has
read the manuscript and has proposed the investigation of the baker's map
(see in the Addendum).

The major part of this work was done while the author was a member of the
Research Group for Statistical Physics of the Hungarian Academy of
Sciences.

%% file: pencata.tex
\appendix
\section*{Appendices}
%====================
\addcontentsline{toc}{section}{Appendices}

\section{\label{Aa} Commutative and approximately finite dimensional
%===================================================================
$C^*$-al\-ge\-bras}
%==================

In this appendix, just for the convenience of the unfamiliar reader, a few
facts are summarized about $C^*$-algebras, which are particularly important
for the investigations of Section~\ref{Snoncom}. All the material covered
here can be found in any standard textbook on $C^*$-algebras, like e.g.
\cite{Dav96} or \cite{WeggOl93}. A concise summary is also contained in
\cite{Land98} or \cite{GLand97}.

\subsection{\label{Aa1} $C^*$-algebras}
%--------------------------------------

A {\it $C^*$-algebra} $\mathcal A$ is a Banach $*$-algebra (in most
cases over the complex number field $\C$) satisfying the so called
$C^*$-algebra equality $\norm{a^* a} = \norm{a}^2$, for all $a \in
{\mathcal A}$. In more details, the fact that $\mathcal A$ is an {\it
involutive} or {\it $*$-algebra}, means that in addition to the linear
algebraic operations there is an involution ${\mathcal A} \to {\mathcal
A}$, $a \mapsto a^*$ given on $\mathcal A$, which is conjugate linear
[i.e., $(\alpha a+\beta b)^* =\Bar{\alpha} a^* +\Bar{\beta} b^*$],
involutive [i.e., $(a^*)^* =a$], and has the property $(ab)^* =b^* a^*$. At
the same time $\mathcal A$ is a {\it Banach algebra}, thus $\mathcal A$ is
a Banach space (i.e., a normed complete vector space), and for the
algebraic product of any two elements $a,b \in \mathcal A$ the inequality
$\norm{ab} \le \norm{a} \norm{b}$ holds. (It follows that the
multiplication is separately continuous in both variables.) Two
$C^*$-algebras $\mathcal A$ and $\mathcal B$ are {\it isomorphic} if there
is a norm preserving linear bijection $\varPhi: {\mathcal A} \to {\mathcal
B}$ between them which commutes with the algebraic multiplication and with
the involution, i.e., $\varPhi(xy) =\varPhi(x) \varPhi(y)$, and
$\varPhi(x^*) =\bigl( \varPhi(x)\bigr)^*$.

The basic example for a $C^*$-algebra is the algebra ${\mathcal
B}(\mathfrak{H})$ of all bounded operators acting on the Hilbert space
$\mathfrak{H}$, with the operator norm as $C^*$-algebra norm and the
operator adjoint as $*$-involution. Moreover, as it was shown by Gelfand
and Naimark, this example is generic in the sense that every abstract
$C^*$-algebra is isometrically $*$-isomorphic to a subalgebra of the
concrete operator algebra ${\mathcal B}(\mathfrak{H})$ for an appropriate
(not necessarily separable) Hilbert space $\mathfrak{H}$.

\subsection{\label{Aa2} Commutative $C^*$-algebras}
%--------------------------------------------------

One of the simplest subcategory of the $C^*$-algebras is the class of
commutative $C^*$-algebras. It can be shown that if $X$ is any locally
compact Hausdorff space, then the algebra $C_0(X)$ of all $X\to \C$
continuous (complex valued) functions vanishing at infinity, with pointwise
addition and multiplication, supremum norm and complex conjugation as
$*$-operation is a commutative $C^*$-algebra. If, in addition $X$ is
compact, then the set $C(X)$ of all continuous functions (that agrees with
$C_0(X)$ for compact $X$) forms a unital $C^*$-algebra, where the unit is
the constant $1\in \C$ function on $X$. Moreover, this example is again
generic, i.e., any commutative (unital) $C^*$-algebra $\mathcal C$ is
isometrically $*$-isomorphic to the function algebra $C_0(X)$ [$C(X)$ in
the unital case], where the locally compact (compact) Hausdorff space $X$
can be naturally constructed from the algebra $\mathcal C$. It means that
the commutative function algebra $C(X)$ encodes all topological information
about the space $X$.

These facts can be expressed in a more abstract and concise way by stating
that $C$ (resp. $C_0$) is a contravariant invertible functor from the
category of compact (resp. locally compact Hausdorff) topological spaces
with continuous (proper) maps to the category of unital (resp. non unital)
commutative $C^*$-algebras with $*$-preserving algebra morphisms. On a
continuous map $\varPhi: X\to Y$ between two topological spaces the effect
of the functor $C$ (or $C_0$) is defined by the pull back operation
$C(\varPhi) =\varPhi^* : C(Y) \to C(X)$, $f\mapsto f\circ \varPhi$.

Thus the topological study of locally compact (compact) Hausdorff spaces is
tantamount to the algebraic study of commutative (unital) $C^*$-algebras.
Given a pure topological statement, one can at will interpret it as an
algebraic statement concerning commutative $C^*$-algebras and vice versa.
Observations of this type constitute the basic philosophy of {\it
noncommutative geometry}, which tries to generalize geometric concepts,
statements and theories, interpreted in the language of commutative
algebras, to noncommutative algebras. Of course, in the latter case the
direct classical geometrical picture is completely missing.

A nice demonstration of these guiding principles is the emergence of
algebraic $K$-theory from topological $K$-theory which is the subject of
Appendix~\ref{Ab}.

\subsection{\label{Aa3} Finite dimensional $C^*$-algebras}
%---------------------------------------------------------

The simplest noncommutative $C^*$-algebras are the finite dimensional ones.
It can be shown that they always have the form ${\mathcal M}_{n_1} \oplus
{\mathcal M}_{n_2} \oplus \dots \oplus {\mathcal M}_{n_k}$, i.e., they are
direct sums of full matrix algebras ${\mathcal M}_{n_i}$ of $n_i \times
n_i$ complex matrices ($n_i \in \N_+$, $i=1,2,\dots k$). Finite dimensional
$C^*$-algebras are always unital, and they are uniquely characterized (up
to isomorphism) by the finite set $\{ n_1, n_2, \dots n_k \}$ of numbers.
It is also an elementary fact that given a unit preserving $*$-algebraic
morphism $\varPhi: {\mathcal A} \to {\mathcal B}$ between two finite
dimensional $C^*$-algebras ${\mathcal A} =\bigoplus_{i=1}^{k} {\mathcal
M}_{n_i}$ and ${\mathcal B} =\bigoplus_{j=1}^{l} {\mathcal M}_{m_j}$ ($n_i,
m_j \in \N_+$, $i=1,2,\dots k$, $j=1,2,\dots l$), $\varPhi$ can be uniquely
decomposed into the form $\varPhi =\bigoplus_{i=1}^{k} \bigl(
\sum_{j=1}^{l} \varphi_{ij} \bigr)$, where the mappings $\varphi_{ij}
:{\mathcal M}_{n_i} \to {\mathcal M}_{m_j} \subset {\mathcal B}$ have
pairwise orthogonal and commuting ranges, and $\varPhi$ is (up to unitary
equivalence) uniquely determined by the {\it multiplicity matrix}
$A_\varPhi = [a_{ji} ]$ (of size $l\times k$) built up from the partial
multiplicities $a_{ji} \in \N$ of the mappings $\varphi_{ij}$. Since
$\varPhi$ is unital, $A$ satisfies the equation $m_j =\sum_{i=1}^k a_{ji}
n_i$. This decomposition of $\varPhi$ can be graphically visualized by the
so called {\it Bratteli diagram} \cite{Brat72}

\begin{equation}\label{EBratd}
\unitlength 0.8mm
\raisebox{-8mm}{
\begin{picture}(45,22)
\thinlines
\put(10,16){\vector(0,-1){10}}
\put(10,16){\vector(1,-1){10}}
\put(10,16){\vector(3,-1){30}}
\put(20,16){\vector(-1,-1){10}}
\put(20,16){\vector(0,-1){10}}
\put(20,16){\vector(2,-1){20}}
\put(40,16){\vector(-3,-1){30}}
\put(40,16){\vector(-2,-1){20}}
\put(40,16){\vector(0,-1){10}}
\put(2,20){\makebox(0,0){${\mathcal A}:$}}
\put(10,19){\makebox(0,0){$n_1$}}
\put(20,19){\makebox(0,0){$n_2$}}
\put(30,19){\makebox(0,0){$\dots$}}
\put(40,19){\makebox(0,0){$n_k$}}
\put(2,4){\makebox(0,0){${\mathcal B}:$}}
\put(10,3){\makebox(0,0){$m_1$}}
\put(20,3){\makebox(0,0){$m_2$}}
\put(30,3){\makebox(0,0){$\dots$}}
\put(40,3){\makebox(0,0){$m_l$}}
\put(7,11){\makebox(0,0){$a_{11}$}}
\put(43,11){\makebox(0,0){$a_{lk}$}}
\end{picture}}
\end{equation}
where the two rows symbolize the direct sum decomposition of the finite
dimensional algebras $\mathcal A$ and $\mathcal B$, and the arrows,
labeled by the multiplicities $a_{ji}$, represent the partial embeddings
$\varphi_{ij}$. In practice the arrows of zero multiplicity are omitted,
and the multiplicities of low degree are denoted by single, double, triple
arrows instead of labels.

If, for example ${\mathcal A} ={\mathcal M}_1 \oplus {\mathcal M}_2$,
${\mathcal B} ={\mathcal M}_1 \oplus {\mathcal M}_2 \oplus {\mathcal M}_3$,
and the unital homomorphism $\varPhi :{\mathcal A} \to {\mathcal B}$ is
given by

\begin{equation}\label{Ephi}
\varPhi :
\left[\begin{smallmatrix}
	[\begin{smallmatrix} a \end{smallmatrix}] & \\
	& \left[\begin{smallmatrix} b& c\\ d& e \end{smallmatrix}\right]
\end{smallmatrix}\right]
\mapsto
\left[\begin{smallmatrix}
	[\begin{smallmatrix} a \end{smallmatrix}]& & \\
	& \left[\begin{smallmatrix} a& \\ & a \end{smallmatrix}\right] & \\
	& & \left[\begin{smallmatrix} a & \\ & \begin{smallmatrix} b& c \\
d& e \end{smallmatrix} \end{smallmatrix}\right]
\end{smallmatrix}\right]
\end{equation}
then the multiplicity matrix $A_\varPhi$ and the Bratteli diagram are

\begin{align}\label{EBratex}
A_\varPhi &=
\begin{bmatrix}
1& 0\\ 2& 0\\ 1& 1
\end{bmatrix};&&
\unitlength 0.8mm
\raisebox{-8mm}{
\begin{picture}(35,22)
\thinlines
\put(10,16){\vector(0,-1){10}}
\put(10.22,16.22){\vector(1,-1){10}}
\put(9.78,15.78){\vector(1,-1){10}}
\put(10,16){\vector(2,-1){20}}
\put(20,16){\vector(1,-1){10}}
\put(2,19){\makebox(0,0){${\mathcal A}:$}}
\put(10,19){\makebox(0,0){$1$}}
\put(20,19){\makebox(0,0){$2$}}
\put(2,3){\makebox(0,0){${\mathcal B}:$}}
\put(10,3){\makebox(0,0){$1$}}
\put(20,3){\makebox(0,0){$2$}}
\put(30,3){\makebox(0,0){$3$}}
\end{picture}}
\end{align}

\subsection{\label{Aa4} Approximately finite dimensional $C^*$-algebras}
%-----------------------------------------------------------------------

Among infinite dimensional $C^*$-algebras in many respects the simplest
ones are the (unital) {\it approximately finite dimensional} (AF)
$C^*$-algebras, which are defined as the {\it increasing union} ${\mathcal
A} =\varinjlim {\mathcal A}_i =\overline{\bigcup_{i\in \N} {\mathcal A}_i}$
of a (directed) sequence $\{ {\mathcal A}_i , \varPhi_i \}_{i \in \N}$ of
finite dimensional algebras ${\mathcal A}_i$, where the mappings $\varPhi_i
:{\mathcal A}_i \hookrightarrow {\mathcal A}_{i+1}$ ($i \in \N$) are unit
preserving injective homomorphisms. (Most of the results easily carry over
to the nonunital case, where the image $\varPhi_i (\boldsymbol{1}_i )$ of
the unit $\boldsymbol{1}_i \in {\mathcal A}_i$ is a proper projection in
${\mathcal A}_{i+1}$. We, however, ---retaining the definition of Bratteli
\cite{Brat72}--- disregard this case, just for avoiding unnecessary
complications.)

Because of the construction of the AF algebra ${\mathcal A} =\varinjlim
{\mathcal A}_i$, every finite subset $\{a_1, a_2, \dots a_n \} \subset
{\mathcal A}$ of it can be approximated in norm with elements $\{ b_1, b_2,
\dots b_n \} \subset {\mathcal A}_k \subset {\mathcal A}$ from a finite
dimensional subalgebra ${\mathcal A}_k$ in the way that $\norm{a_i -b_i} <
\varepsilon$ for any arbitrarily prescribed $\varepsilon >0$ and $i \in
\{1,2,\dots n\}$. The above statement can also be reversed, namely if in a
separable algebra $\mathcal A$ any finite set of elements can be uniformly
approximated with elements from a finite dimensional subalgebra, then
$\mathcal A$ is approximately finite dimensional \cite{Brat72}.

We remark that given an increasing directed sequence ${\mathcal A}_0
\overset{\varPhi_0}{\hookrightarrow} {\mathcal A}_1
\overset{\varPhi_1}{\hookrightarrow} {\mathcal A}_2 \hookrightarrow \cdots$
of finite dimensional algebras ${\mathcal A}_i$ with unital inclusions
$\varPhi_i$, the AF algebra ${\mathcal A} =\varinjlim {\mathcal A}_i$ is
uniquely determined, but for a given AF algebra $\mathcal A$ the defining
sequence $\{ {\mathcal A}_i, \varPhi_i \}_{i\in \N}$ is far from being
unique.

The simplest way for the graphical representation of the AF algebra
${\mathcal A} =\varinjlim {\mathcal A}_i$ is drawing the Bratteli diagrams
of the subsequent mappings $\varPhi_i : {\mathcal A}_i \hookrightarrow
{\mathcal A}_{i+1}$ in a single chain.

For example, the unital algebra ${\mathcal A} =\C {\mathbf 1} \oplus
{\mathcal K}$ generated by the identity $\mathbf 1$ and the compact
operators ${\mathcal K}$ of a separable Hilbert space $\mathfrak H$ is the
injective limit of the sequence $\{ {\mathcal A}_i =\C{\mathbf 1} \oplus
P_i {\mathcal K} P_i, \varPhi_i \}_{i \in \N}$, where $P_i
:{\mathfrak H} \to {\mathfrak H}$ is the projection onto the closed
subspace of $\mathfrak H$ spanned by the first $i$ vectors of a selected
orthonormal basis of $\mathfrak H$, and $\varPhi_i$ is the natural
inclusion ${\mathcal A}_i \hookrightarrow {\mathcal A}_{i+1}$. Thus the
multiplicity matrix $A_0$ of $\varPhi_0$, $A_i$ of $\varPhi_i$ (for all
$i\ge 1$) and the Bratteli diagram of this AF algebra are clearly
\begin{align}\label{EBrCom}
A_0 &= \begin{bmatrix} 1 \\ 1 \end{bmatrix},&
A_i &= \begin{bmatrix} 1& 0\\ 1& 1 \end{bmatrix};&&
\unitlength 0.7mm \raisebox{-7mm}{
\begin{picture}(90,22)
\thinlines
\put(8,16){\vector(1,0){10}}
\put(8,16){\vector(1,-1){10}}
\put(5,16){\makebox(0,0){$1$}}
\put(24,6){\vector(1,0){10}}
\put(24,16){\vector(1,0){10}}
\put(24,16){\vector(1,-1){10}}
\put(21,16){\makebox(0,0){$1$}}
\put(21,6){\makebox(0,0){$1$}}
\put(40,6){\vector(1,0){10}}
\put(40,16){\vector(1,0){10}}
\put(40,16){\vector(1,-1){10}}
\put(37,16){\makebox(0,0){$1$}}
\put(37,6){\makebox(0,0){$2$}}
\put(56,6){\vector(1,0){10}}
\put(56,16){\vector(1,0){10}}
\put(56,16){\vector(1,-1){10}}
\put(53,16){\makebox(0,0){$1$}}
\put(53,6){\makebox(0,0){$3$}}
\put(72,6){\vector(1,0){10}}
\put(72,16){\vector(1,0){10}}
\put(72,16){\vector(1,-1){10}}
\put(69,16){\makebox(0,0){$1$}}
\put(69,6){\makebox(0,0){$4$}}
\put(85,16){\makebox(0,0){$\dots$}}
\put(85,6){\makebox(0,0){$\dots$}}
\end{picture}}
\end{align}
(In this diagram the mappings are drawn from left to right, for
convenience.)

\nocite{CuKri80}

%% file: pencatb.tex
\section{\label{Ab} $K$-theory of approximately finite dimensional
%=================================================================
$C^*$-algebras}
%==============
\nocite{Ell78}

In this appendix we shall have a quick glance into the algebraic
$K$-theory, focusing the attention on the narrow slice of the theory needed
for the calculation of the {\it scaled dimension group} of approximately
finite dimensional $C^*$-algebras, which is a complete invariant of this
class of algebras, according to the result of G.\ A.\ Elliot \cite{Ell76}.
As a general reference on $K$-theory of $C^*$-algebras we recommend
\cite{WeggOl93} and Chapter~IV of \cite{Dav96} for the case of AF
$C^*$-algebras.

Algebraic $K$-theory has its roots in topological $K$-theory \cite{At67},
and as the sprouting of the algebraic theory provides an excellent example
how the basic philosophy of noncommutative geometry works in practice, we
start our overview ---following the chronological way--- by the definition
of the $K^0 (X)$ group of a topological space $X$, which is the basic
ingredient of topological $K$-theory.

\subsection{\label{Ab1} The topological $K^0$ group}
%---------------------------------------------------

Roughly speaking $K^0$ is a contravariant functor which associates to each
compact topological space $X$ an ordered Abelian group $K^0 (X)$, whose
elements are equivalence classes of formal differences of vector bundles
over $X$. The group $K^0(X)$ is a homotopy invariant of $X$, and it nicely
fits into certain exact sequences, but these properties are beyond our
present scope. The precise definition of $K^0(X)$ is as follows.

Given a compact topological space $X$, let ${\mathcal V}(X)$ denote the set
(of equivalence classes) of all locally trivial complex vector bundles
over the base space $X$. For simplicity the notation $[n]_X$ is used for
the $n$-dimensional trivial bundle $\C^n \times X \xrightarrow{\pi} X$,
where $n\in \N$. From two vector bundles ${\mathcal E} =(E
\xrightarrow{\pi_E} X)$ and ${\mathcal F} =(F \xrightarrow{\pi_F} X)$ over
the same base space $X$ one can construct the Whitney sum ${\mathcal E}
\oplus {\mathcal F} \in {\mathcal V}(X)$, which is essentially the
fiberwise direct sum of the two bundles, i.e., its fiber over the point
$x\in X$ is $E_x \oplus F_x$. This operation $\oplus: {\mathcal V}(x)
\times {\mathcal V}(x) \to {\mathcal V}(x)$ defines a commutative semigroup
structure on ${\mathcal V}(x)$, with zero element (semigroup unit) $[0]_X$.

The topological $K^0 (X)$ group is obtained from the semigroup ${\mathcal
V}(X)$ by the {\it Grothendieck construction}, i.e., basically in the same
way as the additive group of integers $\Z$ is obtained from the additive
semigroup $\N$ of the natural numbers, by considering (equivalence
classes of) formal differences $[a-b] \in \Z$ of natural numbers $a,b \in
\N$. (The set $\Q_+$ of positive rational numbers is also a Grothendieck
group constructed from the multiplicative semigroup $\N_+$ of positive
natural numbers by forming formal fractions $\big[ \frac{p}{q} \big] \in
\Q_+$, $p,q \in \N_+$.)

Generally the Grothendieck construction consists of two steps. Given a
commutative semigroup $(S,+)$ with zero element $0 \in S$, first the factor
semigroup $\tilde{S} =S/\approx$ is created, where two elements $a,b \in S$
are $\approx$-equivalent, $a \approx b$, if and only if there is a third
element $c\in S$ such that $a+c =b+c$. This factorization ensures that
$\tilde{S}$ is a cancellation semigroup. (The commutative semigroup $S$ is
called {\it cancellation semigroup} if it has the {\it cancellation
property}, i.e., whenever $a+c=b+c$ holds for arbitrary elements $a,b,c \in
S$ then $a=b$.)

Second, the Grothendieck group $G$ of $S$ is defined by the
formula $G=\tilde{S}^2 /\sim$. Writing the elements of $\tilde{S}^2$, for
convenience, in the form of formal differences, i.e., $x-y =(x,y) \in
\tilde{S}^2$, the elements $a-b$, $c-d \in \tilde{S}^2$ are
$\sim$-equivalent, $(a-b) \sim (c-d)$, if and only if $a+d =c+b$ holds in
$\tilde{S}$. The group addition in $G$ is defined by $[a-b] +[c-d] =\big[
(a+c)-(b+d) \big]$. It is straightforward to verify that the above given
group operation in $G$ is well defined on the $\sim$-equivalence classes,
and it defines a commutative group structure with unit $[0-0]= [s-s]$ (for
any $s \in \tilde{S}$). The cancellation semigroup $\tilde{S}$ is naturally
considered as a sub-semigroup of the Grothendieck group $G$ defined by the
inclusion $\tilde{S} \hookrightarrow G$, $s \mapsto [s-0]$, and the subset
$\tilde{S} \subset G$ is generating (by the definition of $G$), i.e., $G
=\tilde{S} -\tilde{S}$ holds. It is worth remarking that the cancellation
property of $\tilde{S}$ is needed for the transitivity of the relation
$\sim$.

Our previous examples of the Grothendieck group were a bit untypical, since
both the additive semigroup $\N$ and the multiplicative semigroup $\N_+$
have the cancellation property, so in these cases the first step of the
construction is unnecessary, $\tilde{\N} \cong \N$, $\tilde{\N}_+ \cong
\N_+$.

In general, however, the semigroup ${\mathcal V}(X)$ of vector bundles
does not have the cancellation property, so first the factor semigroup
${\mathcal V} (X) /\approx$ has to be constructed, which is denoted
by $K^{0+} (X)$, and then the group $K^0 (X) =K^{0+} (X) \times K^{0+}
(X) /\sim$ is by definition the Grothendieck group of ${\mathcal V}(X)$.
Thus, putting the two steps of the Grothendieck construction together, the
elements of the group $K^0 (X)$ are equivalence classes of formal
differences of vector bundles ${\mathcal E_i}, {\mathcal F_i} \in
{\mathcal V}(X)$ (here $i=1,2$), two such objects being equivalent,
$({\mathcal E}_1 -{\mathcal F}_1) \sim ({\mathcal E}_2 -{\mathcal F}_2)$,
if and only if there is a third bundle ${\mathcal G} \in {\mathcal V}(X)$
such that ${\mathcal E}_1 \oplus {\mathcal F}_2 \oplus {\mathcal G} \cong
{\mathcal E}_2 \oplus {\mathcal F}_1 \oplus {\mathcal G}$, and the sum of
two classes is $[{\mathcal E}_1 -{\mathcal F}_1] +[{\mathcal E}_2
-{\mathcal F}_2] =\big[ ({\mathcal E}_1 \oplus {\mathcal E}_2)- ({\mathcal
F}_1 \oplus {\mathcal F}_2) \big]$.

(As a counter example for the cancellation property of ${\mathcal V} (X)$
consider the real tangent resp. normal bundle $TS^2 \in {\mathcal V}_\R
(S^2)$ resp. $NS^2 \cong [1]_{S^2}^\R \in {\mathcal V}_\R (S^2)$ as well as
the trivial real one dimensional bundle $[1]_{S^2}^\R$ over the two
dimensional sphere $S^2 \subset \R^3$ as base space. Clearly, $TS^2 \oplus
NS^2 \cong [2]_{S^2}^\R \oplus [1]_{S^2}^\R \cong [3]_{S^2}^\R$ holds, but
$TS^2 \not\cong [2]_{S^2}^\R$.)

A continuous map $\varPhi :X \to Y$ between two (compact) topological
spaces $X$ and $Y$ induces a `pull back' map $\varPhi^* : {\mathcal V}(Y)
\to {\mathcal V}(X)$ between the sets of vector bundles over the two
spaces in reverse order, in such a way that for ${\mathcal
E}=(E\xrightarrow{\pi_E} Y) \in {\mathcal V}(Y)$ the induced pull back
bundle $\varPhi^* {\mathcal E} \in {\mathcal V}(X)$ has the fiber
$E_{\varPhi(x)}$ at the point $x\in X$. It turns out that $\varPhi^*$
commutes with the Whitney sum operation [i.e., $\varPhi^* ({\mathcal E}
\oplus {\mathcal F}) \cong \varPhi^* {\mathcal E} \oplus \varPhi^*
{\mathcal F}$ for ${\mathcal E}, {\mathcal F} \in {\mathcal V}(Y)$],
respects the equivalence classes, and thus it induces a group homomorphism
$K^0 (\varPhi): K^0 (Y) \to K^0 (X)$. It means that $K^0$ is a
contravariant functor from the category of topological spaces to the
category of commutative groups.

\subsection{\label{Ab2} The algebraic $K_0$ group}
%-------------------------------------------------

Now the definition of the (topological) $K^0$ group is rephrased in an
algebraic language and extended from commutative to noncommutative
$C^*$-algebras, according to the basic philosophy of noncommutative
geometry. Again, for the sake of simplicity, we consider only unital
$C^*$-algebras.

As it has been seen in Appendix~\ref{Aa2}, the compact base space
$X$ is replaced with the unital commutative $C^*$-algebra $C(X)$ of
continuous $X\to \C$ functions. The algebraic counterpart of a given
bundle ${\mathcal E}=(E \xrightarrow{\pi_E} X)$ is the set
$\Gamma({\mathcal E})$ of its continuous sections (here $\Gamma({\mathcal
E})=\{ f:X\to E \, |\, f \text{ is continuous, } \pi_E \circ f = \id_X
\}$), which has a $C(X)$-module structure by fiberwise defined operations.
Moreover, the module $\Gamma({\mathcal E})$ is always {\it projective} and
{\it finitely generated}, what is the algebraic equivalence of Swan's
theorem (see \cite{WeggOl93}, Theorem~13.1.6) stating that for every vector
bundle ${\mathcal E} \in {\mathcal V} (X)$ over the compact base space $X$
there is an `orthogonal complement' bundle ${\mathcal F} \in {\mathcal V}
(X)$ such that ${\mathcal E} \oplus {\mathcal F} \cong [n]_X$ is trivial.
This means that the module $\Gamma({\mathcal E})$ is {\it projective} and
{\it finitely generated}, i.e., it can be described as the range of a
($C(X)$-linear) projection $P:\Gamma([n]_X) \to \Gamma([n]_X)$ (which acts
fiberwise on the bundle $[n]_X \cong {\mathcal E} \oplus {\mathcal F}$),
and since $\Gamma([n]_X) \cong \{f:X\to \C^n \, |\, f \text{ is continuous
} \}$, the projection $P$ is a selfadjoint idempotent element of the $n
\times n$ matrix algebra ${\mathcal M}_n \big( C(X)\big) \cong {\mathcal
M}_n \otimes C(X)$ with entries in the function space $C(X)$. (Here
${\mathcal M}_n = {\mathcal M}_n (\C)$ is the usual complex matrix
algebra.) The converse of these statements is also true, i.e., given a
projection (selfadjoint idempotent) $P\in {\mathcal M}_n \big( C(X) \big)$,
there is a vector bundle ${\mathcal E} \in {\mathcal V} (X)$ such that
$\Gamma({\mathcal E}) \cong \Ran P$. (These results are due to Serre and
Swan \cite{Ser57,Swan62}.)

By these observations the geometric structure of a bundle ${\mathcal E}=(E
\xrightarrow{\pi_E} X) \in {\mathcal V}(X)$ is entirely characterized in
purely algebraic terms; the base space $X$ is replaced with the commutative
algebra $C(X)$, and the structure of the bundle is encoded by a
projection $P\in {\mathcal M}_n \big( C(X) \big)$ of the $n\times n$ matrix
algebra over $C(X)$. It can also be shown that two projections $P\in
{\mathcal M}_n \big( C(X)\big)$ and $Q\in {\mathcal M}_m \big( C(X)\big)$
describe isomorphic bundles if and only if they are von Neumann
equivalent \cite{MvN36}, $P\sim Q$, which means that their ranges are
isometric, i.e., there exists a partial isometry $U: {\mathcal M}_n \big(
C(X)\big) \to {\mathcal M}_m \big( C(X)\big)$ with support projection
$P=U^*U$ and range projection $Q=UU^*$ \cite{MvN36}.

In this algebraic formulation the commutativity of the algebra $C(X)$ has
never been explicitely exploited, and in principle nothing prevents the
substitution of a general {\it noncommutative} $C^*$-algebra for the
commutative function algebra $C(X)$. The amazing news of noncommutative
topology is the astounding fact that the constructions and the main results
of topological $K$-theory do survive this rather drastic change of
omitting commutativity of the algebra $C(X)$, and along the guiding lines
of the topological theory a similar, even richer theory can be established
for noncommutative $C^*$-algebras called {\it algebraic $K$-theory}. In the
rest of this appendix the construction of the $K_0({\mathcal C})$ group of
a general $C^*$-algebra ${\mathcal C}$ is surveyed, and then the scaled
dimension group $\big( K_0({\mathcal A}), K_0^+ ({\mathcal A}),
\Gamma({\mathcal A}) \big)$ is introduced for an approximately finite
dimensional algebra $\mathcal A$.

Given a (noncommutative) $C^*$-algebra $\mathcal C$, let ${\mathcal
M}_\infty ({\mathcal C}) =\bigcup_{n=1}^\infty {\mathcal M}_n ({\mathcal
C})$ denote the (non-complete) $*$-algebra of all finite dimensional matrix
algebras with entries in $\mathcal C$ [where ${\mathcal M}_n ({\mathcal C})
= {\mathcal M}_n \otimes {\mathcal C}$ is considered as a subalgebra of
${\mathcal M}_{n+1} ({\mathcal C})$ by the inclusion ${\mathcal M}_n
({\mathcal C}) \hookrightarrow {\mathcal M}_{n+1} ({\mathcal C})$ in the
upper left corner $a \mapsto \diag(a,0)$], and let ${\mathcal P}({\mathcal
C})$ be the set of all projections in ${\mathcal M}_\infty ({\mathcal C})$.
The algebraic counterpart of the set (of isomorphism classes) of vector
bundles is ${\mathcal V}({\mathcal C}) ={\mathcal P}({\mathcal C}) /\sim$,
where $\sim$ is the von Neumann equivalence (see above). With the
addition $\oplus: {\mathcal V}({\mathcal C}) \times {\mathcal V}({\mathcal
C}) \to {\mathcal V}({\mathcal C})$, $[P]\oplus [Q] =\big[
\begin{smallmatrix} P& 0\\ 0& Q \end{smallmatrix} \big]$ and $[0]$ as zero
element ${\mathcal V}({\mathcal C})$ is a commutative semigroup, and the
algebraic $K_0({\mathcal C})$ group is by definition its Grothendieck
extention.

As in the topological theory, ${\mathcal V}({\mathcal C})$ need not have
the cancellation property, thus the first step of the Grothendieck
construction is to introduce the cancellation semigroup $K_0^+ ({\mathcal
C}) ={\mathcal V}({\mathcal C})/ \approx$. We have again $K_0^+ ({\mathcal
C}) \subset K_0({\mathcal C})$ and $K_0({\mathcal C}) =K_0^+ ({\mathcal C})
-K_0^+ ({\mathcal C})$.

A unital $*$-algebraic morphism $\varPhi :{\mathcal C} \to {\mathcal
D}$ between two $C^*$-algebras induces morphisms $\varPhi_* :{\mathcal M}_n
({\mathcal C}) \to {\mathcal M}_n ({\mathcal D})$ between the matrix
algebras (applying $\varPhi$ to each element of the matrices separately),
and from this yields a well defined group homomorphism $K_0 (\varPhi) :K_0
({\mathcal C}) \to K_0 ({\mathcal D})$, $[P-Q] \mapsto [\varPhi_* P
-\varPhi_* Q]$ by which $K_0$ becomes a covariant functor from the
category of $C^*$-algebras to that of Abelian groups. For a commutative
unital $C^*$-algebra ${\mathcal B} \cong C(X)$ isomorphic to the function
space over $X$, the algebraic and topological $K$-groups coincide, in the
sense that $K_0 ({\mathcal B}) \cong K^0 (X)$.

\subsection{\label{Ab3} The scaled dimension group of AF algebras}
%-----------------------------------------------------------------

From this point on we restrict our attention to approximately finite
dimensional $C^*$-algebras denoted by $\mathcal A$. It can be proven that
in this case $K_0^+ ({\mathcal A})$ coincides with ${\mathcal V}({\mathcal
A})$ (thus ${\mathcal V}({\mathcal A})$ has already the cancellation
property), and $K_0^+ ({\mathcal A})$ is generated by the (equivalence
classes of) projections in $\mathcal A$ (\cite{Dav96}, Theorem IV.1.6).
Moreover, the (generating) semigroup $K_0^+ ({\mathcal A}) \subset K_0
({\mathcal A})$ has the property $K_0^+ ({\mathcal A}) \bigcap \big( -K_0^+
({\mathcal A}) \big) =\{0\}$ (\cite{Dav96}, Theorems IV.2.3, IV.2.4), which
means that $K_0^+ ({\mathcal A})$ defines an order on the group $K_0
({\mathcal A})$ by $x\ge y$ if and only if $x-y \in K_0^+ ({\mathcal A})$
for all $x,y \in K_0 ({\mathcal A})$. The ordered group $\big( K_0
({\mathcal A}), K_0^+ ({\mathcal A}) \big)$ is called the {\it dimension
group} of the AF algebra $\mathcal A$. For the sake of brevity and for
convenience, from now $K_0 ({\mathcal A})$ refers to the ordered group of
the AF algebra $\mathcal A$, even if the order structure is not denoted
explicitely.

As examples, we give now the dimension groups of finite dimensional
$C^*$-algebras.

First let ${\mathcal A} ={\mathcal M}_k$ be a full matrix algebra with
entries in $\C$. [For simplicity, ${\mathcal M}_k$ denotes ${\mathcal M}_k
(\C)$.] The algebra ${\mathcal M}_n ({\mathcal A}) \cong {\mathcal M}_{nk}$
is again a full matrix algebra (over $\C$), thus the equivalence classes of
projections in ${\mathcal M}_\infty ({\mathcal A}) \cong {\mathcal
M}_\infty (\C) ={\mathcal M}_\infty $ are labeled by the rank of the
projections, which can be any natural number. It means that
$K_0^+({\mathcal M}_k )\cong \N$, and the dimension group of ${\mathcal
M}_k$ is the additive group of integers with its usual order,
$K_0({\mathcal M}_k) \cong (\Z,\N)$.

Now let ${\mathcal A} =\bigoplus_{i=1}^k {\mathcal M}_{n_i}$ be a general
finite dimensional $C^*$-algebra over $\C$. Every finite dimensional matrix
algebra over $\mathcal A$ splits into the direct sum of $k$ full matrix
algebras with entries in $\C$, because
\begin{equation}\label{Emnadec}
{\mathcal M}_s ({\mathcal A}) \cong
{\mathcal M}_s \otimes \big( {\textstyle\bigoplus_{i=1}^k} {\mathcal
M}_{n_i} \big) \cong {\textstyle\bigoplus_{i=1}^k} {\mathcal M}_s \otimes
{\mathcal M}_{n_i} \cong {\textstyle\bigoplus_{i=1}^k} {\mathcal M}_{sn_i}.
\end{equation}
It means that the projections $P,Q \in {\mathcal M}_s ({\mathcal A})$ have
the decomposition $P= \bigoplus_{i=1}^k P_i$, $Q= \bigoplus_{i=1}^k Q_i$
where $P_i$, $Q_i$ are projections in ${\mathcal M}_{sn_i}$, and $P\sim Q$
if and only if $\Rank P_i =\Rank Q_i$ for all $i=1,2 \ldots k$. Thus $K_0^+
({\mathcal A}) \cong \N^k$, and the ordered dimension group of $\mathcal A$
is $K_0({\mathcal A}) \cong(\Z^k, \N^k)$.

These two examples demonstrate that the dimension group is not a complete
invariant of the finite dimensional $C^*$-algebra ${\mathcal
A}=\bigoplus_{i=1}^k {\mathcal M}_{n_i}$ since it does not `feel' the
dimensions $n_i$ of the matrix algebras ${\mathcal M}_{n_i}$ the algebra
$\mathcal A$ is built of. To overcome this difficulty, the {\it scale}
$\Gamma({\mathcal A})$ is introduced, which is a subset of $K_0^+
({\mathcal A})$, and in the case when $\mathcal A$ is unital it is simply
\begin{equation}\label{Escale}
\Gamma({\mathcal A}) =\big\{ [P] \in K_0^+ ({\mathcal A}) \bigr. \bigl|
[P] \le [ \boldsymbol{1}_{\mathcal A}] \big\},
\end{equation}
where $\boldsymbol{1}_{\mathcal A}$ is the identity (largest projection) of
$\mathcal A$.

The scaled dimension group $\big( K_0 ({\mathcal A}), K_0^+ ({\mathcal
A}), \Gamma({\mathcal A}) \big)$ of the finite algebra ${\mathcal
A}=\bigoplus_{i=1}^k {\mathcal M}_{n_i}$, denoted again sloppily by the
single symbol $K_0 ({\mathcal A})$ is clearly
\begin{subequations}\label{Esdg}
\begin{align}\label{Esdg:a}
 K_0 ({\mathcal A}) &\cong
 \big( \Z^k, \N^k, [(0,0\ldots0),(n_1,n_2\ldots n_k )] \big) \cong \\
\label{Esdg:b} &\cong
\big( {\textstyle\bigoplus_{i=1}^k} \Z[1/n_i],
{\textstyle\bigoplus_{i=1}^k} \N[1/n_i], [(0,0\ldots 0),(1,1\ldots 1)]
\big),
\end{align}
\end{subequations}
where $[(0,0\ldots 0), (n_1,n_2\ldots n_k )]$ denotes the points $(p_1, p_2
\ldots p_k) \in K_0^+ ({\mathcal A})$ for which $0 \le p_i \le n_i$ holds
($i=1,2\ldots k$), and $\Z[q]$ (resp. $\N[q]$) denotes the additive
(semi-) group generated by $\Z$ and $q \in \R_+$ (resp. by $\N$ and $q\in
\R_+$). In \eqref{Esdg:b}, for convenience, the dimension group $\big( K_0
({\mathcal A}), K_0^+ ({\mathcal A}) \big)$ was `scaled' so that the scale
$\Gamma({\mathcal A})$ should have a standard form. It is visible that the
scaled dimension group already distinguishes between different dimensions
$n_i$. Even more is true; the scaled dimension group is a {\it complete}
invariant of AF $C^*$-algebras \cite{Ell76}.

Now, as a first step towards the description of the $K_0$ group of a given
AF algebra let us investigate the effect of the functor $K_0$ on a unital
$*$-algebra homomorphism $\varPhi :{\mathcal A} \to {\mathcal B}$ between
two finite dimensional algebras ${\mathcal A} =\bigoplus_{i=1}^k {\mathcal
M}_{n_i}$ and ${\mathcal B} =\bigoplus_{j=1}^l {\mathcal M}_{m_j}$ with
multiplicity matrix $A_\varPhi =[a_{ji}]$. (The relation $m_j =\sum_{i=1}^k
a_{ji} n_i$ holds between the dimensions, because $\varPhi$ is unit
preserving.)

Since in any finite dimensional (or even AF) algebra the (equivalence
classes of) projections generate the $K_0^+$ semigroup, it is enough to
investigate the effect of $K_0 (\varPhi)$ on the projections. Every
projection $P \in {\mathcal A}$ decomposes in the form $P=
\bigoplus_{i=1}^k P_i$ (where the projections $P_i \in {\mathcal
M}_{n_i}$), according to the direct sum decomposition of the algebra
itself, and the element $[P] \in K_0^+ ({\mathcal A})$ is represented by
the $n$-tuple $(\Rank P_1, \Rank P_2 \ldots \Rank P_k ) \in \N^k$ in the
dimension group \eqref{Esdg:a}. The image of $P$ is also a projection $Q=
\varPhi_* P =\bigoplus_{j=1}^l Q_j$ (where $Q_j$ is a projection in
${\mathcal M}_{m_j}$), and $\Rank Q_j= \sum_{i=1}^k a_{ji} \Rank P_i$.
Since $K_0 (\varPhi )[P]= [\varPhi_* P] =[Q]$, the effect of $K_0(\varPhi)$
on the dimension group is simply described by the multiplication with the
multiplicity matrix $A_\varPhi$, i.e.:
\begin{subequations}\label{EK0F}
\begin{align}\label{EK0F:a}
&K_0(\varPhi):& K_0({\mathcal A}) &\cong \Z^k &\longrightarrow&&
K_0({\mathcal B}) &\cong \Z^l,
\\ \label{EK0F:b}
&&{\mathbf u} &=(u_1,u_2\ldots u_k)& \longmapsto&& A_\varPhi {\mathbf u} &=
\big( {\textstyle \sum_{i=1}^k} a_{ji} u_i \big)_{j=1}^l .
\end{align}
\end{subequations}
We remark that $K_0(\varPhi) \big( K_0^+ ({\mathcal A})\big) \subset K_0^+
({\mathcal B})$, since the entries $a_{ji}$ of the matrix $A_{\varPhi}$ are
nonnegative numbers, $K_0(\varPhi) \big( \Gamma ({\mathcal A})\big)
\subset \Gamma ({\mathcal B})$, and $K_0(\varPhi)\big( [\id_{\mathcal A}]
\big) =[\id_{\mathcal B}]$ since $\varPhi$ is unital, i.e., $m_j
=\sum_{i=1}^k a_{ji} n_i$ holds. Such kind of morphisms between scaled
dimension groups are called {\it positive unital homomorphisms}.

A very nice property of the $K_0$ functor is that it commutes with
direct limit (\cite{Dav96}, Theorem~IV.3.3), thus given an AF algebra
${\mathcal A}=\varinjlim {\mathcal A}_i$ with unital inclusions $\varPhi_i
:{\mathcal A}_i \hookrightarrow {\mathcal A}_{i+1}$, the algebraic $K_0$
group of $\mathcal A$ is $K_0({\mathcal A}) =\varinjlim K_0 ({\mathcal
A}_i)$, where the direct limit of (scaled dimension) groups on the right
hand side is understood with the (positive unital) group
homomorphisms $K_0 (\varPhi_i): K_0 ({\mathcal A}_i)
\to K_0 ({\mathcal A}_{i+1})$.

The direct limit group $K_0 ({\mathcal A}) =\varinjlim K_0 ({\mathcal
A}_i)$ is defined abstractly, up to isomorphism, by its universal property
(see \cite{WeggOl93}, Appendix~L). For calculations, however, it is good to
have a concrete realization of $K_0 ({\mathcal A})$. It can be defined as
\begin{subequations}\label{EilKgen}
\begin{equation}\label{EilKgen_a}
K_0 ({\mathcal A}) =\Big. {\prod_{i\in \N}}' K_0 ({\mathcal A}_i)
\Big/ \sim ,
\end{equation}
where the `primed' product $\prod_{i\in \N}' K_0 ({\mathcal A}_i)$ is the
subset of the infinite direct product space $\prod_{i \in \N} K_0
({\mathcal A}_i)$ consisting of the elements with {\it `predictable
tails'}, i.e.,
\begin{equation}\label{EilKgen_b}
{\prod_{i \in \N}}' K_0({\mathcal A}_i) =\left\{
(x_i )_{i\in \N} \in \prod_{i\in \N} K_0 ({\mathcal A}_i) \left|
\begin{array}{l}
\exists n\in \N, \text{ such that}\\
(\forall j>n) \; K_0 (\varPhi_j) (x_j) =x_{j+1}
\end{array}\right. \right\},
\end{equation}
and two elements $(x_i)_{i\in \N}, (y_i)_{i \in \N} \in \prod_{i \in \N}'
K_0({\mathcal A}_i)$ are $\sim$-equivalent, if their tails coincide, i.e.,
\begin{align}\label{EilKgen_c}
(x_i)_{i\in \N} &\sim (y_i)_{i \in \N} &&\iff &
\exists n\in \N, &\text{ such that } (\forall j>n) \; x_j =y_j .
\end{align}
The definition of the positive elements $K_0^+({\mathcal A})$ resp. the
scale $\Gamma({\mathcal A})$ is now straightforward, they consists of the
series $(x_i)_{i \in \N} \in \prod_{i\in \N}' K_0({\mathcal A}_i)$ whose
tail is positive resp. in the scale, i.e.,
\begin{align}\label{EilKgen_d}
K_0^+ ({\mathcal A}) &= \left\{
\big[ (x_i)_{i \in \N} \big] \in K_0({\mathcal A}) \left|
\begin{array}{l}
\exists n\in \N \text{ such that}\\
(\forall j>n) \; x_j \in K_0^+ ({\mathcal A}_j)
\end{array}\right. \right\},
\\ \label{EilKgen_e}
\Gamma ({\mathcal A}) &= \left\{
\big[ (x_i)_{i \in \N} \big] \in K_0({\mathcal A}) \left|
\begin{array}{l}
\exists n\in \N \text{ such that}\\
(\forall j>n) \; x_j \in \Gamma ({\mathcal A}_j)
\end{array}\right. \right\}.
\end{align}
\end{subequations}

Particularly, if from a threshold index all the homomorphisms $K_0
(\varPhi_i) :K_0({\mathcal A}_i) \hookrightarrow K_0({\mathcal A}_{i+1})$
are {\it injective}, then
\begin{subequations}\label{EinjlimK}
\begin{align}\label{EinjlimK:a}
K_0({\mathcal A}) &=\bigcup_{i=1}^\infty K_0({\mathcal A}_i),&
K_0^+ ({\mathcal A}) &=\bigcup_{i=1}^\infty K_0^+ ({\mathcal A}_i)&
&\text{and}&
\Gamma({\mathcal A}) &=\bigcup_{i=1}^\infty \Gamma({\mathcal A}_i),
\end{align}
where the right hand side of the equations are increasing unions defined by
$K_0(\varPhi_i)$, i.e.,
\begin{align}\label{EinjlimK:b}
K_0({\mathcal A_0})
\xrightarrow[\subset]{K_0(\varPhi_0)} K_0({\mathcal A_1})
\xrightarrow[\subset]{K_0(\varPhi_1)} K_0({\mathcal A_2})
\xrightarrow[\subset]{K_0(\varPhi_2)} K_0({\mathcal A_3})
\hookrightarrow \cdots,
\end{align}
\end{subequations}
and similar inclusions hold for $K_0^+({\mathcal A_i})$ as well as
$\Gamma({\mathcal A_i})$. (This injectivity property of $K_0 (\varPhi_i)$
holds for the algebra associated to the Penrose tilings [see
Statements~\ref{StC*Pen} and \ref{StK0Pen}] and for the example of compact
operators discussed bellow, but it is not true for the algebra associated
to the cat map [see Statements~\ref{StC*cat} and \ref{StK0cat}].)

\begin{figure}
\centerline{\includegraphics[scale=0.5]{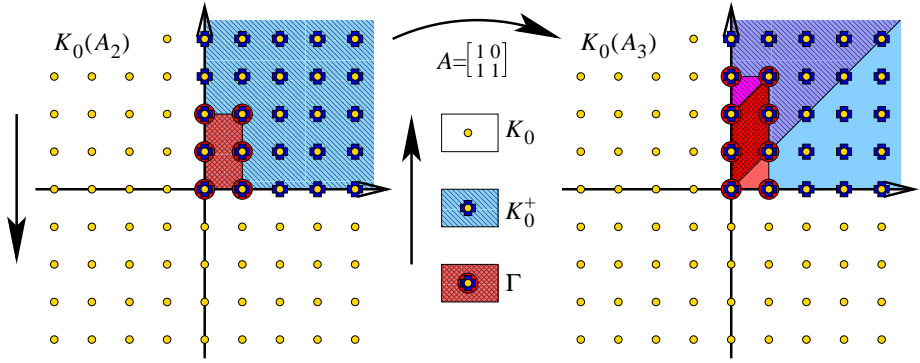}}
\caption{\label{F:sdgh} The shearing transformation between the dimension
groups $K_0 ({\mathcal A}_2)$ and $K_0 ({\mathcal A}_3)$.}
\end{figure}

As a final example we present the scaled dimension group of the unital AF
algebra ${\mathcal A} =\C \boldsymbol{1} \oplus {\mathcal K} =\varinjlim
{\mathcal A}_i$ of compact operators $\mathcal K$ extended with the unit
$\boldsymbol{1}$, defined by the Bratteli diagram~\eqref{EBrCom} at the
end of Section~\ref{Aa4}. According to the previous result \eqref{Esdg:a},
the scaled dimension group of the finite dimensional algebras ${\mathcal
A}_i =\C\boldsymbol{1} \oplus P_i {\mathcal K}P_i \cong {\mathcal M}_1
\oplus {\mathcal M}_i$ (where $P_i$ is a projection of rank $i$) are
\begin{subequations}\label{EK0Ai}
\begin{align}\label{EK0Ai:a}
K_0 ({\mathcal A}_0) &\cong \big( \Z, \N, \{ 0,1\} \big),&&\\
K_0 ({\mathcal A}_i) &\cong
\big( \Z^2, \N^2, \{ 0,1\} \times \{0,1\ldots i\} \big),&&
\text{for $i\ge 1$},
\end{align}
\end{subequations}
and the (scaled dimension) group homomorphisms $K_0 (\varPhi_i): K_0
({\mathcal A}_i) \hookrightarrow K_0 ({\mathcal A}_{i+1})$, which  are all
injective, are defined by the multiplicity matrices $A_0 =\big[
\begin{smallmatrix} 1\\ 1 \end{smallmatrix} \big]$, $A_i = \big[
\begin{smallmatrix} 1&0 \\ 1&1 \end{smallmatrix} \big]$ (for $i\ge 1$)
given in \eqref{EBrCom}. Figure~\ref{F:sdgh} illustrates the homomorphism
$K_0 (\varPhi_2): K_0 ({\mathcal A}_2) \hookrightarrow K_0 ({\mathcal
A}_3)$.

\begin{figure}
\centerline{\includegraphics[scale=0.5]{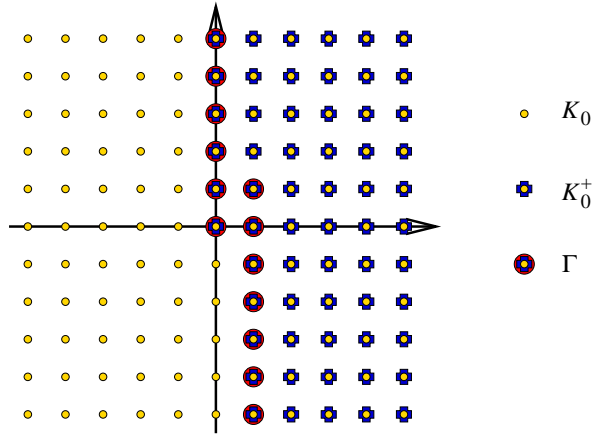}}
\caption{\label{F:sdgcom} The scaled dimension group $K_0 (\C
{\mathbf 1} \oplus {\mathcal K})$ of the algebra generated by the
compact operators ${\mathcal K}$ and the identity ${\mathbf 1}$.}
\end{figure}

Taking the injective limit $\varinjlim K_0 ({\mathcal A}_i)$, according to
the formulas~\eqref{EinjlimK}, we get that
\begin{subequations}\label{EK0Com} \begin{align}\label{EK0Com_a}
K_0 (\C \boldsymbol{1} \oplus {\mathcal K}) &\cong ( \Z^2 , K_0^+,
\Gamma),&& \text{where}
\\ \label{EK0Com_b}
K_0^+ &= \big( \{ 0\} \times \N \big) \cup ( N_+ \times \Z ) && \text{and}
\\ \label{EK0Com_c}
\Gamma &= \big( \{ 0 \} \times \N \big) \cup \big( \{ 1\} \times (1- \N)
\big),&&
\end{align}
\end{subequations}
as it is depicted in Figure~\ref{F:sdgcom}. (Indeed, for $i\ge 2$ the
transformations $A_i$ are $\Z^2 \to \Z^2$ bijections, so the $K_0$ group in
question is $\Z^2$. The positive elements $K_0^+$ are the points which are
shifted into the cone $\N^2 \subset \Z^2$ after sufficiently many
applications of the transformation $A= \left[\begin{smallmatrix} 1& 0\\1& 1
\end{smallmatrix} \right]$, and the scale $\Gamma$ consists of the points
${\mathbf x} \in \Z^2$ for which the $n^{\rm th}$ iterate $A^n {\mathbf x}$
falls into the scale $\{ 0,1\} \times \{0, 1\ldots n+1 \}$ of the group
$K_0 ({\mathcal A}_{n+1})$.)

%\begin{equation}\label{EK0Com}
%K_0 (\C \boldsymbol{1} \oplus {\mathcal K}) \cong \big( \Z^2 , \N^2,
%\{0,1\} \times \N \big). \end{equation}

We remark that the $K_0$ group of the (nonunital AF) algebra $\mathcal K$
of compact operators is $K_0 ({\mathcal K}) \cong (\Z, \Z, \N)$ (see
\cite{Dav96}, Example~IV.3.5), so the unitization brings in an extra
summand $\Z$, what is generally true (\cite{WeggOl93}, Proposition~6.2.2).

%% file: pencat0.bbl
\begin{thebibliography}{MvN36}

\bibitem[AA68]{A_A}
V.~I. Arnold and A.~Avez.
\newblock {\em Ergodic Problems of Classical Mechanics}.
\newblock The Mathematical Physics Monograph Series. W. A. Benjamin, Inc.,
  1968.

\bibitem[Ati67]{At67}
M.F. Atiyah.
\newblock {\em $K$-theory}.
\newblock W.A. Benjamin Inc., New York, Amsterdam, 1967.

\bibitem[AW67]{AdWe67}
R~Adler and B~Weiss.
\newblock Entropy a complete metric invariant for automorphism of the torus.
\newblock {\em Proc. Nat. Acad. Sci. USA}, 57:1573--1576, 1967.

\bibitem[Bow70]{Bow70}
R~Bowen.
\newblock {M}arkov partitions for axiom {$A$} diffeomorphisms.
\newblock {\em Amer. J. Math.}, 92(3):725--747, 1970.

\bibitem[Bra72]{Brat72}
O.~Bratteli.
\newblock Inductive limits of finite dimensional {$C^*$}-algebras.
\newblock {\em Trans. Amer. Math. Soc.}, 171:195--234, 1972.

\bibitem[CFS82]{CoFoSi82}
I.P. Cornfeld, S.V. Fomin, and Ya.G. Sinai.
\newblock {\em Ergodic Theory}, volume 245 of {\em Grundlagen der
  mathemetischen Wissenschaften}.
\newblock Springer-Verlag, New York Heidelbrg Berlin, 1982.

\bibitem[CK80]{CuKri80}
Joachim Cuntz and Wolfgang Krieger.
\newblock A class of ${C}^*$-algebras and topological {M}arkov chains.
\newblock {\em Invent. math.}, 56:251--268, 1980.

\bibitem[Con94]{Con0}
Alain Connes.
\newblock {\em Noncommutative Geometry}.
\newblock Academic Press, Inc., 1994.

\bibitem[Dav96]{Dav96}
Kenneth~R. Davidson.
\newblock {\em $C^*$-Algebras by Example}, volume~6 of {\em Fields Institute
  Monographs}.
\newblock American Mathematical Society, Providence, Rhode Island, 1996.

\bibitem[Ell76]{Ell76}
G.A. Elliott.
\newblock On the classification of inductive limits of sequences of semi-simple
  finite dimensional algebras.
\newblock {\em J. Algebra}, 38:29--44, 1976.

\bibitem[Ell78]{Ell78}
G.A. Elliott.
\newblock On totally ordered groups and {$K_0$}.
\newblock In D.~Handelman and J.~Lawrence, editors, {\em Proc. Ring Theory
  conf.}, volume 734 of {\em Lect. Notes Math.}, pages 1--49, New York, 1978.
  Springer-Verlag.

\bibitem[Gar77]{Gar77}
Martin Gardner.
\newblock Extraordinary nonperiodic tiling that enriches the theory of tiles.
\newblock {\em Scientific American}, pages 110--121, January 1977.

\bibitem[GS89]{GruS89}
B.~Gr\"unbaum and G.C. Shephard.
\newblock {\em Tilings and Patterns}.
\newblock Freeman, New York, 1989.

\bibitem[Lan97]{GLand97}
Giovanni Landi.
\newblock {\em An Introduction to Noncommutative Spaces and their Geometry},
  volume m51 of {\em Lecture Notes in Physics: Monographs}.
\newblock Springer-Verlag, Berlin, Heidelberg, 1997.
\newblock (See also: http://xxx.lanl.gov/abs/hep-th/9701078 ).

\bibitem[Lan98]{Land98}
N.P. Landsman.
\newblock ${C}^*$-{A}lgebras, {H}ilbert ${C}^*$-modules, and {Q}uantum
  {M}echanics.
\newblock http://xxx.lanl.gov/abs/math-ph/9807030, July 1998.

\bibitem[MvN36]{MvN36}
F.J. Murray and J.~von Neumann.
\newblock On rings of operators.
\newblock {\em Ann. Math.}, 37:116--229, 1936.

\bibitem[Pen74]{Pen74}
R.~Penrose.
\newblock The role of aesthetics in pure and applied mathematical research.
\newblock {\em Bull. Inst. Math. Appl.}, 10:266--271, 1974.

\bibitem[Pen78]{Pen78}
R.~Penrose.
\newblock Pentaplexity.
\newblock {\em Eureka}, 39:16--22, 1978.
\newblock Repint: {\it Mathematical Intelligencer} 2(1979), 32--37, and {\it
  Geometrical Combinatorics} F.C.~Holroyd and R.J.~Wilson, eds. Pitman, London,
  1984, pp. 55--65.

\bibitem[Ser58]{Ser57}
J.-P. Serre.
\newblock Modules projectifs et espaces fibr\'es \`a fibre vectorielle.
\newblock {\em S\'eminaire {D}ubreil-{P}isot: alg\`ebre et th\'eorie des
  nombres}, 11(1):531--543, 1957/58.

\bibitem[Sin68a]{Sin68b}
Ya.~G. Sinai.
\newblock Construction of {M}arkov partitions.
\newblock {\em Functional Anal. Appl.}, 2(3):245--253, 1968.

\bibitem[Sin68b]{Sin68a}
Ya.~G. Sinai.
\newblock {M}arkov partitions and {$C$}-diffeomorphisms.
\newblock {\em Functional Anal. Appl.}, 2(1):61--82, 1968.

\bibitem[Swa62]{Swan62}
R.G. Swan.
\newblock Vector bundles and projective modules.
\newblock {\em Trans. Amer. Math. Soc.}, 105:264--277, 1962.

\bibitem[Wal82]{Walt82}
P.~Walters.
\newblock {\em An Introduction to Ergodic Theory}, volume~79 of {\em Graduate
  Texts in Mathematics}.
\newblock Springer-Verlag, New York Heidelberg Berlin, 1982.

\bibitem[WO93]{WeggOl93}
N.E. Wegg-Olsen.
\newblock {\em $K$-Theory and $C^*$-Algebras}.
\newblock Oxford University Press Inc., New York, 1993.

\end{thebibliography}
